\documentclass[9pt,a4paper]{article}
\usepackage[hmarginratio=1:1,vmarginratio=1:1,margin=2cm,top=2cm]{geometry}
\usepackage[utf8]{inputenc}
\usepackage[T1]{fontenc}
\usepackage{textcomp}
\usepackage{graphicx}
\usepackage{subcaption}
\usepackage{amssymb}
\usepackage{float}

\usepackage[numbers]{natbib}
 
\usepackage{caption}
\usepackage{svg}
\usepackage{amsmath}
\usepackage{makecell}
\usepackage{textcomp}
\usepackage{setspace}

\usepackage{lineno}
\usepackage{enumitem}
\usepackage{hyperref}
\usepackage{cleveref}
\usepackage{orcidlink}

\usepackage{authblk}
\usepackage{tabto}
\NumTabs{8}
\usepackage{xr}

\usepackage{tikz}
\usepackage{everypage}
\usepackage{xcolor}

\AddEverypageHook{%
  \begin{tikzpicture}[remember picture, overlay]
    \fill[left color=gray!5,right color=gray!10] 
      ([yshift=0cm]current page.south west) rectangle
      ([xshift=1.2cm,yshift=-0cm]current page.north west);
    \node[rotate=90, anchor=center, text width=\paperheight, align=center, font=\bfseries\small, text=gray]
      at ([xshift=0.6cm]current page.west|-current page.center)
      {This version has been peer-reviewed and accepted for publication by Elsevier}; % Accepted Manuscript (Postprint) - Peer-reviewed version accepted for publication by Elsevier
  \end{tikzpicture}%
}

\usepackage{soul}

\title{\centering A review of ventral hernia biomechanics}

\author[1,2]{Victoria Joppin \orcidlink{0000-0003-2789-6647}}
\author[1]{Catherine Masson \orcidlink{0000-0003-3578-9067}}
\author[2]{David Bendahan \orcidlink{0000-0002-1502-0958}}
\author[1,3]{Thierry Bege \orcidlink{0000-0002-0775-3035}}

\affil[1]{Univ Gustave Eiffel, Aix-Marseille Univ, LBA, F-13016 Marseille, France}
\affil[2]{Aix Marseille Univ, CNRS, CRMBM UMR 7339, Marseille France}
\affil[3]{Department of General Surgery, Aix-Marseille Univ, North Hospital, APHM, Marseille, France}

\newcommand{\captionArticleReviewHerniaone}{
    Definition of the stresses applied on the abdominal wall
}
\newcommand{\figureStressForces}{
\begin{figure}[H]
\centering
\includegraphics[width=0.75\textwidth]{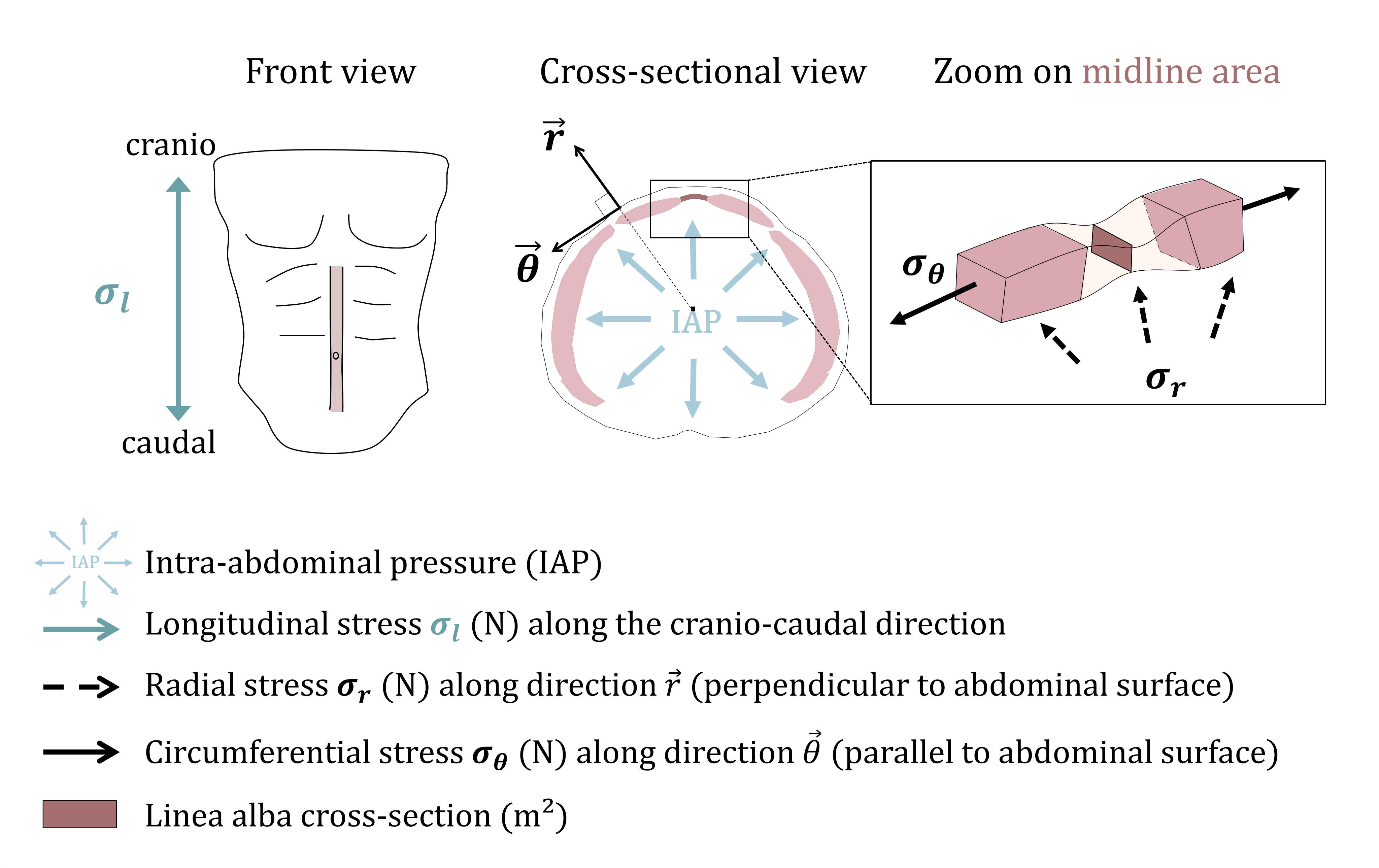}
\captionsetup{justification=centering, format=plain}
\caption[Abdominal wall stress elements]{\captionArticleReviewHerniaone}
\label{fig:stress_and_forces}
\end{figure}
}

\newcommand{\captionArticleReviewHerniatwo}{
    Scheme of different \textit{ex vivo} and \textit{in vivo} experimental tests used for abdomen study.
    \\
    \textbf{\textit{Ex vivo/bench} tests}: Different mechanical tests setup, example of the \guillemotleft Abdoman\guillemotright{} model from Kroese \textit{et al.}, 2017 \cite{kroeseAbdoMANArtificialAbdominal2017} \\
    \textbf{\textit{In vivo} tests}:
    \textbf{a)} Dynamic MRI during contraction
    \textbf{b)} EMG sensor \cite{kychotEnglishNeedleEMG2009}
    \textbf{c)} Ultrasound of diastasis recti \cite{haggstromUltrasonographyDiastasisRecti2018}
    \textbf{d)} MRI (left) and elastography (right) of healthy kidney \cite{mrelastoMRElastographyKidney2023}
}
\newcommand{\figureExpeNumericTests}{
\begin{figure}[H]
\centering
\includegraphics[width=0.85\textwidth]{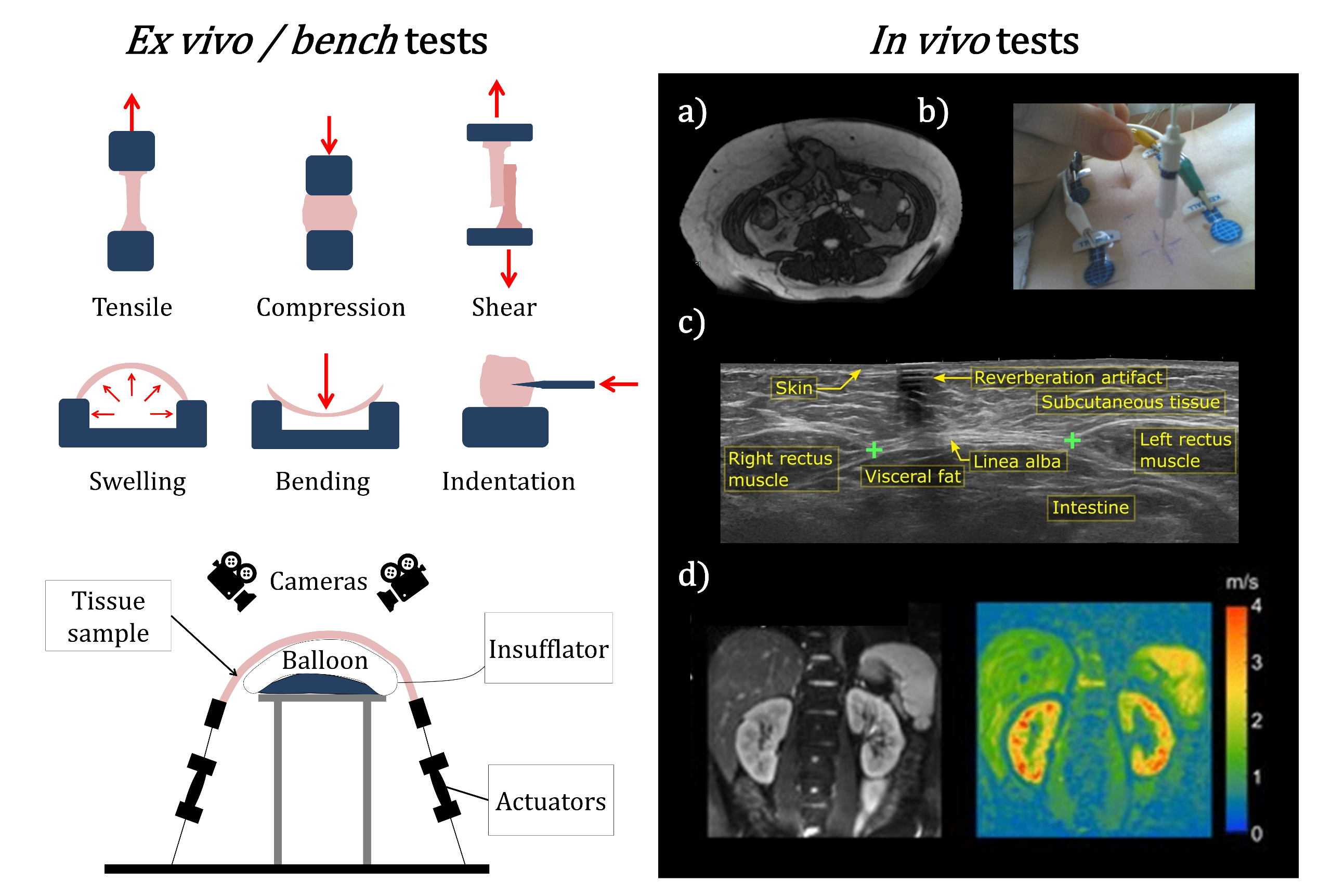}
\captionsetup{justification=centering, format=plain}
\caption[Experimental tests for abdomen study]{\captionArticleReviewHerniatwo}
\label{fig:mechanical_tests_ex_vivo}
\end{figure}
}

\newcommand{\captionArticleReviewHerniathree}{
    \textbf{Legend}: Midline zone of abdominal wall in different stages: \textbf{a)} healthy, \textbf{b)} diastasis and \textbf{c)} hernia occurrence
}
\newcommand{\figureHerniaDefinition}{
\begin{figure}[H]
\centering
\includegraphics[width=0.99\textwidth]{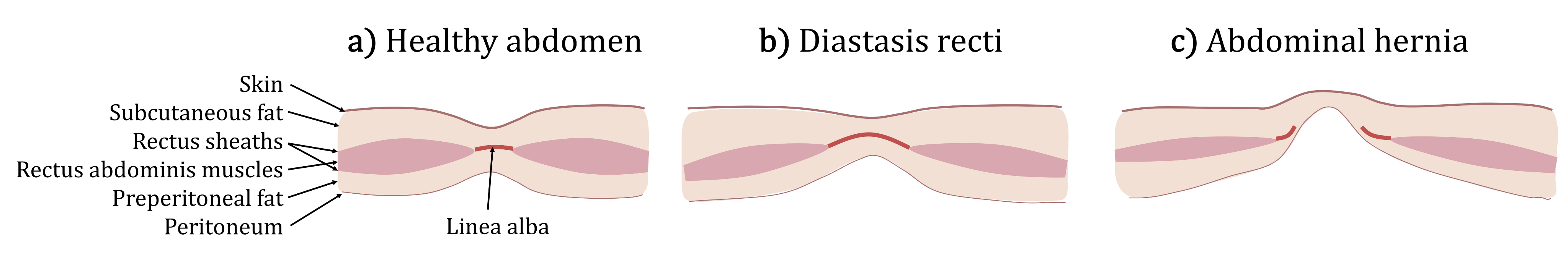}
\captionsetup{justification=centering, format=plain}
\caption[Definition of an abdominal hernia]{\captionArticleReviewHerniathree}
\label{fig:hernia_abdominal_wall}
\end{figure}
}

\newcommand{\captionArticleReviewHerniafour}{
    Scheme of suture with stitch spacing and bite size
}
\newcommand{\figureSutureMeijer}{
\begin{figure}[H]
\centering
\includegraphics[width=0.7\textwidth]{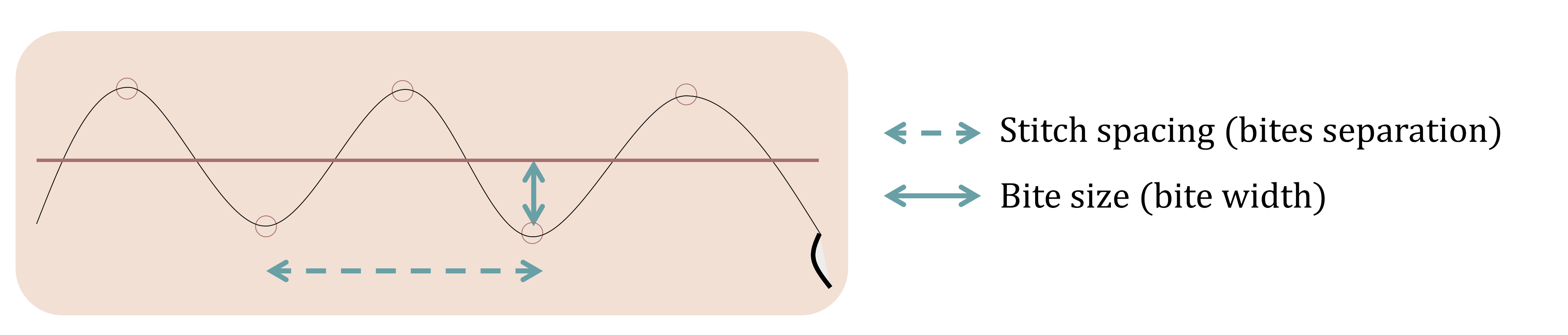}
\captionsetup{justification=centering, format=plain}
\caption[Wound suturing techniques]{\captionArticleReviewHerniafour}
\label{fig:suture_length_Meijer}
\end{figure}
}

\newcommand{\captionArticleReviewHerniafive}{
    Numerical model's displacements obtained during coughing motion in three settings: healthy, with hernia and repaired abdominal walls, adapted from Todros \textit{et al.} \cite{todrosComputationalModelingAbdominal2018}
}
\newcommand{\figureNumericalModelTodros}{
\begin{figure}[H]
\centering
\includegraphics[width=0.8\textwidth]{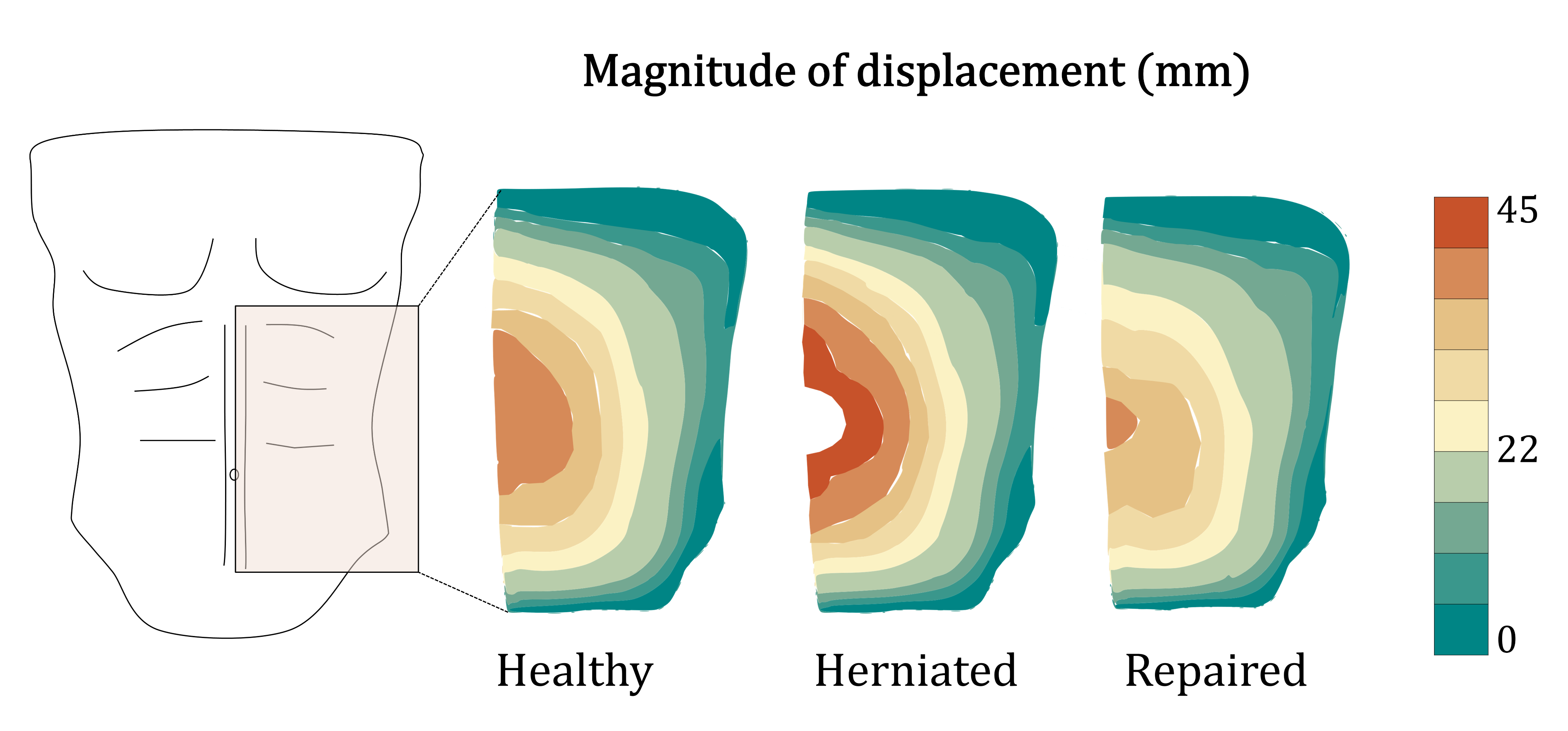}
\captionsetup{justification=centering, format=plain}
\caption[Numerical model of healthy, with hernia and repaired abdominal wall]{\captionArticleReviewHerniafive}
\label{fig:numerical_model_Todros}
\end{figure}
}

\begin{document}

\date{Publication date: August 20th, 2025}

\maketitle

% \tableofcontents

\section*{Authors}

\noindent Victoria Joppin \orcidlink{0000-0003-2789-6647} \tab \tab victoria.joppin@proton.me

\noindent Catherine Masson \orcidlink{0000-0003-3578-9067} \tab catherine.masson@univ-eiffel.fr

\noindent David Bendahan \orcidlink{0000-0002-1502-0958} \tab david.bendahan@univ-amu.fr

\noindent Thierry Bege \orcidlink{0000-0002-0775-3035} \tab \tab thierry.bege@ap-hm.fr

\subsection*{Corresponding author}

\noindent Victoria Joppin \orcidlink{0000-0003-2789-6647} \tab victoria.joppin@proton.me

\noindent Laboratoire de Biomécanique Appliquée, UMRT24 Université Gustave Eiffel - Aix Marseille Université

\noindent Faculté des Sciences Médicales et Paramédicales - Secteur Nord

\noindent 51 Boulevard Pierre Dramard, F-13016 Marseille, France

\vspace{1cm}

% \noindent \textbf{Word count}

% \noindent Abstract: 346

% \noindent Manuscript: 8646

\section*{Abstract}

\noindent Despite advancements in surgical techniques, hernia recurrence rates remain high, underscoring the need for improved understanding of abdominal wall behaviour. While surgeons are aware of many factors contributing to hernia occurrence (e.g obesity, smoking, surgical technique or site infection), it would be of interest to consider it as a biomechanical pathology. Indeed, an abdominal hernia arises from an imbalance between abdominal wall deformability and applied forces. This review article discusses how biomechanics offer a quantitative framework for assessing healthy and damaged tissue behaviour, guiding personalised surgical strategies throughout the pre-, intra-, and post-operative periods.

\vspace{0.2cm}

\noindent The abdominal wall is a dynamic, load-bearing structure, continuously subjected to intra-abdominal pressure and mechanical stress. Its biomechanical properties, including elasticity and resistance to loading forces, dictate its function and response to surgical intervention. The linea alba is the stiffest component experiencing the highest stress, while the abdominal wall’s anisotropic nature influences deformation patterns. Various experimental and computational methods enable biomechanical characterisation.

\vspace{0.2cm}

\noindent Hernias represent mechanical failures at anatomical weak points. While surgeons qualitatively evaluate abdominal wall's biomechanics by estimating deformation and closure forces, functional imaging (elastography, dynamic acquisitions) could provide objective biomechanical insights. Hernia formation alters abdominal wall biomechanics, inducing greater mobility and elasticity.

\vspace{0.2cm}

\noindent Surgical repair fundamentally alters the biomechanics of the abdominal wall. The choice of defect's suturing technique, mesh properties, placement, overlap and fixation methods (e.g. suture, tacks) significantly influence mechanical outcomes. Surgical repair tends to restore physiological biomechanics by re-establishing force transmission and hernia-induced excessive mobility. Suturing techniques, mesh selection and placement influence mechanical outcomes. However, optimal results require implants with mechanical properties mimicking native tissue. Lightweight meshes (<70 $g/m^2$) placed in a retrorectus position, combined with a small-bite suture technique, have been associated with lower recurrence rates and improved post-operative function.

\vspace{0.2cm}

\noindent By bridging biomechanics with surgical practice, this review highlights how mechanical principles shape hernia formation, diagnosis, and repair. A deeper integration of biomechanical principles into surgical decision-making could refine hernia management and lead to patient-specific, mechanics-informed strategies. For surgeons, this knowledge is not just academic - it is practical and can make a difference to patient outcomes.

\noindent \textbf{Keywords}

\noindent Abdominal wall; Hernia repair; Elasticity; Stress; Biomechanics; In vivo

%%%%%%%%%%%%%%%%%%%%%%%%%%%%%%%%%%
\vspace{0.5cm}

\noindent \textbf{Abbreviations}

\noindent LM: lateral muscles (grouping the internal and external obliques and the transversus abdominis); RA: rectus abdominis; MRI: magnetic resonance imaging; CT: computed tomography; IAP: intra-abdominal pressure; IAV: intra-abdominal volume; UTS: ultimate tensile strength; IO: Internal oblique; EO: External oblique; TA: Transversus abdominis; CNN: Convolutional neural network; EHS: European Hernia Society; MAST: mesh-aponeurosis scar tissue; GRIP: Gained Resistance of the Repair to Pressure; COPD: chronic obstructive pulmonary disease; ASA: American Society of Anesthesiologists

%%%%%%%%%%%%%%%%%%%%%%%%%%%%%%%%%%%%%%
\vspace{0.5cm}

\noindent \textbf{Symbols}

\noindent E: Young modulus or elastic modulus (Pa); $\sigma$: stress (Pa)

\noindent $\sigma_{r}$: radial stress (Pa); $\sigma_{\theta}$: circumferential stress (Pa)

\noindent $\sigma_{elastic}$: elastic limit (Pa); $\sigma_{rupture}$: tensile strength (Pa)

\vspace{0.1cm}

\noindent $\varepsilon$: strain (-)

\vspace{0.1cm}

\noindent $F$: force (N); $S$: surface ($m^2$)

\noindent $\Delta l$: elongation (m); $l$: tissue length (m); $l_0$: tissue length at rest (m); 

\noindent $C$: abdominal compliance ($mL/mmHg$)

\section{Rethinking abdominal wall physiology in a biomechanical context}

%% ---------------------------------------- %%
\subsection{The abdominal cavity: a pressurised, deformable structure}

\noindent The abdominal wall is a remarkable and complex structure, both anatomically and biomechanically. Composed of different layers of muscle and connective tissue, it operates as a multi-material composite system \cite{greviousStructuralFunctionalAnatomy2006}. It encloses the \textbf{abdominal cavity}, which is a pressurised and deformable structure that undergoes continuous deformation during daily activities \cite{campbellVariationsIntraabdominalPressure1953}. It is bounded superiorly by the diaphragm, inferiorly by the pelvic cavity, posteriorly by the spine and back muscles, and anterolaterally by the abdominal wall \cite{joyceAbdominalCavityAnatomy2022}. Its biomechanics are determined by a \textbf{complex interplay} of muscle contractions, intra-abdominal pressure (IAP), and structural constraints imposed by surrounding tissues and organs \cite{todrosNumericalModellingAbdominal2020}. 

\vspace{0.2cm}

\noindent This musculo-aponeurotic complex plays a critical role in posture, trunk movement and load-bearing \cite{hodgesVivoMeasurementEffect2001}. The antero-lateral abdominal wall is highly \textbf{deformable} and is influenced by muscular contractions that actively modulate its \textbf{elasticity} \cite{tranAbdominalWallMuscle2016,pavanEffectsMuscularContraction2019}. This deformation is constrained by attachments to relatively immobile structures such as the spine, and mobile structures such as the ribs and diaphragm \cite{detroyerMechanicsRespiratoryMuscles2011}. Abdominal deformation occurs continuously following the rhythm of breathing \cite{campbellVariationsIntraabdominalPressure1953}, and more markedly during abdominal muscle \textbf{contraction} such as coughing, defecation, Valsalva maneuver or trunk movement \cite{jourdanDynamicMRIQuantificationAbdominal2022}. Connective tissues such as fascia and aponeuroses maintain structural integrity and transmit forces \cite{wilkeNotMerelyProtective2018}.

\vspace{0.2cm}

\noindent The abdominal wall \textbf{regulates} the IAP, which is essential for supporting the viscera and preventing herniation \cite{malbrainResultsInternationalConference2006,novakIntraabdominalPressureCorrelates2021}. The abdominal wall can deform \textbf{passively} due to an external impact or the pressure exerted by abdominal content \cite{malbrainResultsInternationalConference2006}, or \textbf{actively} through muscle contraction \cite{misuriVivoUltrasoundAssessment1997}. The relationship between IAP and abdominal deformation has been shown to be highly individual-specific, with IAP correlating more closely with rectus muscle movement during breathing and lateral muscle movement during coughing and the Valsalva maneuver \cite{joppinBetterUnderstandingAbdominal2025}. This \textbf{dynamic adaptability} is critical for functions such as urination, defecation, and childbirth.

\noindent During inhalation, the diaphragm acts as a mobile piston, descending and compressing the abdominal cavity \cite{wilsonEffectBreathingIntraabdominal1933}, increasing IAP without abdominal wall contraction \cite{campbellVariationsIntraabdominalPressure1953,hodgesVivoMeasurementEffect2001}. Active muscle contractions, like Valsalva maneuver, exert significant aponeurotic \textbf{tension} on rectus sheaths \cite{lyonsMechanicalCharacterisationPorcine2014,luoStudyRelationshipsDiastasis2023} and linea alba \cite{jourdanCombinedExperimentalNumerical2025}. The lateral muscles move inward, narrowing the waist, while the rectus abdominis pushes outward, increasing the antero-posterior diameter as demonstrated by MRI scans \cite{jourdanDynamicMRIQuantificationAbdominal2022}, image correlation \cite{todros3DSurfaceImaging2019} and numerical studies \cite{todrosNumericalModellingAbdominal2020}. Strong abdominal muscles significantly raise IAP \cite{misuriVivoUltrasoundAssessment1997,jourdanExplorationsCombineesBiomecaniques2021}, with the transversus abdominis being the primary generator of IAP \cite{axerCollagenFibersLinea2001a,misuriVivoUltrasoundAssessment1997}. Rapid contractions cause \textbf{spikes in IAP} \cite{soucasseBetterUnderstandingDaily2022,blazekSystematicReviewIntraabdominal2019}, increasing the stiffness of rectus sheaths \cite{benabdelounisEffectTwoLoading2013} and the risk of hernia \cite{lienContractionAbdominalWall2015}.

%% ---------------------------------------- %%
\subsection{Biomechanics notions applied to abdominal wall}

\noindent Biomechanical characterisation of abdominal wall tissues provides an \textbf{objective framework} for understanding how a tissue responds to an applied force, which is essential for understanding its behaviour in different pathophysiological situations. The abdominal wall biomechanics is complex because its overall response is not merely the sum of its components \cite{hernandezMechanicalHistologicalCharacterization2011}, but rather results from their intricate interactions, creating a dynamic system that responds to mechanical forces in ways that are not yet fully understood.

\noindent In this context, stress and strain are fundamental concepts for quantifying the mechanical behaviour of tissues under applied loads and displacements, such as normal physiological or pathological conditions (hernia, trauma). A thorough introduction to these principles is essential to highlight their importance in hernia research and surgical decision-making. For surgeons, this knowledge is not just academic - it is practical and can make a difference to patient outcomes.

%% ---------------------------------------- %%
\subsubsection{Elasticity, the key parameter of the abdominal wall behaviour}

\noindent \textbf{Stress} ($\sigma$) is defined as force (F) per unit area (S) as shown in \autoref{eq:eq_stress}. The concept of stress is of paramount importance as forces cannot fully describe the failure of a material. Indeed, a given force applied to a complex structure such as the abdominal wall will not result in a uniform stress distribution throughout the whole structure. The thinner the structure, the smaller is the area (S) on which the force is applied, the higher the stress. This explains why the linea alba, approximately 1.5 mm thick \cite{axerCollagenFibersLinea2001} and located in the midline area, is a key area of biomechanical interest. As it is the stiffest abdominal structure \cite{hollinskyMeasurementTensileStrength2007,tranContributionSkinRectus2014}, the mechanical stress will be mainly concentrated in this area, playing a central role in the mechanical response of the abdomen \cite{hernandez-gasconUnderstandingPassiveMechanical2013}. As well, it has been shown that anterior rectus sheaths play an important role in the abdominal wall structural response \cite{tranContributionSkinRectus2014}. IAP serves as a valuable indicator of the mechanical stresses applied to the abdominal wall \cite{konerdingMaximumForcesActing2011,novakIntraabdominalPressureCorrelates2021,karkhanehyousefiPatientspecificComputationalSimulations2023}. 

\noindent The abdominal wall is subject to complex mechanical stresses which can be decomposed into three components according to their direction, as shown in \autoref{fig:stress_and_forces}. The circumferential stress ($\sigma_{\theta}$) is parallel to the abdominal wall surface and stretches the wall transversely, particularly at weak points such as the linea alba. Longitudinal stress ($\sigma_l$) increases when the abdominal wall is stretched and compressed in the cranio-caudal direction, for example during cranio-caudal flexion or extension activity. The radial stress ($\sigma_r$) is applied to the inner surface, perpendicular to the abdominal wall surface. 

\begin{equation}
    \sigma = \frac{F}{S} \quad \text{(Pa = $\frac{N}{m^2}$)}
    \label{eq:eq_stress}
\end{equation}

\noindent A material can deform in response to applied loads in one or more directions. Elongation $\Delta l$ relates to the change in tissue dimension in one direction, called $l$, with respect to its initial size. It is quantified by $\Delta l = l - l_0$ (m).

\noindent \textbf{Strain} ($\varepsilon$) is a dimensionless measure of internal tissue deformation in a given direction, normalised by its initial dimension (\autoref{eq:eq_strain}). It represents the relative elongation of a material, its ability to stretch. \textbf{Elastic strain} is fully recoverable, meaning that the tissue deforms under stress and return to its original size and shape once the stress is removed.

\begin{equation}
    \varepsilon = \frac{l - l_0}{l_0}
    \label{eq:eq_strain}
\end{equation}

%%% FIGURE CALLING %%%
\figureStressForces

\noindent At the heart of the abdominal wall's biomechanical behaviour lies the concept of tissue \textbf{elasticity}. Elasticity can be considered as the tissue's \guillemotleft springiness\guillemotright{}, with the \textbf{elastic modulus} $E$ quantifying stiffness. Also called Young's modulus, it represents the slope of the linear elastic region of the stress-strain curve, governed by the Hooke's law \autoref{eq:eq_Young} \cite{roederChapter3Mechanical2013}. Higher $E$ values would indicate less elastic, stiffer structures. According to \autoref{eq:eq_Young}, for the same amount of strain $\Delta \varepsilon$, stiffer tissues (higher $E$) will experience greater amount of stress $\Delta \sigma$.

\begin{equation}
    E = \frac{\Delta \sigma}{\Delta \varepsilon} (Pa)
    \label{eq:eq_Young}
\end{equation}

\noindent Elasticity is not a uniform property in all tissues; it varies significantly depending on the type of tissue and its composition \cite{deekenMechanicalPropertiesAbdominal2017}. For example, muscle tissue is highly elastic and capable of significant deformation \cite{cardosoExperimentalStudyHuman2012}, whereas connective tissue is stiffer and more resistant to stretching \cite{astrucCharacterizationAnisotropicMechanical2018}.

\noindent Understanding these mechanical characteristics is key to appreciating how the abdominal wall adapts to mechanical demands and why certain tissues are more prone to injury or failure under excessive loads.  This concept of elasticity also has direct clinical implications in abdominal wall repair, where a mismatch in elasticity between native tissue and surgical materials can lead to complications such as recurrent herniation, chronic pain, or impaired abdominal wall function. This is discussed in more detail in \autoref{section_3}.

\noindent The mechanical properties of materials should always be characterised using stress $\sigma$ and strain $\varepsilon$, instead of force and displacement, in order to avoid the influence of specimen size and facilitate interstudy comparison.

\vspace{0.2cm}

\noindent Abdominal \textbf{compliance} $C$ is another quantity that reflects the overall abdomen's elasticity and is defined as the ratio of the intra-abdominal volume (IAV) change to the IAP change \cite{kirkpatrickIntraabdominalHypertensionAbdominal2013} (\autoref{eq:eq_compliance}). More compliant abdominal walls expand more easily under IAP. Normal compliance is between 250 and 450 mL/mmHg \cite{malbrainRoleAbdominalCompliance2014}.

\begin{equation}
    C = \frac{\Delta IAV}{\Delta IAP} (\frac{mL}{mmHg})
    \label{eq:eq_compliance}
\end{equation}

%% ---------------------------------------- %%
\vspace{0.3cm}

\noindent The abdominal wall has an \textbf{anisotropic} nature, meaning it does not stretch equally in all directions. Its mechanical properties differ depending on the \textbf{direction} of the applied load. Many studies on excised tissue samples have shown that the linea alba \cite{gravelAnisotropyHumanLinea2005,forstemannForcesDeformationsAbdominal2011}, rectus sheaths \cite{hollinskyMeasurementTensileStrength2007} and full-thickness abdominal wall \cite{jungeElasticityAnteriorAbdominal2001,astrucCharacterizationAnisotropicMechanical2018}
are stiffer in the transverse direction (medial-lateral axis) than in the longitudinal direction (cranio-caudal axis). For the lateral muscles, it seems to depend on the muscles considered, with the anisotropy of the composite being less pronounced than that obtained when the muscles were studied separately \cite{hernandezMechanicalHistologicalCharacterization2011}.

\noindent This has also been shown in studies of the entire torso, the midline deforms twice as much in the longitudinal direction as in the transverse direction at the umbilicus level \cite{szymczakInvestigationAbdomenSurface2011}. 
It has also been shown that the abdominal wall stretching capacity depends both on the direction (longitudinal or transverse) and on the location on the abdomen, reaching up to 32\% in upper midline along longitudinal direction, to 7\% in the lower abdomen along transverse direction \cite{smietanskiBiomechanicsFrontAbdominal2012}. As IAP increases, the abdomen's curvature changes more in the longitudinal direction, while the transverse curvature remains relatively stable due to the higher stiffness in this plane \cite{songElasticityLivingAbdominal2006}.

\noindent This anisotropy is due to the orientation of the collagen fibers in the aponeuroses, which are predominantly transverse \cite{axerCollagenFibersLinea2001a,levillainContributionCollagenElastin2016}, making them stiffer along the main fiber orientation \cite{kirilovaExperimentalStudyMechanical2011,martinsMechanicalCharacterizationConstitutive2012,lyonsMechanicalCharacterisationPorcine2014}.

%% ---------------------------------------- %%
\subsubsection{Advanced biomechanical concepts}

\noindent The preceding section has laid the groundwork for understanding the fundamental biomechanical principles, including stress, strain, Young's modulus, compliance and anisotropy. These concepts, rooted in linear elasticity, provide a useful starting point for describing the mechanical behaviour of abdominal tissues. However, they have limitations in characterizing it, assuming a constant relationship between stress and strain (i.e., a constant and fixed Young's modulus). In practice, the abdominal wall is a complex multi-layered structure subjected to large deformations under physiological conditions where linear models become insufficient \cite{smietanskiBiomechanicsFrontAbdominal2012,lubowieckaNovelVivoApproach2022}. Consequently, more advanced constitutive model is needed.

\noindent Experimental studies have consistently shown that abdominal wall tissues exhibit nonlinear stress-strain behaviour. Particularly pronounced in collagen-rich structures like linea alba \cite{forstemannForcesDeformationsAbdominal2011,levillainContributionCollagenElastin2016}, rectus sheath \cite{martinsMechanicalCharacterizationConstitutive2012}, and fascia \cite{kirilovaExperimentalStudyMechanical2011}, but also in abdominal muscles \cite{hernandezMechanicalHistologicalCharacterization2011,cardosoExperimentalStudyHuman2012,krienerMechanicalCharacterizationHuman2023}. Abdominal wall tissues also have a time-dependent mechanical behaviour, called \textbf{viscoelasticity}. It has been shown that for the same strain, high-speed muscle contractions increase tissue stiffness and, consequently, the resulting stress \cite{benabdelounisEffectTwoLoading2013}.
Compliance analysis during laparoscopic procedures reveals hyperelastic behaviour, where small pressure increases initially cause large displacements, followed by a plateau \cite{songElasticityLivingAbdominal2006,mcdougallLaparoscopicPneumoperitoneumImpact1994} where more pressure does not increase significantly the working space and only increases stresses applied to the tissues \cite{ottAbdominalComplianceLaparoscopy2019}. This observation has led to the development of \guillemotleft low-impact laparoscopy\guillemotright{} protocols, optimising working pressure to minimise physiological disruption \cite{malbrainResultsInternationalConference2006,beckerComplianceAbdominalWall2017} such as cardiorespiratory changes \cite{casatiCardiorespiratoryChangesGynaecological1997} or intracranial pressure \cite{kamineEffectAbdominalInsufflation2014}. 

\noindent The typical stress-strain curve of abdominal wall tissues features a distinct “toe” region up to ~10–15\% strain, where the tissue stretches without much resistance due to the straightening of wavy collagen fibers \cite{holzapfelBiomechanicsSoftTissue2001,krienerMechanicalCharacterizationHuman2023}. This is followed by a linear stress-strain region governed by the Hooke's law until ~30–40\% strain, where the material deforms proportionally to the applied stress. Beyond this, if the stress exceeds the \textbf{elastic limit} ($\sigma_{elastic}$, or yield strength) of the tissue, it undergoes plastic, i.e. irreversible, deformation and will not fully recover when the load is removed \cite{holzapfelBiomechanicsSoftTissue2001}. If the stress continues to increase and exceeds the \textbf{tensile strength} (or break stress, ultimate tensile strength (UTS), often denoted $\sigma_{rupture}$ or $\sigma_{max}$), the tissue can no longer withstand and resist the stress and begins to fail. Ultimately, complete failure occurs \cite{holzapfelBiomechanicsSoftTissue2001}.

\noindent These different mechanical properties (Young's modulus $E$, $\sigma_{elastic}$, $\sigma_{rupture}$) define the unique \guillemotleft \textbf{mechanical profile}\guillemotright{} of each individual. \autoref{tab:stiffness_abdominal_wall_components} shows the elastic moduli and tensile strength of various abdominal wall components from different studies, highlighting their anisotropic nature. If not specified, the values refer to human tissue.

\begin{table}[H]
\renewcommand{\arraystretch}{2}
\centering
\begin{center}
\resizebox{\columnwidth}{!}{%
\begin{tabular}{ |c|c|c| } 
\hline
\textbf{} & \textbf{Young modulus $E$ (MPa)} & \textbf{Tensile strength $\sigma_{rupture}$ (MPa)} \\
\hline
\textbf{Component / Anatomical plane} & \begin{tabular}{p{3cm}|p{3cm}}
    \centering Transverse & \centering Sagittal
\end{tabular} & \begin{tabular}{p{3cm}|p{3cm}}
    \centering Transverse & \centering Sagittal
\end{tabular} \\
\hline
\makecell{\textbf{Linea alba} \\ } & \begin{tabular}{p{3cm}|p{3cm}} 
    \centering \makecell{0.47** \cite{forstemannForcesDeformationsAbdominal2011} \\ 4.80* \cite{astrucCharacterizationAnisotropicMechanical2018} \\ 72* \cite{cooneyUniaxialBiaxialTensile2016} \\ 335** \cite{cooneyUniaxialBiaxialTensile2016}} &
    \centering \makecell{0.19** \cite{forstemannForcesDeformationsAbdominal2011} \\ 1.20* \cite{astrucCharacterizationAnisotropicMechanical2018} \\ 8* \cite{cooneyUniaxialBiaxialTensile2016} \\ 23**\cite{cooneyUniaxialBiaxialTensile2016}} % 2016 Cooney: human, excludes the toe region // Astruc: human, Uniaxial E0, tangent modulus for small deformations - E' tangent modulus for large deformations. Here: E' // Forstemann 2011 5 kPa = 37.5 mmHg (extrapolation) Biaxial E'' 0.47 MPa
\end{tabular} & 
\begin{tabular}{p{3cm}|p{3cm}}
    \centering 10* \cite{hollinskyMeasurementTensileStrength2007} & \centering 4.50* \cite{hollinskyMeasurementTensileStrength2007} % Hollinsky: human, Uniaxial, In the epigastric region the linea alba ruptured at a mean horizontal load of 10.0 (SD 3.4) N/mm^2 and at a mean vertical load of 4.5 (SD 2.0) N/mm^2 in the linea alba
\end{tabular} \\
\hline
\makecell{\textbf{Anterior rectus sheath aponeurosis}} & \begin{tabular}{p{3cm}|p{3cm}} % Young modulus
    \centering 11.00* \cite{astrucCharacterizationAnisotropicMechanical2018} & \centering 1.20* \cite{astrucCharacterizationAnisotropicMechanical2018} % E tangent modulus for large deformations
\end{tabular} & % Tensile strength
\begin{tabular}{p{3cm}|p{3cm}}
    \centering \makecell{8.10* \cite{hollinskyMeasurementTensileStrength2007} \\ 12.19* (Rats) \cite{anurovBiomechanicalCompatibilitySurgical2012}} & %  $\frac{32.9 N/cm}{0.27mm} = 12.19$*
    \centering \makecell{3.40* \cite{hollinskyMeasurementTensileStrength2007} \\ 4.89* (Rats)  \cite{anurovBiomechanicalCompatibilitySurgical2012}} % $\frac{13.2 N/cm}{0.27mm} = 4.89$*
    % Anurov 2012: rats breaking strength, The anterior layer thickness was, on average, 0.27 ± 0.05 mm. Formula (MPa): xx (N/cm) * 0.1 / thickness (mm) // Hollinsky: Human cadavers Uniaxial, Epigastric anterior rectus sheath Horizontal 8.1 N/mm^2, Vertical 3.4
\end{tabular} \\
\hline
\makecell{\textbf{Posterior rectus sheath aponeurosis}} & \begin{tabular}{p{3cm}|p{3cm}} % Young modulus
    \centering 15* \cite{astrucCharacterizationAnisotropicMechanical2018} & \centering 0.82* \cite{astrucCharacterizationAnisotropicMechanical2018} % E tangent modulus for large deformations
\end{tabular} &
\begin{tabular}{p{3cm}|p{3cm}}
\centering \makecell{5.60* \cite{hollinskyMeasurementTensileStrength2007} \\ 9.44* (Rats) \cite{anurovBiomechanicalCompatibilitySurgical2012}} & % $\frac{8.5 N/cm}{0.09mm} = 9.44$*
\centering \makecell{1.90* \cite{hollinskyMeasurementTensileStrength2007} \\ 1.56* (Rats) \cite{anurovBiomechanicalCompatibilitySurgical2012}} % $\frac{1.4 N/cm}{0.09mm} = 1.56$*
 % Anurov 2012: Rats uniaxial breaking strength, the posterior layer thickness was much smaller, being 0.09 ± 0.03 mm // // Hollinsky human : Uniaxial, Epigastric posterior rectus sheath Horizontal 5.6 N/mm^2, Vertical 1.9
\end{tabular} \\
\hline
\makecell{\textbf{Fascia transversalis}} & \begin{tabular}{p{3cm}|p{3cm}}
    \centering ~2* \cite{kureshiMatrixMechanicalProperties2008} & \centering ~1.50* \cite{kureshiMatrixMechanicalProperties2008}
\end{tabular} & 
\begin{tabular}{p{3cm}|p{3cm}}
    \centering ~0.7* \cite{kureshiMatrixMechanicalProperties2008} & \centering ~1* \cite{kureshiMatrixMechanicalProperties2008} % Kureshi: Human cadavers 20 herniated fascia transversalis HTF, 4 controls (non-HFT). uniaxial test For NHTF, no numeric values on the text but read on the Fig5...
\end{tabular} \\
\hline
\makecell{\textbf{Rectus abdominis muscle}  \\ (fibers direction*) \cite{cardosoExperimentalStudyHuman2012}} & 0.52 & 0.23 \\
\hline
\makecell{\textbf{External oblique} \\ (fibers direction*) \cite{cardosoExperimentalStudyHuman2012} } & 1.00 & 0.57 \\
\hline
\makecell{\textbf{Internal oblique} \\ (fibers direction*) \cite{cardosoExperimentalStudyHuman2012}} & 0.65 & 0.39 \\
\hline
\makecell{\textbf{Transverse abdominis} \\ (fibers direction*) \cite{cardosoExperimentalStudyHuman2012} } & 1.03 & 0.73 \\
\hline
\makecell{\textbf{Abdominal skin} \\ Shear wave elastography \cite{yangDeterminationNormalSkin2018}} & 0.0095 & - \\ % Yang: Human Mean skin thickness and elastic modulus values from 90 healthy volunteers were evaluated with B-mode ultrasonography and shear wave elastography (SWE) 9.5 kPa for the abdominal wall
\hline
\makecell{\textbf{\textit{In vivo} full abdominal wall} \\ \cite{songMechanicalPropertiesHuman2006}} & \begin{tabular}{p{3cm}|p{3cm}}
    \centering 0.042 & \centering 0.022 % Song in vivo (more biaxial) until 12 mmHg = 1.6 KPa
\end{tabular} & - \\
\hline
\end{tabular}
}
\end{center}
\captionsetup{justification=centering, format=plain}
\caption[Stiffness of abdominal wall components]{Young modulus $E$ and tensile strength of different abdominal wall components according to the considered direction \\ NB: If not specified, the values refer to human tissue. \\ * indicate uniaxial tests, ** indicate biaxial tests}
\label{tab:stiffness_abdominal_wall_components}
\end{table}

%% ---------------------------------------- %%
\subsection{Tools for biomechanical analysis: experiments and simulations}

\noindent \textbf{Biomechanical testing} assesses the mechanical properties of biological tissues or implants. These tests vary based on their methodology, environment, and clinical relevance, with a combination of approaches often needed for a comprehensive understanding of tissue mechanics.

\vspace{0.2cm}

\noindent \textbf{\textit{Ex vivo}} uses freshly excised human or animal tissue to evaluate tensile strength, elasticity, and mesh integration under controlled conditions. Experiments include tensile, compression, shear, and indentation tests (\autoref{fig:mechanical_tests_ex_vivo}) at multiple scales, from isolated tissue aponeuroses to full abdominal wall structure. Such tests are particularly valuable for investigating mechanical properties that may influence surgical outcomes, such as tensile strength and elasticity, or for assessing mesh integration into explanted tissues. Tensile tests can be uni-axial (one direction of loading) or bi-axial, as abdominal wall behaves differently with two \cite{cooneyUniaxialBiaxialTensile2016} or various \cite{jungeElasticityAnteriorAbdominal2001,gravelAnisotropyHumanLinea2005} directions of loading due to anisotropy. \textit{Ex vivo} tests can be performed \textit{in situ}, where tissues remain in their anatomical location (e.g., cadaveric studies of abdominal deformation \cite{leruyetDifferencesBiomechanicsAbdominal2020}) or \textit{ex situ} where samples are tested outside their original environment (e.g., harvested tissue sample \cite{krienerMechanicalCharacterizationHuman2023}). Bench testing refers to \textit{ex vivo} or synthetic models testing performed on standardised setups to reproduce abdominal wall contraction \cite{siassiDevelopmentDynamicModel2014}, study bio-materials integration \cite{lyonsBiomechanicalAbdominalWall2015} or hernia suture modalities \cite{kroeseAbdoMANArtificialAbdominal2017}.

\vspace{0.2cm}

%% ---------------------------------------- %%
\noindent \textbf{\textit{In vivo}} evaluates biomechanical behaviour in living organisms (humans or animals) under physiological conditions (\autoref{fig:mechanical_tests_ex_vivo}). While medical imaging is widely used and will be detailed further in \autoref{subsection_diagnosis_functional}, digital image correlation measures abdominal wall deformation in real-time using high-speed cameras. For example, it has been used in different \textit{in situ} studies to assess how the abdominal wall deforms under load \cite{lubowieckaNovelVivoApproach2022,songMechanicalPropertiesHuman2006}, revealing significant linea alba displacement \cite{todros3DSurfaceImaging2019} and strain \cite{szymczakInvestigationAbdomenSurface2011} in its central region.

\noindent Additional tools can be used to assess abdominal muscle strength, such as surface electromyography \cite{royComparativeStudyEvaluate2023}, dynamometry to assess changes after surgery \cite{strigardGiantVentralHernia2016,crissFunctionalAbdominalWall2014} or the impact of a prior training \cite{ahmedEffectPreoperativeAbdominal2018}, strain gauge \cite{sanchezarteagaImpactIncisionalHernia2024} or combination of these tools to assess relationship between strength and morphometric factors \cite{garciamorianaEvaluationRectusAbdominis2023}. 

\noindent To complement \textit{in vivo} biomechanical evaluations and to quantify mechanical stress, monitoring IAP is critical. These methods can be direct \cite{cobbNormalIntraabdominalPressure2005} or indirect measurements of IAP, and can be carried out under different conditions (e.g. rest, Valsalva maneuver). Manometry, pressure sensors \cite{soucasseBetterUnderstandingDaily2022, liaoIngestibleElectronicsContinuous2020}, less-invasive tools \cite{tayebiConciseOverviewNoninvasive2021} or indirect measurements \cite{vincentAbdominalWallMovements2023,liDevelopmentFeasibilityCredibility2024} such as strain gauges, ultrasound, or image correlation techniques allow IAP measurement. These measurements provide insight into how IAP influences abdominal wall displacement during physiological activities \cite{joppinBetterUnderstandingAbdominal2025, szepietowskaFullfieldVivoExperimental2023}. 

\noindent Abdominal wall tension assessment complements these measurements by quantifying the force required for defect closure. Intraoperative tension measurement provides patient-specific data, critical since hernia width alone does not reliably predict closure forces \cite{hopeRationaleTechniqueMeasuring2018}, reinforcing the need for direct biomechanical evaluation \cite{tenzelTensionMeasurementsAbdominal2019}. Recent studies correlate closure tension with BMI \cite{millerPhysiologicTensionAbdominal2023} and IAP \cite{novakIntraabdominalPressureCorrelates2021}.

\vspace{0.2cm}

%% ---------------------------------------- %%
\noindent \textbf{Theoretical models} use mathematical equations to describe biomechanical behaviour, such as Laplace's law for abdominal wall mechanics or analytical models of suture tension and wound closure forces \cite{dudleyLayeredMassClosure1970,konerdingMaximumForcesActing2011}.

\vspace{0.2cm}

\noindent \textbf{Numerical models} have emerged as a complementary tool to experimental testing, overcoming the challenge of capturing the interplay of multiple factors such as tissue heterogeneity, dynamic loading and patient-specific variations. Although invaluable, experiments are often time-consuming, expensive and ethically constrained. Numerical models, particularly finite element analysis (FEA) \cite{zienkiewicz1StandardDiscrete2025}, simulate complex abdominal wall mechanics alone \cite{hernandez-gasconUnderstandingPassiveMechanical2013} and its interaction with a surgical implant \cite{hernandez-gasconComputationalFrameworkModel2014}.

\noindent \textbf{Non-linear models} are empirical or theoretical mathematical descriptions of physical behaviour employed to fit experimental data. They enable the extraction of parameters characterizing tissue properties, useful to inform theoretical and computational models \cite{lohrIntroductionOgdenModel2022}.
Hyperelastic models such as the Neo-Hookean, Mooney-Rivlin, Ogden, and Fung-type models are commonly used for soft tissues \cite{holzapfelBiomechanicsSoftTissue2001}. Hernández-Gascón \textit{et al.} applied a Neo-Hookean formulation to model the diaphragm and pelvis of their numerical model of abdomen \cite{hernandez-gasconUnderstandingPassiveMechanical2013}, while an Ogden model was employed by Lyons \textit{et al.} for the linea alba matrix \cite{lyonsMechanicalCharacterisationPorcine2014}. Similarly, Grasa \textit{et al.} and Pavan \textit{et al.} modelled the active and passive mechanical properties of abdominal muscles to study the effects of intra-abdominal pressure and muscle activation \cite{grasaActiveBehaviorAbdominal2016,pavanEffectsMuscularContraction2019}.

\noindent Furthermore, by imposing experimental pressures or displacements to a numerical model, \textbf{inverse modelling} allows to adjust the material parameters and calibrate the model's constants to match the simulated deformations with experimental response \cite{kauerInverseFiniteElement2002}. This has been done using indentation tests \cite{remusSoftTissueMaterial2024}, insufflation with image correlation \cite{simon-allueMechanicalCharacterizationAbdominal2017} and tensile tests \cite{cooneyUniaxialBiaxialMechanical2015}.

\noindent Numerical models allow to predict stress and deformation distribution \cite{pacheraNumericalInvestigationHealthy2016,jourdanNumericalInvestigationFinite2024}, simulate muscle contraction \cite{grasaActiveBehaviorAbdominal2016,todrosNumericalModellingAbdominal2020}, and implant behaviour under varying IAP conditions \cite{zamkowskiBiomechanicalCausesFailure2023} such as the influence of implant's anisotropy \cite{hernandez-gasconMechanicalResponseHerniated2013}, implant fixation \cite{guerinImpactDefectSize2013,todrosBiomechanicsSurgicalMesh2020}, implant placement \cite{karrechBiomechanicalStabilityHerniadamaged2023}, and the resulting abdominal wall mobility \cite{hernandez-gasconMechanicalBehaviourSynthetic2011,heNumericalMethodGuiding2020}. 

\noindent These models allow for \guillemotleft what-if\guillemotright{} scenarios such as the effect of aponeurosis thickness or stiffness \cite{jourdanNumericalInvestigationFinite2024}, without requiring physical experiments. Importantly, numerical models can be tailored to the patient, offering the potential for specific predictions and personalised surgical planning. However, numerical models will never perfectly reproduce the \guillemotleft real life\guillemotright{} as they rely on simplifying assumptions, such as uniform material properties, to remain computationally feasible. These simplifications affect prediction accuracy, requiring extensive validation before clinical application \cite{VV40AssessingCredibilityComputational}.

%%% FIGURE CALLING %%%
\figureExpeNumericTests

%% ---------------------------------------- %%
\subsection*{Key points}
\addcontentsline{toc}{subsection}{Key points}
\begin{itemize}
    \item The abdominal wall is a complex, dynamic structure of muscle and connective tissue that is essential for trunk stability, organ protection and intra-abdominal pressure regulation.
    \item The forces acting on the abdominal wall originate from intra-abdominal pressure, which is increased by muscle activation.
    \item The elasticity of the abdominal wall is a key biomechanical concept linking stress and strain.
    \item The linea alba is the stiffest structure of the abdominal wall and experiences the highest levels of stress.
    \item The abdominal wall is anisotropic, deforming less in the transverse direction than in the sagittal direction, resulting in high circumferential stress.
    \item Biomechanical characterisation methods (\textit{ex vivo}, \textit{in vivo}, numerical) allow the mechanical behaviour of the abdominal wall to be characterised.
\end{itemize}

\section{Abdominal hernia, a biomechanical pathology}

%% ---------------------------------------- %%
\subsection{Biomechanical definition of hernia}

\noindent Despite its robust design, the abdominal wall is susceptible to pathologies such as hernias. While surgeons are familiar with hernias from anatomical and biological perspectives, they are fundamentally multifactorial in origin \cite{urschelEtiologyLateDeveloping1988,kroesePrimaryIncisionalVentral2018}. In addition to patient-specific comorbidities \cite{parkerIdentifyingPredictorsVentral2021}, surgical variables \cite{al-mansourAssociationHerniaspecificProcedural2024}, biological processes \cite{franzBiologyHerniasAbdominal2006}, it would also be of interest to consider hernias as a \textbf{biomechanical pathology}. In fact, hernias result from a biomechanical \textbf{imbalance} between the stresses (related to intra-abdominal pressure and muscle activation) and the tissue's ability to resist them \cite{franzBiologyHerniasAbdominal2006}.

\noindent During physical activity, IAP increases \cite{cobbNormalIntraabdominalPressure2005} and leads to deformation \cite{smietanskiBiomechanicsFrontAbdominal2012} and stress  \cite{vanramshorstNoninvasiveAssessmentIntraabdominal2011} in the abdominal wall. Under normal conditions, the abdominal wall returns to its \textbf{original state} once the stress is removed (\autoref{eq:eq_Young}). However, if the stress exceeds the tissue elastic limit $\sigma_{elastic}$, irreversible damage occurs, such as microtears or diastasis recti \cite{wernerDiastasisRectiAbdominisdiagnosis2019}. It has been studied that the primary mechanism of permanent set in biological tissues is the rupture of proteoglycan bridges between adjacent collagen fibrils, which allows for relative sliding of the fibrils and plastic deformation of the matrix material in the direction of the fibers \cite{gasserRateindependentElastoplasticConstitutive2002,wernerDiastasisRectiAbdominisdiagnosis2019}. However, this condition illustrated in \autoref{fig:hernia_abdominal_wall}-b, could also be caused by the fatigue phenomenon \cite{martinFatigueDamageCollagenous2015,wernerDiastasisRectiAbdominisdiagnosis2019} due to sustained IAP during pregnancy \cite{motaDiastasisRectiAbdominis2015,cavalliPrevalenceRiskFactors2021} or obesity \cite{gueroultLineaAlba3D2024}. If the stress increases until reaching the ultimate tensile strength $\sigma_{rupture}$, the abdominal wall ruptures, thereby resulting in a \textbf{hernia} as illustrated in \autoref{fig:hernia_abdominal_wall}-c. 

\noindent While biological phenomenon should also be taken into account into hernia incidence, this explanation highlights the fundamental biomechanical principles underlying hernia formation.  

%%% FIGURE CALLING %%%
\figureHerniaDefinition

\noindent Hernias are broadly categorised into two types based on their \textbf{etiology} \cite{kroesePrimaryIncisionalVentral2018,verstoepHerniaWidthExplains2021}. This review will focus on ventral hernias, including primary and incisional hernias, occurring in the anterior abdominal wall. Other types of hernias such as inguinal, femoral, Spigelian, parastomal, and hiatal hernias will not be discussed.

\begin{itemize}
    \item \textbf{Primary hernias} occur spontaneously, often due to inherent weaknesses in the abdominal wall, unrelated to previous surgery. 
    \item \textbf{Incisional hernias} occur at the site of a previous surgical incision. A scar in the musculo-aponeurotic complex creates a discontinuity in the mechanical properties and acts as a point of structural weakness where mechanical stress is concentrated \cite{simon-allueMechanicalBehaviorSurgical2018}. Surgical complications or sub-optimal healing can leave weak spots in the tissue, allowing internal organs or tissues to protrude \cite{romainRecurrenceElectiveIncisional2020,parkerIdentifyingPredictorsVentral2021}. Incisional hernias are a common post-operative complication \cite{romainRecurrenceElectiveIncisional2020} and can occur months \cite{burgerIncisionalHerniaEarly2005} or even years after surgery \cite{bhardwajYearOverYearVentralHernia2024,finkIncisionalHerniaRate2014}.
\end{itemize}

\noindent It has been shown that it is \textbf{inappropriate} to \textbf{pool} these two types of hernia in studies, as their patient and hernia characteristics \cite{kroesePrimaryIncisionalVentral2018}, surgical management \cite{stirlerLaparoscopicRepairPrimary2014,kurianLaparoscopicRepairPrimary2010}, complication rates \cite{kockerlingPooledDataAnalysis2015}, chronic pain and recurrence rates \cite{parkerIdentifyingPredictorsVentral2021,kroesePrimaryIncisionalVentral2018} \textbf{differ} significantly.

\noindent Consequently, the \textbf{European Hernia Society} has developed separate classifications and guidelines for each type of hernia \cite{muysomsClassificationPrimaryIncisional2009,kroesePrimaryIncisionalVentral2018}. In addition to comorbidities and wound healing specific to the patient, the width of the hernia appears to be a key factor influencing outcomes, rather than whether it is primary or incisional \cite{verstoepHerniaWidthExplains2021,juvanyResultsProspectiveCohort2022}. Indeed, larger hernias being associated with higher risk of early complications \cite{al-mansourAssociationHerniaspecificProcedural2024}, readmission \cite{helgstrandNationwideProspectiveStudy2013}, recurrence rates \cite{linRiskFactorsRecurrence2024} and increased stress on the mesh \cite{guerinImpactDefectSize2013,simon-allueProsthesesSizeDependency2016}. However, incisional hernias tend to be larger due to their etiology and the impact of defect size is still controversial. Some studies have found no correlation between defect size and recurrence risk \cite{luijendijkComparisonSutureRepair2000, porukEffectHerniaSize2016}, while other have conclude that defect size is a risk factor \cite{hesselinkEvaluationRiskFactors1993,al-mansourAssociationHerniaspecificProcedural2024}.

%% ---------------------------------------- %%
\subsection{Diagnosis of hernias: based on functionality using medical imaging} \label{subsection_diagnosis_functional}

\noindent Hernia diagnosis relies primarily on physical examination \cite{gutierrezdelapenaValueCTDiagnosis2001,baucomProspectiveEvaluationSurgeon2014}, where surgeons detect a bulge that expands with exertion, emphasising the dynamic nature of hernias. Mechanical phenomena are qualitatively assessed by observing abdominal wall deformation during coughing or estimating the force required to close a defect. However, those assessments lack objective quantification. Medical imaging complements clinical evaluation by providing quantitative data for diagnosis and treatment planning.

\vspace{0.2cm}

\noindent \textbf{Computed tomography (CT)} remains the gold standard para-clinical tool for hernia detection \cite{muysomsEuropeanHerniaSociety2015} and  post-operative follow-up \cite{gutierrezdelapenaValueCTDiagnosis2001,gossiosValueCTLaparoscopic2003}. It provides high-resolution anatomical details, including defect size or hernia sac volume \cite{duCTmeasuredHerniaParameters2023}, and additional metrics such as abdominal volume, visceral and subcutaneous fat \cite{wintersPreoperativeCTScan2019,bellioPreoperativeAbdominalComputed2019}, and muscle dimensions \cite{jourdanAbdominalWallMorphometric2020,duCTmeasuredHerniaParameters2023} which could refine this morphometric evaluation. However, standard radiological reports often omit these important features \cite{kushnerIdentifyingCriticalComputed2021} and are not always associated with a reliable diagnosis of ventral hernia recurrence \cite{holihanUseComputedTomography2016}. CT is also poor at accurately identifying the type of previous mesh \cite{messerCanSurgeonsAccurately2024}. While traditionally used for anatomical assessment, more functional pre-operative CT imaging has been shown to significantly improve the understanding of ventral hernia defects, thereby facilitating the development of efficient surgical plans \cite{readImagingInsightsAbdominal2022}. The use of Valsalva maneuver is a biomechanically-based diagnosis. By increasing the IAP, it amplifies hernia protrusion \cite{plumbContemporaryImagingRectus2021}, mimicking real-life stress conditions \cite{vossAutomatedIncisionalHernia2020}, improving defect assessment \cite{bellioPreoperativeAbdominalComputed2019,jaffeMDCTAbdominalWall2005} and patient-specific compliance evaluation \cite{kallinowskiCTAbdomenValsalva2019}. While continuous dynamic CT scans are not feasible due to radiation exposure, alternative imaging modalities such as ultrasound and MRI may provide valuable pre- or post-operative information \cite{readImagingInsightsAbdominal2022,dallaudiereDynamicMagneticResonance2024}.

\vspace{0.2cm}

\noindent \textbf{Ultrasound} provides real-time bedside assessment of the hernia, making it an effective alternative to CT for pre-operative classification and measurement of the abdominal wall defect or hernia sac volume \cite{qiuMeasurementsAbdominalWall2023}. Ultrasound has also been shown to be effective in detecting early signs of recurrence, such as a large distance between the rectus abdominis muscles \cite{harlaarDevelopmentIncisionalHerniation2017,wangDeepLearningbasedApproach2023}. 
Shear-wave elastography, a specific ultrasound technique, can quantify tissue stiffness (\autoref{eq:eq_Young}) and is effective in assessing anisotropic properties in healthy subjects \cite{vergariEffectBreathingVivo2023} and hernia-induced changes in shear wave propagation speed within abdominal muscles \cite{wangUseShearWave2020}. It has also been used to show that diastasis recti is associated with reduced stiffness and increased distortion of the linea alba during abdominal crunches \cite{beamishDifferencesLineaAlba2019}. Furthermore, ultrasound can reliably detect and quantify implanted mesh stiffness \cite{chaudhryCharacterizationVentralIncisional2017}.

\vspace{0.2cm}

\noindent \textbf{Magnetic resonance imaging (MRI)} is much less commonly used in clinical practice than CT scan or ultrasound, despite its excellent soft tissue contrast \cite{plumbContemporaryImagingRectus2021} and ability to detect muscle atrophy, denervation \cite{rodriguez-acevedoFunctional3DVRImaging2020}, and interaction of abdominal structures \cite{randallNovelDiagnosticAid2017}. Dynamic MRI, or cine-MRI, extends its utility by capturing real-time movement and deformation of the abdominal wall during breathing or exertion \cite{blanchardDynamicLiftingTask2016,jourdanDynamicMRIQuantificationAbdominal2022}. It also supports various post-operative assessments, such as abdominal adhesions by quantifying movement restrictions \cite{randallNovelDiagnosticAid2017,yaseminAssessmentDiagnosticEfficacy2020}, evaluating the effect of abdominal binders \cite{miyamotoFastMRIUsed2002}, as well as monitoring mesh positioning, integration, migration or folding \cite{fischerFunctionalCineMRI2007,ciritsisTimeDependentChangesMagnetic2014}. Magnetic resonance elastography (MRE) further enables non-invasive tissue stiffness mapping by analysing the propagation of mechanical waves, offering insights into muscle and connective tissue elasticity.

\vspace{0.3cm}

\noindent \textbf{Quantitative analysis} of medical imaging remains underused, although it is crucial for monitoring hernia progression and assessing pre-operative and post-operative conditions. Artificial intelligence (AI)-based methods, particularly convolutional neural networks (CNNs) enable precise segmentation of abdominal muscles, fat, and hernia-related structures in CT \cite{westonAutomatedAbdominalSegmentation2019,parkDevelopmentValidationDeep2020}, MRI \cite{kwayAutomatedSegmentationVisceral2021,joppinAutomaticMuscleSegmentation2024}, and ultrasound \cite{wangDeepLearningbasedApproach2023}. Machine learning applications also aid in surgical step recognition, outcome prediction, and complication risk assessment \cite{limaMachineLearningDeep2024}. However, automated abdominal wall analysis remains an emerging field, requiring further development to reach its full clinical potential \cite{borgaMRIAdiposeTissue2018}.

%% ---------------------------------------- %%
\subsection{Impact of a hernia on the mechanical behaviour}

\noindent The mechanical behaviour of the abdominal wall is significantly \textbf{altered} in the presence of a hernia. A hernia increases the \textbf{mobility} of the abdominal wall and amplifies its overall \textbf{elasticity}, resulting in greater displacement and deformation compared to a healthy or surgically repaired wall \cite{todrosComputationalModelingAbdominal2018,podwojewskiMechanicalResponseAnimal2013,podwojewskiMechanicalResponseHuman2014,kallinowskiBiomechanicalInfluencesMeshRelated2021}. Using shear-wave elastography, a higher shear wave speed was reported in patients with incisional hernia compared to healthy controls \cite{wangUseShearWave2020}. Despite this increased elasticity, specific regions such as the oblique muscles, experience fiber stiffening and \textbf{atrophy} \cite{dubayIncisionalHerniationInduces2007}. The internal oblique and transversus abdominis were also found to be significantly \textbf{thinner} in patients with hernia compared to healthy controls \cite{wangUseShearWave2020}. In addition, the lateral abdominal wall shortens \cite{dubayIncisionalHerniationInduces2007}, contributing to the widening of the aponeurotic orifice and increasing stress on the midline.

\vspace{0.2cm}

\noindent The \textbf{strength} of the abdominal muscles - including the rectus abdominis, external and internal obliques - is significantly \textbf{reduced} in patients with incisional hernias \cite{royComparativeStudyEvaluate2023}. Studies using isokinetic dynamometry \cite{strigardGiantVentralHernia2016,garciamorianaEvaluationRectusAbdominis2023} and strain gauge methods \cite{sanchezarteagaImpactIncisionalHernia2024} show a \textbf{negative correlation} between the \textbf{width} of the hernia defect and the functional \textbf{strength} of the abdominal wall.

\vspace{0.2cm}

\noindent Hernia orifice size and hernia sac volume evolve over time and increase significantly during events such as laparoscopic insufflation or voluntary muscle contraction \cite{bellioPreoperativeAbdominalComputed2019,jaffeMDCTAbdominalWall2005,qandeelRelationshipVentralHernia2016}. CT imaging has been used to quantify abdominal metrics at rest and during \textbf{Valsalva} straining \cite{jaffeMDCTAbdominalWall2005,bellioPreoperativeAbdominalComputed2019}. Jaffe \textit{et al.} observed significant \textbf{increases} in the antero-posterior abdominal diameter, \textbf{hernia sac diameter}, and \textbf{width of the fascial defect} during the Valsalva maneuver, with more than half of the hernias becoming more prominent \cite{jaffeMDCTAbdominalWall2005}. Bellio \textit{et al.} reported a 23\% increase in the cross-sectional area of the hernia orifice and an \textbf{83\% increase in the volume of the hernia sac} during Valsalva straining, with great variability depending on the patient and the location of the hernia \cite{bellioPreoperativeAbdominalComputed2019}.

\noindent These temporary changes due to increased IAP can provide valuable insight into the mechanical behaviour of the hernia and its potential for growth over time. Indeed, while some hernias grow progressively, others remain stable. This insufficiently studied phenomenon is likely to be related to underlying biomechanical parameters. Taking these factors into account could help clinicians decide whether surgical intervention is necessary and how to approach treatment.

\vspace{0.2cm}

\noindent Hernias also have an impact on \textbf{IAP}, which tends to be \textbf{lower} in hernia patients, both at rest and during exercise \cite{shafikDirectMeasurementIntraabdominal1997}.
Finally, hernias can have a significant impact on quality of life due to pain and functional impairment \cite{smithHealthrelatedQualityLife2022}. 

%% ---------------------------------------- %%
\subsection*{Key points}
\addcontentsline{toc}{subsection}{Key points}
\begin{itemize}
    \item A hernia is a biomechanical event resulting from an \textbf{imbalance} between \textbf{loading} forces (related to intra-abdominal pressure and muscle activation) and \textbf{tissue resistance} that disrupts the abdominal wall integrity.
    \item Understanding the progressive mechanical deterioration that leads to hernia formation may improve prevention strategies.
    \item Imaging-based \textbf{functional diagnosis} (e.g. shear-wave elastography, dynamic MRI etc.) provides biomechanical insights beyond anatomy.
    \item Abdominal hernias occur mainly at geometric discontinuities (e.g. umbilicus or scars).
    \item Once a hernia has formed, the stress tends to be concentrated on the hernia orifice, promoting its enlargement, but this phenomenon remains to be studied in more detail.
    \item A hernia \textbf{increases} the \textbf{mobility} and the overall \textbf{elasticity} of the abdominal wall and \textbf{reduces intra-abdominal pressure}.
\end{itemize}

\section{Surgical restoration of the abdominal wall} \label{section_3}

\noindent Recent advancements propose the use of \textbf{biomechanically optimised} hernia repairs, tailored to a patient's unique abdominal wall properties \cite{leschSTRONGHOLDFirstyearResults2024,nesselThreeyearFollowupGrip2024}. This approach demonstrated reduced hernia recurrence rates and post-operative pain among patients, underscoring the value of integrating biomechanical assessments into surgical planning, marking a significant step toward personalised surgical care.

%% ---------------------------------------- %%
\subsection{Mechanical benefits of sutures}

\noindent In ventral hernia repair, \textbf{suturing} is the first step in abdominal wall closure, whether after laparotomy or laparoscopy. Although closure might seem straightforward, the way it is performed appears to \textbf{contribute to early repair failure} \cite{burgerIncisionalHerniaEarly2005,harlaarDevelopmentIncisionalHerniation2017}. The long-term success of closure relies on biomechanical principles and can be improved on a biomechanical basis \cite{kallinowskiBiomechanicsAppliedIncisional2021}. \textbf{Biomechanical stability} of a ventral hernia repair can be defined as the ability of the repair to \textbf{withstand physiological forces} and maintain structural integrity \textbf{over time}. 

\noindent Historically, rigid, thick sutures, full-thickness (from skin to peritoneum) or separate stitches have been used, but have been largely abandoned due to their association with high rates of incisional hernia \cite{dienerElectiveMidlineLaparotomy2010}. The \guillemotleft \textbf{small bites}\guillemotright{} continuous suture technique with a slowly absorbable suture represents a major biomechanical advance in hernia repair. It has been shown to significantly reduce the risk of recurrence \cite{millbournEffectStitchLength2009, deerenbergSmallBitesLarge2015} and is now recommended in closure guidelines \cite{deerenbergUpdatedGuidelineClosure2022} (\autoref{fig:suture_length_Meijer}). This technique uses fine threads, reduced stitch spacing (bites separation), and appropriate tissue bite size (or bite width, i.e. the thickness of tissue trapped in the suture from the wound) which significantly reduces suture tension \cite{varshneySixfoldSutureWound1999,hoerInfluenceSutureTechnique2001}. A ratio between suture length over incision length of at least four has been shown to significantly decrease the recurrence rate of incisional hernias \cite{varshneySixfoldSutureWound1999,millbournEffectStitchLength2009,meijerPrinciplesAbdominalWound2016}.

%%% FIGURE CALLING %%%
\figureSutureMeijer

\noindent Biomechanical studies corroborate these findings by demonstrating that smaller stitch spacing (e.g., 5 mm) \cite{harlaarSmallStitchesSmall2009} and larger tissue bites (e.g., 16 mm) \textbf{reduce strain} and \textbf{enhance rupture resistance} \cite{cooneyOptimizedWoundClosure2018,deerenbergSmallBitesLarge2015}. The increased number of suture points distributes tension more evenly across the abdominal wall \cite{karkhanehyousefiPatientspecificComputationalSimulations2023}, mitigating the risk of ischemia and scar dehiscence \cite{cooneyOptimizedWoundClosure2018}.

\noindent The mechanical properties of the suture type (non-absorbable, slowly absorbable or quickly absorbable) for closing the midline fascia should be taken into account. Absorbable sutures lose their strength over time, increasing the risk of breaking before the fascia has healed well enough \cite{henriksenGuidelinesTreatmentUmbilical2020,deerenbergUpdatedGuidelineClosure2022}. It takes over a year to complete the complex process of fascial healing, starting from the eighth postoperative day, fascia rapidly regains strength, reaching about 50\% of its original strength by two months \cite{rathHealingLaparotomiesReview1998,deerenbergUpdatedGuidelineClosure2022}. Fast-absorbing multifilament sutures retain only 25\% of their initial strength at four weeks and are fully absorbed by 8-10 weeks \cite{vasanthanComparingSutureStrengths2009,deerenbergUpdatedGuidelineClosure2022}, while slowly absorbable monofilament sutures maintain over 50\% strength at six weeks and absorb within 6-8 months \cite{deerenbergUpdatedGuidelineClosure2022}. Non-absorbable sutures retain their strength indefinitely, however their use do not reduce the risk of re-operation \cite{christoffersenLowerReoperationRate2013}, incisional hernia or burst abdomen compared to absorbable sutures, and are associated with prolonged wound pain and suture sinus formation \cite{jeroukhimovReducedPostoperativeChronic2014,patelClosureMethodsLaparotomy2017}. 

\vspace{0.2cm}

\noindent The location and orientation of the surgical incision also influence repair outcomes \cite{greenallMidlineTransverseLaparotomy1980}. \textbf{Large incisions} increase the recurrence risk \cite{romainRecurrenceElectiveIncisional2020}. \textbf{Transverse incisions}, when feasible, are preferred to midline incisions, as they reduce hernia recurrence rates by half \cite{lehuunhoIncidencePreventionVentral2012, leeIncisionalHerniaMidline2018}. This benefit is attributed to the \textbf{anisotropic} mechanical properties of the abdominal wall, which exhibit greater elasticity along the sagittal plane. Consequently, transverse scars experience \textbf{lower transverse stress} than cranio-caudal scars and are more resistant to rupture \cite{brownTransverseVersesMidline2005,grantcharovVerticalComparedTransverse2001}.

%% ---------------------------------------- %%
\subsection{Mechanical benefits of implants}

\subsubsection{Why do we use implants?}

\noindent A surgical closure must withstand IAP which fluctuates throughout daily activities. The mechanical cause of abdominal wound bursting has long been studied \cite{jenkinsBurstAbdominalWound1976,dudleyLayeredMassClosure1970}. Dehiscence results from a poorly performed closure that creates \textbf{areas of high stress}. Due to the high recurrence rate following suture-alone repair \cite{luijendijkComparisonSutureRepair2000,burgerLongtermFollowupRandomized2004}, reinforcement with a \textbf{prosthetic implant} is now recommended for the repair of incisional hernias larger than 2 cm \cite{liangVentralHerniaManagement2017,jinkaClinicallyAppliedBiomechanics2024}. Implants are also being considered for \textbf{prophylactic} use in primary closure to prevent incisional hernias \cite{pereira-rodriguezDefiningHighRiskPatients2023,sadavaSyntheticMeshContaminated2022}. 

\noindent Implants are designed as reticular mesh structures and are intended to protect and reinforce the weakened abdominal wall. During the acute post-operative phase, the implant plays a crucial mechanical role in reducing stresses and strains applied to the hernia closure \cite{jourdanCombinedExperimentalNumerical2025}. As tissue integration starts, the mesh provides a surface for aponeurosis scar tissue to adhere to the mesh pore and thread via connective tissue fibrosis reactions \cite{zogbiRetractionFibroplasiaPolypropylene2010,idreesSurgicalMeshesSearch2018,rossiPeritonealAdhesionsType2017}. A strong mesh aponeurosis scar tissue (MAST) complex strengthens the weakened abdominal tissue and takes approximately 6 months to regain 70-80\% of the \textbf{mechanical strength} of native tissue \cite{wolosonBiochemistryImmunologyTissue2001,earleProstheticMaterialInguinal2008}. Compared to sutures alone, they \textbf{mitigate stiffening} by slowing the deposition of type III collagen \cite{dubayMeshIncisionalHerniorrhaphy2006}.

\noindent This is critical in preventing post-operative complications, including recurrence and chronic pain \cite{nesselThreeyearFollowupGrip2024}. The \textbf{GRIP} (Gained Resistance of the Repair to Pressure) concept developed by Kallinowski \textit{et al.} \cite{kallinowskiAssessingGRIPVentral2018} is a biomechanical framework designed to improve the durability of ventral hernia repairs. This concept encompasses mesh size, overlap, fixation methods and the resistance of the repair to cyclic loading with repeated IAP fluctuations.

%% ---------------------------------------- %%
\subsubsection{What are key factors of mesh-based repair?}

\noindent Mesh properties, including tensile characteristics \cite{brownWhichMeshHernia2010,conzeBiocompatibilityBiomaterialsClinical1999}, material composition, pore size, and weave structure, influence the mechanical behaviour of the repaired abdominal wall \cite{todrosSyntheticSurgicalMeshes2017a}. More than 150 types exist \cite{deekenMechanicalPropertiesAbdominal2017}  with various mechanical tests used for characterization \cite{deekenPhysicomechanicalEvaluationAbsorbable2011,deekenMechanicalPropertiesAbdominal2017}. Implants should meet specific \textbf{mechanical thresholds} to \textbf{mimic} the behaviour of the healthy abdominal wall \cite{lubowieckaVivoPerformanceIntraperitoneal2020,szymczakInvestigationAbdomenSurface2011}.

\vspace{0.2cm}

\noindent \textbf{Elasticity:} Mesh elasticity (governed by its elastic modulus $E$) and resistance to rupture (governed by its tensile strength $\sigma_{\text{rupture}}$) are key properties for implant selection \cite{brownWhichMeshHernia2010,conzeBiocompatibilityBiomaterialsClinical1999}. Mismatch between the implant and abdominal wall properties \cite{szymczakModelingFasciameshSystem2010} can result in the implant deforming less than the abdominal wall \cite{jourdanCombinedExperimentalNumerical2025}, stress concentrations at the implant-tissue interface \cite{jungeElasticityAnteriorAbdominal2001,hernandez-gasconMechanicalBehaviourSynthetic2011,tomaszewskaPhysicalMathematicalModelling2013}, causing important shear stress leading to abnormal connective tissue formation \cite{anurovBiomechanicalCompatibilitySurgical2012}, compromising the repair durability \cite{simon-allueMechanicalCharacterizationAbdominal2017, simon-allueMechanicalBehaviorSurgical2018,chandaBiomechanicalModelingProsthetic2018}. 

\noindent The optimal mesh elasticity remains unclear and likely varies per patient. Overly \textbf{stiff} meshes will not accommodate natural deformations of the abdominal wall, \textbf{limit mobility}, thereby increasing foreign body sensation and generating \textbf{high stresses} \cite{hernandez-gasconUnderstandingPassiveMechanical2013,simon-allueProsthesesSizeDependency2016,simon-allueMechanicalBehaviorSurgical2018}. On the other hand, highly \textbf{elastic} meshes are more prone to rupture \cite{blazquezhernandoRoturasMallaCausa2015} with excessive deformation under increased IAP, leading to complications such as \textbf{bulging} \cite{deerenbergMeshExpansionCause2016}, especially in large incisional hernia repairs \cite{dubayMeshIncisionalHerniorrhaphy2006,guerinImpactDefectSize2013,simon-allueProsthesesSizeDependency2016}. Mesh size also increases the overall stiffness of the repaired abdominal wall \cite{leruyetDifferencesBiomechanicsAbdominal2020}. Slightly \textbf{anisotropic} implants better match the anisotropic properties of the abdominal wall \cite{jinkaClinicallyAppliedBiomechanics2024,hernandez-gasconLongtermAnisotropicMechanical2012,jungeElasticityAnteriorAbdominal2001}, especially for large defects \cite{simon-allueProsthesesSizeDependency2016}. 

\noindent Implant stiffness evolves during the healing process \cite{franzBiologyHerniasAbdominal2006}, as host tissue growth leads to further stiffening \cite{todrosSyntheticSurgicalMeshes2017}. Repeated stress causes strain hardening \cite{liCharacterizingExVivo2014} and polymer realignment \cite{velayudhanEvaluationDynamicCreep2009} further increasing stiffness \cite{tomaszewskaMechanicsMeshImplanted2018,liCharacterizingExVivo2014,deerenbergMeshExpansionCause2016,eliasonEffectRepetitiveLoading2011}.

\vspace{0.2cm}

\noindent \textbf{Resistance to rupture:} Textiles are currently optimised for tensile strength, but often fail to provide adhesion, dynamic grip and strain resistance to impulse impacts \cite{kallinowskiGripConceptIncisional2021}.
\textbf{Non-absorbable} synthetic meshes, mainly made of various polymers, offer great mechanical strength \cite{wangseeHerniaMeshHernia2020}. However, when placed intraperitoneally, they are associated with either a higher risk of adhesion formation, inflammation, and pain \cite{kalabaDesignStrategiesApplications2016,wangseeHerniaMeshHernia2020}. \textbf{Absorbable} synthetic meshes reduce inflammation but offer limited long-term mechanical strength \cite{wangseeHerniaMeshHernia2020}, which may lead to hernia recurrence. \textbf{Biologic} meshes, derived from allografts and xenografts, have lower mechanical strength \cite{wangseeHerniaMeshHernia2020}, a sub-optimal benefit-to-cost ratio \cite{schneebergerCostUtilityAnalysisBiologic2019}, and have been identified as a significant prognostic factor for hernia recurrence \cite{parkerIdentifyingPredictorsVentral2021,mazzolapolidefigueiredoBiologicSyntheticMesh2023}. However, this biomechanics-based comparison should be treated with caution, given that the clinical applications of non-absorbable, absorbable, and biological meshes differ (e.g., prophylactic, contaminated field etc) \cite{deerenbergUpdatedGuidelineClosure2022,sadavaSyntheticMeshContaminated2022,bittnerUpdateGuidelinesLaparoscopic2019}.

\noindent \textbf{Lightweight implants} (<70 $g/m^2$) \cite{codaClassificationProstheticsUsed2012} with \textbf{large pores} (>1mm) \cite{brownWhichMeshHernia2010} reduce inflammation \cite{klosterhalfenLightweightLargePorous2005}, scar tissue, risk of implant retraction, adhesions, post-operative pain and improve patient quality of life \cite{brownWhichMeshHernia2010,yuMechanicalPropertiesWarpknitted2021}. From a biomechanical point of view, these meshes cause less foreign body reaction thanks to their reduced amount of materials and their \textbf{good elasticity}, ranging from 20–35\% at the minimum physiologic tensile strength of 16 N/cm \cite{conzeBiocompatibilityBiomaterialsClinical1999}. Studies have indicated that meshes with low resistance and high elasticity are better choice for enhancing abdominal wall \textbf{mobility} \cite{kalabaDesignStrategiesApplications2016}. However, their lower tensile strength than heavy meshes may lead to central mesh failure in some cases, i.e. an obvious central defect in the mesh with intact areas around \cite{petroCentralFailuresLightweight2015,warrenPatternsRecurrenceMechanisms2017}. Despite this, studies showed that they are still able to withstand pressures above 170 mmHg \cite{idreesSurgicalMeshesSearch2018} and maintain sufficient mechanical strength \cite{kalabaDesignStrategiesApplications2016,wangseeHerniaMeshHernia2020}.

\vspace{0.2cm}

\noindent \textbf{Size of implant:} Hernia size influences required mesh overlap for biomechanical success \cite{kallinowskiAssessingGRIPVentral2018,nesselThreeyearFollowupGrip2024}. A larger overlap reduces abdominal wall deformation \cite{leruyetDifferencesBiomechanicsAbdominal2020} and stress \cite{guerinImpactDefectSize2013}, distributes tension more evenly along the suture line \cite{harlaarSmallStitchesSmall2009,cooneyOptimizedWoundClosure2018} and reduces recurrence rates \cite{hautersAssessmentPredictiveFactors2017}. The mesh should ideally be at least twice the defect area \cite{qiuMeasurementsAbdominalWall2023}, but evidence on ideal size remains insufficient, with EHS and AHS (European and American Hernia Societies) providing recommendations on overlap \cite{henriksenGuidelinesTreatmentUmbilical2020}. In preperitoneal mesh repair for open umbilical and epigastric hernia repair, an overlap of 2 cm is suggested for defects of 0-1 cm, and 3 cm for defects of 1-4 cm. For laparoscopic umbilical or epigastric hernia repairs, an overlap of at least 5 cm is suggested \cite{henriksenGuidelinesTreatmentUmbilical2020}.

\vspace{0.2cm}

\noindent \textbf{Implant positioning:} In ventral hernia repair, mesh placement can be categorized based on its anatomical location relative to the peritoneum \cite{parkerInternationalClassificationAbdominal2020}. The intraperitoneal approach involves placing the mesh directly inside the abdominal cavity, in contact with the viscera, as seen in the intraperitoneal onlay mesh (IPOM) technique. Conversely, the extraperitoneal approach positions the mesh outside the peritoneal cavity, such as in the retromuscular (sublay), preperitoneal (underlay) or subcutaneous (onlay) planes, thereby avoiding direct contact with intra-abdominal organs \cite{parkerInternationalClassificationAbdominal2020}. There is no consensus on the optimal placement, which depends on the size and location of the hernia \cite{karrechBiomechanicalStabilityHerniadamaged2023}. Karrech \textit{et al.} suggest critical hernia defect sizes, below which surgical repair may be unnecessary. These thresholds are 4.1 cm for hernias located in the rectus abdominis area and between 5.2 to 8.2 cm for other anterior abdominal muscles. Studies indicate \textbf{retrorectus} placement provides the most effective and durable repair, followed by underlay, onlay and inlay placements \cite{albinoDoesMeshLocation2013,holihanVentralHerniaRepair2017,kallinowskiDynamicIntermittentStrain2015}. Retrorectus and preperitoneal positions induce less stress on the linea alba \cite{karrechBiomechanicalStabilityHerniadamaged2023}, correlating with lower recurrence rates \cite{sosinPerfectPlaneSystematic2018,parkerIdentifyingPredictorsVentral2021}. Mesh positioning also depends on the material selection, onlay and gap bridging configurations are unsuitable for biologic meshes due to its lack of mechanical strength \cite{sosinPerfectPlaneSystematic2018}. Fascial closure alongside mesh placement distributes forces more effectively resulting in stronger repair \cite{parkerIdentifyingPredictorsVentral2021}.

\vspace{0.2cm}

\noindent \textbf{Closure force and suture tension:} Suture tension during closure is another critical factor. Optimally, abdominal closure should be tailored to restore native physiologic tension \cite{millerPhysiologicTensionAbdominal2023}. \textbf{Excessive tension} may cause ischemia and tissue necrosis \cite{rocaSurgicalDynamometerSimultaneously2018}, leading to mechanical failure and discomfort. \textbf{Insufficient tension} can fail to counteract the abdominal wall stresses leading to repair failure. Notably, suture tension decreases significantly within 24 hours post-operatively, suggesting that 50\% of the initial suture tension may be an unnecessary excess \cite{schachtruppImplantableSensorDevice2016}. However, the optimal closure tension remains undefined \cite{pereiraHowItUsing2024}.

\noindent Mesh fixation methods vary, including sutures, permanent or absorbable tacks, and synthetic glue. Despite limited long-term follow-up data, non-absorbable tacks are most commonly used due to strength and ease of application when fixing intraperitoneal
mesh \cite{rezazahiriAbdominalWallHernia2018}. Tack fixation strategies, notably the "single crown" and "double crown" techniques, involve placing tacks in one or two concentric circles to secure the mesh. The double crown method has been associated with shorter operative times and reduced immediate postoperative pain compared to combined suture and tack approaches, without increasing recurrence rates \cite{muysomsRandomizedClinicalTrial2013}.
However, clinical outcomes (pain and recurrence) do not significantly differ between absorbable and non-absorbable tacks \cite{khanAbsorbableNonabsorbableTacks2018} and the various fixation methods \cite{harslofEffectFixationDevices2018,calpinEvaluatingMeshFixation2024} and a heterogeneous literature \cite{henriksenGuidelinesTreatmentUmbilical2020}.
\textbf{Minimising relative motion} (sliding) between the abdominal wall and the mesh is also critical for successful healing and reduced post-operative pain \cite{kallinowskiBiomechanicalInfluencesMeshRelated2021,jourdanCombinedExperimentalNumerical2025}.

\noindent Techniques such as \textbf{component separation} improve the feasibility of closing large defects without excessive tension \cite{afifiQuantitativeAssessmentTension2017}. It does not significantly alter fascial area \cite{lisieckiAbdominalWallDynamics2015} or abdominal wall biomechanics \cite{daesChangesAbdominalWall2022}, as some muscles atrophy while others (rectus abdominis and obliques) undergo compensatory muscular hypertrophy \cite{daesChangesAbdominalWall2022}, preserving overall abdominal wall physiology and integrity. Pre-operative CT scans measurements can predict the need for component separation \cite{loveComputedTomographyImaging2021,duCTmeasuredHerniaParameters2023}.

\noindent Recently, the Fasciotens\textregistered{} medical device has been designed for large ventral hernias or open abdomen management \cite{Fasciotens}. By applying vertical traction to stretch the fascia, it prevents the retraction of lateral muscles and gradually lengthen the tissues, reducing closure tension \cite{heesPreventionFascialRetraction2020}. This method offers potential advantages over traditional component separation \cite{niebuhrIntraoperativeFasciaTension2021}.

\vspace{0.2cm}

\noindent \textbf{Defect size}, although important for EHS classification \cite{muysomsClassificationPrimaryIncisional2009}, \textbf{does not correlate with closure force} \cite{hopeRationaleTechniqueMeasuring2018}, underscoring the need for patient-specific biomechanical assessment. Experimental \cite{guerinImpactDefectSize2013} and numerical studies \cite{simon-allueProsthesesSizeDependency2016,hernandez-gasconUnderstandingPassiveMechanical2013} show that defect size affects stress and displacement in a repaired abdominal wall. Surgeon's estimation of closure force does not correlate with objective dynamometer-based measurements \cite{hopeRationaleTechniqueMeasuring2018}, advocating for quantitative decision-making tools \cite{hopeRationaleTechniqueMeasuring2018}.

\noindent Experimental \textit{in vitro} models such as the \guillemotleft Abdoman\guillemotright{} \cite{kroeseAbdoMANArtificialAbdominal2017} (\autoref{fig:mechanical_tests_ex_vivo}) replicate dynamic mesh behaviour using various surgical closures on synthetic, porcine, and human tissues, under simulated muscle contraction and IAP changes. This enables assessment of mesh performance and closure forces under conditions closer to physiological reality, which differs from static model results \cite{siassiDevelopmentDynamicModel2014,kroeseAbdoMANArtificialAbdominal2017}. 

\vspace{0.3cm}

\noindent Overall, hernia repair success depends in part on the mechanical stress intensity and duration \cite{kallinowskiBiomechanicalInfluencesMeshRelated2021}, as well as specific requirements that the mesh must meet \cite{deekenPhysicomechanicalEvaluationAbsorbable2011, liuRegulatoryScienceHernia2021} to mimic the behaviour of a healthy abdominal wall \cite{lubowieckaVivoPerformanceIntraperitoneal2020,szymczakInvestigationAbdomenSurface2011}.

%% ---------------------------------------- %%
\subsection{Surgery-induced biomechanical changes in the abdominal wall}

\noindent The mechanical behaviour of the repaired abdominal wall can be assessed by both experimental \cite{harlaarSmallStitchesSmall2009,kroeseAbdoMANArtificialAbdominal2017,lyonsBiomechanicalAbdominalWall2015} and numerical studies \cite{todrosComputationalModelingAbdominal2018,heNumericalMethodGuiding2020,guerinImpactDefectSize2013,simon-allueProsthesesSizeDependency2016,simon-allueMechanicalBehaviorSurgical2018,tomaszewskaMechanicsMeshImplanted2018}. The mechanical properties of a repaired abdominal wall differ significantly from those of a healthy abdominal wall, and the mechanical response of the composite mesh-tissue lies between the two individual responses of the mesh or the tissue alone \cite{chandaBiomechanicalModelingProsthetic2018}.

\vspace{0.2cm}

\noindent \textbf{Impact on mobility:} Implants \textbf{partially restore} the  mechanical properties of the abdominal wall by reducing hernia-induced excessive mobility and strain \cite{podwojewskiMechanicalResponseHuman2014,kallinowskiBiomechanicalInfluencesMeshRelated2021,hernandez-gasconMechanicalBehaviourSynthetic2011,todrosComputationalModelingAbdominal2018,heNumericalMethodGuiding2020}. Furthermore, the post-repair abdominal wall exhibits reduced mobility compared to healthy controls \cite{leruyetDifferencesBiomechanicsAbdominal2020}. This mobility is influenced by the elasticity of the mesh used in the repair \cite{heNumericalMethodGuiding2020}. As shown in \autoref{fig:numerical_model_Todros}, numerical models highlight that abdominal walls with hernia have the greatest displacement during coughing, followed by healthy walls, with repaired walls having the least displacement \cite{todrosComputationalModelingAbdominal2018}. 

\noindent \textbf{Impact on stiffness:} Dubay \textit{et al.} demonstrated, using a rodent model, that an abdominal wall repaired with mesh exhibited greater strain and reduced stiffness compared to a suture-only repair \cite{dubayMeshIncisionalHerniorrhaphy2006}. In contrast, Le Ruyet \textit{et al.}, in their study on human cadavers, reported opposite findings, showing increased stiffness and decreased strain and displacement when mesh reinforcement was used \cite{leruyetDifferencesBiomechanicsAbdominal2020}. After repair, the abdominal wall tends to be \textbf{stiffer} than in healthy condition \cite{todrosComputationalModelingAbdominal2018}. This increased stiffness often \textbf{correlates with patient discomfort} and post-operative pain \cite{weltyFunctionalImpairmentComplaints2001}. Mesh-repaired abdominal walls are also \textbf{less resistant to rupture} \cite{dubayMeshIncisionalHerniorrhaphy2006,dubayIncisionalHerniationInduces2007}.

%%% FIGURE CALLING %%%
\figureNumericalModelTodros

\noindent \textbf{Impact on functional outcomes:} Using isokinetic and isometric measurements of the rectus muscle, Criss \textit{et al.} showed that hernia repair with linea alba restoration improved abdominal wall functionality \cite{crissFunctionalAbdominalWall2014}. A protein-supplemented diet and pre-operative training have been shown to have a significant impact on abdominal muscles strengthening after ventral hernia repair \cite{crocettiDietaryProteinSupplementation2020,ahmedEffectPreoperativeAbdominal2018} and a reduced rate of complications and early recurrence \cite{liangModifyingRisksVentral2018,pezeshkComplexAbdominalWall2015}. As seen on CT scans, hernia surgery alters the abdominal wall morphometry by reducing the abdomen anterior-posterior diameter, area and circumference \cite{lisieckiAbdominalWallDynamics2015}.

\vspace{0.2cm}

\noindent \textbf{Impact on intra-abdominal pressure:} Implant-based defect closure raises the IAP by approximately 2.7 mmHg \cite{angeliciMeasurementIntraabdominalPressure2016}, and \textbf{up to 4 mmHg} for defects larger than 10 cm and with significant hernia sac \cite{espinosa-de-los-monterosImmediateChangesIntraabdominal2022}. These increases must be balanced against the risk of intra-abdominal hypertension \cite{malbrainResultsInternationalConference2006}.

%% ---------------------------------------- %%
\subsection{Rethinking hernia recurrence in a biomechanical context} \label{subsection_risk_factors_occurrence}

\noindent Despite advances in minimally invasive techniques \cite{lujanLaparoscopicOpenGastric2004,barbarosComparisonLaparoscopicOpen2007,lehuunhoIncidencePreventionVentral2012}, robotics \cite{pereiraRoboticAbdominalWall2022} and mesh implants \cite{schumpelickPraeperitonealeNetzplastikReparation1996,anthonyFactorsAffectingRecurrence2000}, hernia recurrence rates still reach up to 70\% within five years \cite{bhardwajYearOverYearVentralHernia2024}.

\noindent The incidence and recurrence of hernia are influenced by many intricated factors related to the patient comorbidities (obesity, diabetes, smoking), biological phenomenon (wound healing, infection) and surgical procedure (such as poor mesh fixation). Biomechanics can provide a helpful framework for understanding these multifactorial influences, why and how hernias develop and fail after repair, with the important role of mechanical stress having long been recognised \cite{urschelEffectMechanicalStress1988}. 

%% ---------------------------------------- %%
\subsubsection{Patient-specific parameters}

\noindent \textbf{Sex:} Males tend to be more prone to primary hernias and females to incisional hernias \cite{parkerIdentifyingPredictorsVentral2021,verstoepHerniaWidthExplains2021,kroesePrimaryIncisionalVentral2018,stirlerLaparoscopicRepairPrimary2014,kurianLaparoscopicRepairPrimary2010,maGlobalRegionalNational2023}. Males have been shown to typically have greater muscle mass and visceral fat than females \cite{jourdanAbdominalWallMorphometric2020,rankinAbdominalMuscleSize2006}, whereas females have greater subcutaneous fat volume \cite{schlosserOutcomesSpecificPatient2020}. Female patients have higher post-operative rates of wound complications, readmission, and immediate and chronic post-operative pain \cite{schlosserOutcomesSpecificPatient2020,coxPredictiveModelingChronic2016,craigPrevalencePredictorsHernia2016}.

\vspace{0.2cm}

\noindent \textbf{Age-related changes:} Older adults have reduced tissue elasticity, slower healing rates after surgery or injury \cite{holtEffectAgeWound1992,goswamiWoundHealingGolden2022}, and reduced muscle mass and function \cite{jourdanAbdominalWallMorphometric2020,garciamorianaEvaluationRectusAbdominis2023,rankinAbdominalMuscleSize2006}. These characteristics are predisposing factors for hernia, particularly in people over the age of 50 \cite{helgstrandNationwideProspectiveStudy2013,lehuunhoIncidencePreventionVentral2012}. Ageing increases visceral fat and lengthens aponeurosis and abdominal muscles (except the transversus abdominis) \cite{jourdanAbdominalWallMorphometric2020}. These changes weaken the ability of the abdominal wall to withstand mechanical stress.

\vspace{0.2cm}

\noindent \textbf{Individual mechanical properties:} There is an increased risk of hernia in patients with aortic aneurysms, who often have collagen dysfunction. This highlights the importance of considering individual tissue properties \cite{antoniouAbdominalAorticAneurysm2011}. While sex and BMI have no direct correlation with abdominal wall mechanical properties \cite{hollinskyMeasurementTensileStrength2007,martinsMechanicalCharacterizationConstitutive2012,rathAbdominalLineaAlba1996,rathSheathRectusAbdominis1997,vanramshorstNoninvasiveAssessmentIntraabdominal2011}, they do influence IAP \cite{cobbNormalIntraabdominalPressure2005,sanchezWhatNormalIntraabdominal2001} and abdominal wall anatomy \cite{jourdanAbdominalWallMorphometric2020}. Collagen metabolism and hereditary factors also play a role in hernia development and recurrence \cite{franzBiologyHerniasAbdominal2006,jansenBiologyHerniaFormation2004,henriksenCollagenTurnoverProfile2015}. 

%% ---------------------------------------- %%
\subsubsection{Mechanical stress}

\noindent \textbf{Intra-abdominal pressure:} Physical activity increases the IAP, causing repetitive strain on the abdominal muscles. Umbilical hernia, often congenital, may develop later due to increased IAP, particularly from pregnancy \cite{al-khanMeasurementIntraabdominalPressure2011} or obesity \cite{sanchezWhatNormalIntraabdominal2001}. Pregnancy stretches the abdominal wall, causing diastasis recti and weakening the central fascia \cite{gueroultLineaAlba3D2024}, further predisposing individuals to hernia formation \cite{al-khanMeasurementIntraabdominalPressure2011}. Increased BMI is a well-documented risk factor of hernia \cite{oweiImpactBodyMass2017,gignouxIncidenceRiskFactors2021,parkerIdentifyingPredictorsVentral2021,bhardwajYearOverYearVentralHernia2024,vansilfhoutRecurrentIncisionalHernia2021,al-mansourAssociationHerniaspecificProcedural2024}, as BMI $>30$ is associated with increased abdominal fat which raises IAP \cite{sanchezWhatNormalIntraabdominal2001} and impairs wound healing \cite{tastaldiEffectIncreasingBody2019}. Additionally, obesity alters abdominal muscle morphology, leading to thinner obliques and thicker transversus abdominis \cite{jourdanAbdominalWallMorphometric2020}, reflecting adaptations to increased visceral mass and IAP \cite{cobbNormalIntraabdominalPressure2005,sanchezWhatNormalIntraabdominal2001}. Some studies suggest that visceral fat volume is a better predictor of hernia recurrence than BMI, although BMI remains correlated with both subcutaneous and visceral fat \cite{frommerPreoperativeCTImaging2024,yamamotoVisceralObesitySignificant2018,wintersPreoperativeCTScan2019}. Chronic high IAP also induces a shift in rectus abdominis muscle fibers towards endurance-oriented type I, potentially reducing overall abdominal wall strength \cite{kotidisEffectChronicallyIncreased2011}. Cyclic loading contributes to pathological changes in cellular function and tissue structure \cite{franzBiologyHerniasAbdominal2006}, though a minimum level of mechanical stress is necessary to stimulate fibroblast activity and collagen synthesis for proper healing \cite{franzBiologyHerniasAbdominal2006}.  

\vspace{0.2cm}

\noindent The \textbf{tensile forces} on sutures are directly influenced by IAP. Elevated IAP results in a significant immediate increase in midline suture tension, but leads to a sustained tension loss once IAP returns to normal \cite{schachtruppInfluenceElevatedIntraabdominal2019}, thereby increasing the risk of repair failure \cite{konerdingMaximumForcesActing2011, schachtruppInfluenceElevatedIntraabdominal2019}. Pre- and post-operative IAP variations have been statistically associated with recurrence \cite{muresanHerniaRecurrenceLong2016}. Muscle contraction plays a role in incisional hernia formation \cite{lienContractionAbdominalWall2015}, muscle relaxants administered pre-operatively appear to be useful \cite{ibarra-hurtadoUseBotulinumToxin2009,motzChemicalComponentsSeparation2018,whitehead-clarkeUseBotulinumToxin2021}, reducing closing forces, thereby decreasing the risk of recurrence and defect size in subsequent hernia repair \cite{lienContractionAbdominalWall2015}. Abdominal binders help manage IAP-induced stress by redistributing forces across the abdominal wall, but must balance rigidity and comfort, as overly rigid designs can increase IAP \cite{zhangEffectDifferentTypes2016}.

\vspace{0.2cm}

\noindent \textbf{Fatigue rupture:} Hernias can also develop due to fatigue rupture \cite{dubayMeshIncisionalHerniorrhaphy2006}, a lesser-studied mechanism where tissue experiences \textbf{progressive damage}, loss of integrity and ability to withstand stress when exposed to \textbf{cyclic} (repeated) loading, inherent to daily life activities. Rupture can occur suddenly, without prior permanent strain and even below their rupture threshold $\sigma_{rupture}$. Chronic coughing accentuated by smoking and chronic obstructive pulmonary disease (COPD), in addition to being established hernia risk factors and disrupting the healing process \cite{sorensenSmokingRiskFactor2005,gignouxIncidenceRiskFactors2021,parkerIdentifyingPredictorsVentral2021,al-mansourAssociationHerniaspecificProcedural2024}, can lead to progressive tissue fatigue.

%% ---------------------------------------- %%
\subsubsection{Early signs of recurrence}

\noindent It has been reported that 56$\%$ of recurrences occur within the first year and 79$\%$ within two years after laparotomy \cite{gignouxIncidenceRiskFactors2021}. Early signs of recurrence often manifest within one month post-surgery, occurring in approximately \textbf{18\% of patients} \cite{al-mansourAssociationHerniaspecificProcedural2024}, and may take up to five years to be clinically diagnosed \cite{romainRecurrenceElectiveIncisional2020,dubayMeshIncisionalHerniorrhaphy2006,lujanLaparoscopicOpenGastric2004}. Indicators include fluid collection in the hernia sac \cite{loftusCTEvidenceFluid2017} and widening of the linea alba \cite{burgerIncisionalHerniaEarly2005,harlaarDevelopmentIncisionalHerniation2017,pollockEarlyPredictionLate1989}. Patients with diastasis or a wider linea alba have a higher risk of recurrence \cite{kohlerSuturedRepairPrimary2015} due to greater distortion of the linea alba \cite{beamishDifferencesLineaAlba2019}. 

\noindent Various radiographic parameters, including hernia volume, subcutaneous fat volume, and defect size, are increasingly being used to predict post-operative complications and assess the likelihood of requiring intraoperative intervention \cite{elfanagelyComputedTomographyImage2020}.

%% ---------------------------------------- %%
\subsection*{Key points}
\addcontentsline{toc}{subsection}{Key points}
\begin{itemize}
    \item Ventral hernia repair changes the biomechanics of the abdominal wall.
    \item Surgical technique (suturing technique, choice of mesh and sizing, positioning and fixation methods) significantly influences mechanical outcomes.
    \item Mesh implantation partially restores the mechanical properties of the abdominal wall by increasing its stiffness.
    \item The success of a repair depends on how well the implant integrates and restores native function.
    \item The implant should have mechanical properties similar to those of the abdominal wall to optimise repair success and reduce post-operative pain (anisotropic elasticity, deformation). 
    \item The most elastic direction of the mesh should be aligned with the patient's cranio-caudal direction.
    \item When possible, a small bite suture technique, lightweight mesh with retrorectus placement is associated with a lower risk of recurrence.
\end{itemize}

\section*{Conclusion}
% \addcontentsline{toc}{section}{Conclusion}

\noindent The abdominal wall cannot be considered as a simple structural barrier but rather as a dynamic, load-bearing system and hernia formation should be seen as a mechanical failure. Hernia repair still represents a significant surgical challenge due to high recurrence rates and the variability in patient outcomes. In that respect, patient-specific biomechanical assessment should be integrated in the clinical decision-making process.

\noindent This review provided key details indicating that the optimisation of hernia repair strategies should consider a biomechanical approach incorporating functional imaging, elasticity analysis, and patient-specific modelling.

\vspace{0.5cm}

\noindent In the pre-operative phase, these tools should help in selecting the appropriate implant to reduce mechanical discrepancies between the abdominal wall and the implant. During surgery, they could objectively inform the decision on whether to use component separation techniques, the number of fixation devices, assist in the implant positioning, and guide the choice of implant overlap and closure type. In the post-operative phase, these tools could improve patient recovery monitoring and enable the early detection of recurrences.

\noindent By embedding biomechanical principles into routine hernia surgery, surgeons can reduce recurrence, minimise complications, and ultimately improve patient quality of life.

%%%%%%%%%%%%%%%%%%%%%%%%%%%%%%%%%%

\section*{Citation Diversity Statement}

\noindent Recent work in several fields of science has identified a bias in citation practices such that papers from women and other minority scholars are under-cited relative to the number of such papers in the field \cite{mitchell2013gendered,dion2018gendered,caplar2017quantitative, maliniak2013gender, Dworkin2020.01.03.894378, bertolero2021racial, wang2021gendered, chatterjee2021gender, fulvio2021imbalance}. Here we sought to proactively consider choosing references that reflect the diversity of the field in thought, form of contribution, gender, race, ethnicity, and other factors.

\noindent This statement is based on the 304 references cited in this paper.

\vspace{0.3cm}

\noindent First, we obtained the predicted gender of the first and last author of each reference by using databases that store the probability of a first name being carried by a woman \cite{Dworkin2020.01.03.894378,zhou_dale_2020_3672110}. By this measure (and excluding self-citations to the first and last authors of our current paper), our references contain 8.4\% woman(first)/woman(last), 15.47\% man/woman, 21.63\% woman/man, and 54.5\% man/man. This method is limited in that a) names, pronouns, and social media profiles used to construct the databases may not, in every case, be indicative of gender identity and b) it cannot account for intersex, non-binary, or transgender people. 

\vspace{0.3cm}

\noindent Second, we obtained predicted racial/ethnic category of the first and last author of each reference by databases that store the probability of a first and last name being carried by an author of color \cite{ambekar2009name, sood2018predicting}. By this measure (and excluding self-citations), our references contain 17.58\% author of color (first)/author of color(last), 15.90\% white author/author of color, 17.98\% author of color/white author, and 48.54\% white author/white author. This method is limited in that a) names and Florida Voter Data to make the predictions may not be indicative of racial/ethnic identity, and b) it cannot account for Indigenous and mixed-race authors, or those who may face differential biases due to the ambiguous racialization or ethnicization of their names. We look forward to future work that could help us to better understand how to support equitable practices in science.

\section*{Acknowledgments}

\noindent The authors would like to thank Arthur Jourdan for his insightful comments on this review. This research was supported by Université Gustave Eiffel.

\vspace{0.2cm}

%%%%%%%%%%%%%%%%%%%%%%%%%%%%%%
\noindent This version of the article has been accepted for publication, after peer review and is subject to Elsevier's \href{https://www.elsevier.com/about/policies-and-standards/sharing#3-accepted-manuscript}{accepted manuscript terms of use}, but is not the published journal article (PJA) and does not reflect post-acceptance improvements, or any corrections.

\noindent The PJA is available online at: https://doi.org/10.1016/j.jmbbm.2025.107159. This accepted manuscript is licensed under the terms of the \href{https://creativecommons.org/share-your-work/cclicenses/}{CC BY-NC-ND} license.

\section*{Statements and Declarations}

\noindent \textbf{Conflict of interest}: The authors declare that they have no conflict of interest.

\section*{CRediT authorship contribution statement}

\noindent \textbf{Victoria Joppin:} Methodology, Conceptualization, Writing - original draft. \textbf{Catherine Masson:} Writing - review \& editing, Methodology, Conceptualization. \textbf{David Bendahan:} Writing - review \& editing, Methodology, Conceptualization. \textbf{Thierry Bège:} Writing - review \& editing, Methodology, Conceptualization.

\newpage

\bibliography{clean_and_diversity_merged.bib}

\begin{thebibliography}{316}
\providecommand{\natexlab}[1]{#1}
\providecommand{\url}[1]{\texttt{#1}}
\expandafter\ifx\csname urlstyle\endcsname\relax
  \providecommand{\doi}[1]{doi: #1}\else
  \providecommand{\doi}{doi: \begingroup \urlstyle{rm}\Url}\fi

\bibitem[Grevious et~al.(2006)Grevious, Cohen, Shah, and
  Rodriguez]{greviousStructuralFunctionalAnatomy2006}
Mark~A. Grevious, Mimis Cohen, Samir~R. Shah, and Pedro Rodriguez.
\newblock Structural and {{Functional Anatomy}} of the {{Abdominal Wall}}.
\newblock \emph{Clinics in Plastic Surgery}, 33\penalty0 (2):\penalty0
  169--179, April 2006.
\newblock ISSN 00941298.
\newblock \doi{10.1016/j.cps.2005.12.005}.

\bibitem[Campbell and
  Green(1953)]{campbellVariationsIntraabdominalPressure1953}
Edward James~Moran Campbell and J.~H. Green.
\newblock The variations in intra-abdominal pressure and the activity of the
  abdominal muscles during breathing; a study in man.
\newblock \emph{The Journal of Physiology}, 122\penalty0 (2):\penalty0
  282--290, November 1953.
\newblock ISSN 0022-3751.

\bibitem[Joyce et~al.(2022)Joyce, Murphy, Maher, and
  O'Connor]{joyceAbdominalCavityAnatomy2022}
Stella Joyce, Kevin~P. Murphy, Michael~M. Maher, and Owen~J. O'Connor.
\newblock Abdominal cavity: Anatomy, structural anomalies, and hernias.
\newblock In \emph{Yamada's {{Textbook}} of {{Gastroenterology}}}, chapter~8,
  pages 131--145. John Wiley \& Sons, Ltd, 2022.
\newblock ISBN 978-1-119-60020-6.
\newblock \doi{10.1002/9781119600206.ch8}.

\bibitem[Todros et~al.(2020)Todros, {de Cesare}, Concheri, Natali, and
  Pavan]{todrosNumericalModellingAbdominal2020}
Silvia Todros, Niccol{\`o} {de Cesare}, Gianmaria Concheri, Arturo~N. Natali,
  and Piero~G. Pavan.
\newblock Numerical modelling of abdominal wall mechanics: {{The}} role of
  muscular contraction and intra-abdominal pressure.
\newblock \emph{Journal of the Mechanical Behavior of Biomedical Materials},
  103:\penalty0 103578, March 2020.
\newblock ISSN 1751-6161.
\newblock \doi{10.1016/j.jmbbm.2019.103578}.

\bibitem[Hodges et~al.(2001)Hodges, Cresswell, Daggfeldt, and
  Thorstensson]{hodgesVivoMeasurementEffect2001}
Paul~W. Hodges, Andrew~G. Cresswell, Karl Daggfeldt, and Alf Thorstensson.
\newblock In vivo measurement of the effect of intra-abdominal pressure on the
  human spine.
\newblock \emph{Journal of Biomechanics}, 34\penalty0 (3):\penalty0 347--353,
  March 2001.
\newblock ISSN 0021-9290.
\newblock \doi{10.1016/S0021-9290(00)00206-2}.

\bibitem[Tran et~al.(2016)Tran, Podwojewski, Beillas, Ottenio, Voirin,
  Turquier, and Mitton]{tranAbdominalWallMuscle2016}
Doris Tran, Florence Podwojewski, P.~Beillas, M.~Ottenio, David Voirin,
  Fr{\'e}d{\'e}ric Turquier, and David Mitton.
\newblock Abdominal wall muscle elasticity and abdomen local stiffness on
  healthy volunteers during various physiological activities.
\newblock \emph{Journal of the Mechanical Behavior of Biomedical Materials},
  60:\penalty0 451--459, July 2016.
\newblock ISSN 1751-6161.
\newblock \doi{10.1016/j.jmbbm.2016.03.001}.

\bibitem[Pavan et~al.(2019)Pavan, Todros, Pachera, Pianigiani, and
  Natali]{pavanEffectsMuscularContraction2019}
Piero~G. Pavan, Silvia Todros, Paola Pachera, Silvia Pianigiani, and Arturo~N.
  Natali.
\newblock The effects of the muscular contraction on the abdominal
  biomechanics: A numerical investigation.
\newblock \emph{Computer Methods in Biomechanics and Biomedical Engineering},
  22\penalty0 (2):\penalty0 139--148, January 2019.
\newblock ISSN 1025-5842.
\newblock \doi{10.1080/10255842.2018.1540695}.

\bibitem[De~Troyer and Boriek(2011)]{detroyerMechanicsRespiratoryMuscles2011}
Andr{\'e} De~Troyer and Aladin~M. Boriek.
\newblock Mechanics of the {{Respiratory Muscles}}.
\newblock In \emph{Comprehensive {{Physiology}}}, pages 1273--1300. John Wiley
  \& Sons, Ltd, 2011.
\newblock ISBN 978-0-470-65071-4.
\newblock \doi{10.1002/cphy.c100009}.

\bibitem[Jourdan et~al.(2022)Jourdan, Rapacchi, Guye, Bendahan, Masson, and
  B{\`e}ge]{jourdanDynamicMRIQuantificationAbdominal2022}
Arthur Jourdan, Stanislas Rapacchi, Maxime Guye, David Bendahan, Catherine
  Masson, and Thierry B{\`e}ge.
\newblock Dynamic-{{MRI}} quantification of abdominal wall motion and
  deformation during breathing and muscular contraction.
\newblock \emph{Computer Methods and Programs in Biomedicine}, 217:\penalty0
  106667, April 2022.
\newblock ISSN 1872-7565.
\newblock \doi{10.1016/j.cmpb.2022.106667}.

\bibitem[Wilke et~al.(2018)Wilke, Schleip, Yucesoy, and
  Banzer]{wilkeNotMerelyProtective2018}
Jan Wilke, Robert Schleip, Can~A. Yucesoy, and Winfried Banzer.
\newblock Not merely a protective packing organ? {{A}} review of fascia and its
  force transmission capacity.
\newblock \emph{Journal of Applied Physiology}, 124\penalty0 (1):\penalty0
  234--244, January 2018.
\newblock ISSN 8750-7587.
\newblock \doi{10.1152/japplphysiol.00565.2017}.

\bibitem[Malbrain et~al.(2006)Malbrain, Cheatham, Kirkpatrick, Sugrue, Parr,
  De~Waele, Balogh, Lepp{\"a}niemi, Olvera, Ivatury, D'Amours, Wendon, Hillman,
  Johansson, Kolkman, and Wilmer]{malbrainResultsInternationalConference2006}
Manu Malbrain, Michael~L. Cheatham, Andrew Kirkpatrick, Michael Sugrue, Michael
  Parr, Jan De~Waele, Zsolt Balogh, Ari Lepp{\"a}niemi, Claudia Olvera, Rao
  Ivatury, Scott D'Amours, Julia Wendon, Ken Hillman, Kenth Johansson, Karel
  Kolkman, and Alexander Wilmer.
\newblock Results from the {{International Conference}} of {{Experts}} on
  {{Intra-abdominal Hypertension}} and {{Abdominal Compartment Syndrome}}.
  {{I}}. {{Definitions}}.
\newblock \emph{Intensive Care Medicine}, 32\penalty0 (11):\penalty0
  1722--1732, November 2006.
\newblock ISSN 0342-4642.
\newblock \doi{10.1007/s00134-006-0349-5}.

\bibitem[Novak et~al.(2021)Novak, Jacisko, Busch, Cerny, Stribrny, Kovari,
  Podskalska, Kolar, and Kobesova]{novakIntraabdominalPressureCorrelates2021}
Jakub Novak, Jakub Jacisko, Andrew Busch, Pavel Cerny, Martin Stribrny, Martina
  Kovari, Patricie Podskalska, Pavel Kolar, and Alena Kobesova.
\newblock Intra-abdominal pressure correlates with abdominal wall tension
  during clinical evaluation tests.
\newblock \emph{CLINICAL BIOMECHANICS}, 88, August 2021.
\newblock ISSN 0268-0033.
\newblock \doi{10.1016/j.clinbiomech.2021.105426}.

\bibitem[Misuri et~al.(1997)Misuri, Colagrande, Gorini, Iandelli, Mancini,
  Duranti, and Scano]{misuriVivoUltrasoundAssessment1997}
Giorgio Misuri, S.~Colagrande, M.~Gorini, I.~Iandelli, M.~Mancini, R.~Duranti,
  and Giorgio Scano.
\newblock In vivo ultrasound assessment of respiratory function of abdominal
  muscles in normal subjects.
\newblock \emph{European Respiratory Journal}, 10\penalty0 (12):\penalty0
  2861--2867, December 1997.
\newblock ISSN 0903-1936, 1399-3003.

\bibitem[Joppin et~al.(2025)Joppin, Jourdan, Bendahan, Soucasse, Guye, Masson,
  and B{\`e}ge]{joppinBetterUnderstandingAbdominal2025}
Victoria Joppin, Arthur Jourdan, David Bendahan, Andr{\'e}a Soucasse, Maxime
  Guye, Catherine Masson, and Thierry B{\`e}ge.
\newblock Towards a better understanding of abdominal wall biomechanics:
  {{{\emph{In}}}}{\emph{ vivo}} relationship between dynamic intra-abdominal
  pressure and magnetic resonance imaging measurements.
\newblock \emph{Clinical Biomechanics}, 121:\penalty0 106396, January 2025.
\newblock ISSN 0268-0033.
\newblock \doi{10.1016/j.clinbiomech.2024.106396}.

\bibitem[Wilson(1933)]{wilsonEffectBreathingIntraabdominal1933}
William Wilson.
\newblock Effect of breathing on the intra-abdominal pressure.
\newblock \emph{The Journal of Physiology}, 79\penalty0 (4):\penalty0 481--486,
  October 1933.
\newblock ISSN 0022-3751.
\newblock \doi{10.1113/jphysiol.1933.sp003061}.

\bibitem[Lyons et~al.(2014)Lyons, Winter, and
  Simms]{lyonsMechanicalCharacterisationPorcine2014}
Mathew Lyons, Des~C. Winter, and Ciaran~K. Simms.
\newblock Mechanical characterisation of porcine rectus sheath under uniaxial
  and biaxial tension.
\newblock \emph{Journal of Biomechanics}, 47\penalty0 (8):\penalty0 1876--1884,
  June 2014.
\newblock ISSN 0021-9290.
\newblock \doi{10.1016/j.jbiomech.2014.03.009}.

\bibitem[Luo et~al.(2023)Luo, Leng, Li, Li, Ma, and
  Yang]{luoStudyRelationshipsDiastasis2023}
Xiang Luo, Xiaohui Leng, Zhenhong Li, Yuefeng Li, Rui Ma, and Kun Yang.
\newblock The study of the relationships between diastasis recti abdominis and
  transversus abdominal: A scoping review, November 2023.
\newblock ISSN 2693-5015.

\bibitem[Jourdan et~al.(2025)Jourdan, Vegleur, Bodner, Rousset, Passot, and
  Ruyet]{jourdanCombinedExperimentalNumerical2025}
Arthur Jourdan, Anthony Vegleur, Jeff Bodner, Pascal Rousset, Guillaume Passot,
  and Anicet~Le Ruyet.
\newblock A combined experimental and numerical approach to evaluate hernia
  mesh biomechanical stability {\emph{in situ}}.
\newblock \emph{Medical Engineering \& Physics}, 135:\penalty0 104271, January
  2025.
\newblock ISSN 1350-4533.
\newblock \doi{10.1016/j.medengphy.2024.104271}.

\bibitem[Todros et~al.(2019)Todros, {de Cesare}, Pianigiani, Concheri, Savio,
  Natali, and Pavan]{todros3DSurfaceImaging2019}
Silvia Todros, Niccol{\`o} {de Cesare}, Silvia Pianigiani, Gianmaria Concheri,
  Gianpaolo Savio, Arturo~N. Natali, and Piero~G. Pavan.
\newblock {{3D}} surface imaging of abdominal wall muscular contraction.
\newblock \emph{Computer Methods and Programs in Biomedicine}, 175:\penalty0
  103--109, July 2019.
\newblock ISSN 0169-2607.
\newblock \doi{10.1016/j.cmpb.2019.04.013}.

\bibitem[Jourdan(2021)]{jourdanExplorationsCombineesBiomecaniques2021}
Arthur Jourdan.
\newblock \emph{Explorations Combin{\'e}es Biom{\'e}caniques et Physiologiques
  de La Paroi Abdominale in Vivo}.
\newblock These de doctorat, Aix-Marseille, December 2021.

\bibitem[Axer et~al.(2001{\natexlab{a}})Axer, v~Keyserlingk, and
  Prescher]{axerCollagenFibersLinea2001a}
Hubertus Axer, Diedrich~Graf v~Keyserlingk, and Andreas Prescher.
\newblock Collagen {{Fibers}} in {{Linea Alba}} and {{Rectus Sheaths}}: {{II}}.
  {{Variability}} and {{Biomechanical Aspects}}.
\newblock \emph{Journal of Surgical Research}, 96\penalty0 (2):\penalty0
  239--245, April 2001{\natexlab{a}}.
\newblock ISSN 0022-4804, 1095-8673.
\newblock \doi{10.1006/jsre.2000.6071}.

\bibitem[Soucasse et~al.(2022)Soucasse, Jourdan, Edin, Gillion, Masson, and
  Bege]{soucasseBetterUnderstandingDaily2022}
Andrea Soucasse, Arthur Jourdan, Lauriane Edin, Jean~F. Gillion, Catherine
  Masson, and Thierry Bege.
\newblock A better understanding of daily life abdominal wall mechanical
  solicitation: {{Investigation}} of intra-abdominal pressure variations by
  intragastric wireless sensor in humans.
\newblock \emph{Medical Engineering \& Physics}, 104:\penalty0 103813, June
  2022.
\newblock ISSN 1350-4533.
\newblock \doi{10.1016/j.medengphy.2022.103813}.

\bibitem[Blazek et~al.(2019)Blazek, Stastny, Maszczyk, Krawczyk, Matykiewicz,
  and Petr]{blazekSystematicReviewIntraabdominal2019}
Dusan Blazek, Petr Stastny, Adam Maszczyk, Magdalena Krawczyk, Patryk
  Matykiewicz, and Miroslav Petr.
\newblock Systematic review of intra-abdominal and intrathoracic pressures
  initiated by the {{Valsalva}} manoeuvre during high-intensity resistance
  exercises.
\newblock \emph{Biology of Sport}, 36\penalty0 (4):\penalty0 373--386, 2019.
\newblock ISSN 0860-021X, 2083-1862.
\newblock \doi{10.5114/biolsport.2019.88759}.

\bibitem[Ben~Abdelounis et~al.(2013)Ben~Abdelounis, Nicolle, Ott{\'e}nio,
  Beillas, and Mitton]{benabdelounisEffectTwoLoading2013}
Houcine Ben~Abdelounis, St{\'e}phane Nicolle, M{\'e}lanie Ott{\'e}nio, Philippe
  Beillas, and David Mitton.
\newblock Effect of two loading rates on the elasticity of the human anterior
  rectus sheath.
\newblock \emph{Journal of the Mechanical Behavior of Biomedical Materials},
  20:\penalty0 1--5, April 2013.
\newblock ISSN 1751-6161.
\newblock \doi{10.1016/j.jmbbm.2012.12.002}.

\bibitem[Lien et~al.(2015)Lien, Hu, Wollstein, Franz, Patel, Kuzon, and
  Urbanchek]{lienContractionAbdominalWall2015}
Samuel~C. Lien, Yaxi Hu, Adi Wollstein, Michael~G. Franz, Shaun~P. Patel,
  William~M. Kuzon, and Melanie~G. Urbanchek.
\newblock Contraction of abdominal wall muscles influences size and occurrence
  of incisional hernia.
\newblock \emph{Surgery}, 158\penalty0 (1):\penalty0 278--288, July 2015.
\newblock ISSN 1532-7361.
\newblock \doi{10.1016/j.surg.2015.01.023}.

\bibitem[Hern{\'a}ndez et~al.(2011)Hern{\'a}ndez, Pe{\~n}a, Pascual,
  Rodr{\'i}guez, Calvo, Doblar{\'e}, and
  Bell{\'o}n]{hernandezMechanicalHistologicalCharacterization2011}
Bel{\'e}n Hern{\'a}ndez, E.~Pe{\~n}a, G.~Pascual, M.~Rodr{\'i}guez, B.~Calvo,
  M.~Doblar{\'e}, and Juan~M. Bell{\'o}n.
\newblock Mechanical and histological characterization of the abdominal muscle.
  {{A}} previous step to modelling hernia surgery.
\newblock \emph{Journal of the Mechanical Behavior of Biomedical Materials},
  4\penalty0 (3):\penalty0 392--404, April 2011.
\newblock ISSN 1751-6161.
\newblock \doi{10.1016/j.jmbbm.2010.11.012}.

\bibitem[Axer et~al.(2001{\natexlab{b}})Axer, v.~Keyserlingk, and
  Prescher]{axerCollagenFibersLinea2001}
Hubertus Axer, Diedrich~Graf v.~Keyserlingk, and Andreas Prescher.
\newblock Collagen {{Fibers}} in {{Linea Alba}} and {{Rectus Sheaths}}: {{I}}.
  {{General Scheme}} and {{Morphological Aspects}}.
\newblock \emph{Journal of Surgical Research}, 96\penalty0 (1):\penalty0
  127--134, March 2001{\natexlab{b}}.
\newblock ISSN 0022-4804.
\newblock \doi{10.1006/jsre.2000.6070}.

\bibitem[Hollinsky and Sandberg(2007)]{hollinskyMeasurementTensileStrength2007}
Christian Hollinsky and Simone Sandberg.
\newblock Measurement of the tensile strength of the ventral abdominal wall in
  comparison with scar tissue.
\newblock \emph{Clinical Biomechanics}, 22\penalty0 (1):\penalty0 88--92,
  January 2007.
\newblock ISSN 0268-0033.
\newblock \doi{10.1016/j.clinbiomech.2006.06.002}.

\bibitem[Tran et~al.(2014)Tran, Mitton, Voirin, Turquier, and
  Beillas]{tranContributionSkinRectus2014}
Doris Tran, David Mitton, David Voirin, Fr{\'e}d{\'e}ric Turquier, and Philippe
  Beillas.
\newblock Contribution of the skin, rectus abdominis and their sheaths to the
  structural response of the abdominal wall ex vivo.
\newblock \emph{Journal of Biomechanics}, 47\penalty0 (12):\penalty0
  3056--3063, September 2014.
\newblock ISSN 0021-9290.
\newblock \doi{10.1016/j.jbiomech.2014.06.031}.

\bibitem[{Hern{\'a}ndez-Gasc{\'o}n}
  et~al.(2013{\natexlab{a}}){Hern{\'a}ndez-Gasc{\'o}n}, Mena, Pe{\~n}a,
  Pascual, Bell{\'o}n, and
  Calvo]{hernandez-gasconUnderstandingPassiveMechanical2013}
Bel{\'e}n {Hern{\'a}ndez-Gasc{\'o}n}, A~Mena, E~Pe{\~n}a, G~Pascual, J~M
  Bell{\'o}n, and B~Calvo.
\newblock Understanding the passive mechanical behavior of the human abdominal
  wall.
\newblock \emph{Annals of biomedical engineering}, 41\penalty0 (2):\penalty0
  433--444, February 2013{\natexlab{a}}.
\newblock ISSN 1573-9686.
\newblock \doi{10.1007/s10439-012-0672-7}.

\bibitem[Konerding et~al.(2011)Konerding, Bohn, Wolloscheck, Batke, Holste,
  Wohlert, Trzewik, F{\"o}rstemann, and
  Hartung]{konerdingMaximumForcesActing2011}
Moritz~A. Konerding, Michael Bohn, Tanja Wolloscheck, Boris Batke, J{\"o}rg-L.
  Holste, Stephen Wohlert, J{\"u}rgen Trzewik, Thorsten F{\"o}rstemann, and
  Christoph Hartung.
\newblock Maximum forces acting on the abdominal wall: Experimental validation
  of a theoretical modeling in a human cadaver study.
\newblock \emph{Medical Engineering \& Physics}, 33\penalty0 (6):\penalty0
  789--792, July 2011.
\newblock ISSN 1873-4030.
\newblock \doi{10.1016/j.medengphy.2011.01.010}.

\bibitem[Karkhaneh~Yousefi et~al.(2023)Karkhaneh~Yousefi, Pierrat, Le~Ruyet,
  and Avril]{karkhanehyousefiPatientspecificComputationalSimulations2023}
Ali~Akbar Karkhaneh~Yousefi, Baptiste Pierrat, Anicet Le~Ruyet, and
  St{\'e}phane Avril.
\newblock Patient-specific computational simulations of wound healing following
  midline laparotomy closure.
\newblock \emph{Biomechanics and Modeling in Mechanobiology}, 22\penalty0
  (5):\penalty0 1589--1605, October 2023.
\newblock ISSN 1617-7940.
\newblock \doi{10.1007/s10237-023-01708-3}.

\bibitem[Roeder(2013)]{roederChapter3Mechanical2013}
Ryan~K. Roeder.
\newblock Chapter 3 - {{Mechanical Characterization}} of~{{Biomaterials}}.
\newblock In Amit Bandyopadhyay and Susmita Bose, editors,
  \emph{Characterization of {{Biomaterials}}}, pages 49--104. Academic Press,
  Oxford, January 2013.
\newblock ISBN 978-0-12-415800-9.
\newblock \doi{10.1016/B978-0-12-415800-9.00003-6}.

\bibitem[Deeken and Lake(2017)]{deekenMechanicalPropertiesAbdominal2017}
Corey~R. Deeken and Spencer~P. Lake.
\newblock Mechanical properties of the abdominal wall and biomaterials utilized
  for hernia repair.
\newblock \emph{Journal of the Mechanical Behavior of Biomedical Materials},
  74:\penalty0 411--427, October 2017.
\newblock ISSN 1751-6161.
\newblock \doi{10.1016/j.jmbbm.2017.05.008}.

\bibitem[Cardoso(2012)]{cardosoExperimentalStudyHuman2012}
Maria Helena~Sequeira Cardoso.
\newblock Experimental {{Study}} of the {{Human Anterolateral Abdominal Wall}}.
\newblock page 104, July 2012.

\bibitem[Astruc et~al.(2018)Astruc, De~Meulaere, Witz, Nov{\'a}{\v c}ek,
  Turquier, Hoc, and Brieu]{astrucCharacterizationAnisotropicMechanical2018}
Laure Astruc, Maurice De~Meulaere, Jean-Fran{\c c}ois Witz, Vit Nov{\'a}{\v
  c}ek, Fr{\'e}d{\'e}ric Turquier, Thierry Hoc, and Mathias Brieu.
\newblock Characterization of the anisotropic mechanical behavior of human
  abdominal wall connective tissues.
\newblock \emph{Journal of the Mechanical Behavior of Biomedical Materials},
  82:\penalty0 45--50, June 2018.
\newblock ISSN 1751-6161.
\newblock \doi{10.1016/j.jmbbm.2018.03.012}.

\bibitem[Kirkpatrick et~al.(2013)Kirkpatrick, Roberts, De~Waele, Jaeschke,
  Malbrain, De~Keulenaer, Duchesne, Bjorck, Leppaniemi, Ejike, Sugrue,
  Cheatham, Ivatury, Ball, Reintam~Blaser, Regli, Balogh, D'Amours, Debergh,
  Kaplan, Kimball, Olvera, and {Pediatric Guidelines Sub-Committee for the
  World Society of the Abdominal Compartment
  Syndrome}]{kirkpatrickIntraabdominalHypertensionAbdominal2013}
Andrew~W. Kirkpatrick, Derek~J. Roberts, Jan De~Waele, Roman Jaeschke, Manu L.
  N.~G. Malbrain, Bart De~Keulenaer, Juan Duchesne, Martin Bjorck, Ari
  Leppaniemi, Janeth~C. Ejike, Michael Sugrue, Michael Cheatham, Rao Ivatury,
  Chad~G. Ball, Annika Reintam~Blaser, Adrian Regli, Zsolt~J. Balogh, Scott
  D'Amours, Dieter Debergh, Mark Kaplan, Edward Kimball, Claudia Olvera, and
  {Pediatric Guidelines Sub-Committee for the World Society of the Abdominal
  Compartment Syndrome}.
\newblock Intra-abdominal hypertension and the abdominal compartment syndrome:
  Updated consensus definitions and clinical practice guidelines from the
  {{World Society}} of the {{Abdominal Compartment Syndrome}}.
\newblock \emph{Intensive Care Medicine}, 39\penalty0 (7):\penalty0 1190--1206,
  July 2013.
\newblock ISSN 1432-1238.
\newblock \doi{10.1007/s00134-013-2906-z}.

\bibitem[Malbrain et~al.(2014)Malbrain, De~Laet, De~Waele, Sugrue, Schachtrupp,
  Duchesne, Van~Ramshorst, De~Keulenaer, Kirkpatrick, {Ahmadi-Noorbakhsh},
  Mulier, Pelosi, Ivatury, Pracca, David, and
  Roberts]{malbrainRoleAbdominalCompliance2014}
Manu Malbrain, Inneke De~Laet, Jan~J. De~Waele, Michael Sugrue, Alexander
  Schachtrupp, Juan Duchesne, Gabrielle Van~Ramshorst, Bart De~Keulenaer,
  Andrew~W. Kirkpatrick, Siavash {Ahmadi-Noorbakhsh}, Jan Mulier, Paolo Pelosi,
  Rao Ivatury, Francisco Pracca, Marcelo David, and Derek~J. Roberts.
\newblock The role of abdominal compliance, the neglected parameter in
  critically ill patients - a consensus review of 16. {{Part}} 2: Measurement
  techniques and management recommendations.
\newblock \emph{Anaesthesiology Intensive Therapy}, 46\penalty0 (5):\penalty0
  406--432, 2014.
\newblock ISSN 1731-2531.
\newblock \doi{10.5603/AIT.2014.0063}.

\bibitem[Gr{\"a}{$\beta$}el et~al.(2005)Gr{\"a}{$\beta$}el, Prescher, Fitzek,
  v.~Keyserlingk, and Axer]{gravelAnisotropyHumanLinea2005}
David Gr{\"a}{$\beta$}el, Andreas Prescher, Sabine Fitzek, Diedrich~Graf
  v.~Keyserlingk, and Hubertus Axer.
\newblock Anisotropy of human linea alba: {{A}} biomechanical study.
\newblock \emph{Journal of Surgical Research}, 124\penalty0 (1):\penalty0
  118--125, March 2005.
\newblock ISSN 0022-4804.
\newblock \doi{10.1016/j.jss.2004.10.010}.

\bibitem[F{\"o}rstemann et~al.(2011)F{\"o}rstemann, Trzewik, Holste, Batke,
  Konerding, Wolloscheck, and
  Hartung]{forstemannForcesDeformationsAbdominal2011}
Thorsten F{\"o}rstemann, Juergen Trzewik, J.~Holste, B.~Batke, M.~A. Konerding,
  Tanja Wolloscheck, and Christoph Hartung.
\newblock Forces and deformations of the abdominal wall---{{A}} mechanical and
  geometrical approach to the linea alba.
\newblock \emph{Journal of Biomechanics}, 44\penalty0 (4):\penalty0 600--606,
  February 2011.
\newblock ISSN 0021-9290.
\newblock \doi{10.1016/j.jbiomech.2010.11.021}.

\bibitem[Junge et~al.(2001)Junge, Klinge, Prescher, Giboni, Niewiera, and
  Schumpelick]{jungeElasticityAnteriorAbdominal2001}
Karsten Junge, Uwe Klinge, A.~Prescher, P.~Giboni, M.~Niewiera, and Volker
  Schumpelick.
\newblock Elasticity of the anterior abdominal wall and impact for reparation
  of incisional hernias using mesh implants.
\newblock \emph{Hernia: The Journal of Hernias and Abdominal Wall Surgery},
  5\penalty0 (3):\penalty0 113--118, September 2001.
\newblock ISSN 1265-4906.
\newblock \doi{10.1007/s100290100019}.

\bibitem[Szymczak et~al.(2011)Szymczak, Lubowiecka, Tomaszewska, and
  {\'S}mieta{\'n}ski]{szymczakInvestigationAbdomenSurface2011}
Czes{\l}aw Szymczak, Izabela Lubowiecka, Agnieszka Tomaszewska, and Maciej
  {\'S}mieta{\'n}ski.
\newblock Investigation of abdomen surface deformation due to life excitation:
  {{Implications}} for implant selection and orientation in laparoscopic
  ventral hernia repair.
\newblock \emph{Clinical Biomechanics}, 27\penalty0 (2):\penalty0 105--110,
  2011.
\newblock ISSN 0268-0033.
\newblock \doi{10.1016/j.clinbiomech.2011.08.008}.

\bibitem[{\'S}mieta{\'n}ski et~al.(2012){\'S}mieta{\'n}ski, Bury, Tomaszewska,
  Lubowiecka, and Szymczak]{smietanskiBiomechanicsFrontAbdominal2012}
Maciej {\'S}mieta{\'n}ski, Kamil Bury, Agnieszka Tomaszewska, Izabela
  Lubowiecka, and Czes{\l}aw Szymczak.
\newblock Biomechanics of the front abdominal wall as a potential factor
  leading to recurrence with laparoscopic ventral hernia repair.
\newblock \emph{Surgical Endoscopy}, 26\penalty0 (5):\penalty0 1461--1467, May
  2012.
\newblock ISSN 1432-2218.
\newblock \doi{10.1007/s00464-011-2056-8}.

\bibitem[Song et~al.(2006{\natexlab{a}})Song, Alijani, Frank, Hanna, and
  Cuschieri]{songElasticityLivingAbdominal2006}
Chengli Song, Afshin Alijani, Tim Frank, George Hanna, and Alfred Cuschieri.
\newblock Elasticity of the living abdominal wall in laparoscopic surgery.
\newblock \emph{Journal of Biomechanics}, 39\penalty0 (3):\penalty0 587--591,
  January 2006{\natexlab{a}}.
\newblock ISSN 00219290.
\newblock \doi{10.1016/j.jbiomech.2004.12.019}.

\bibitem[Levillain et~al.(2016)Levillain, Orhant, Turquier, and
  Hoc]{levillainContributionCollagenElastin2016}
Aur{\'e}lie Levillain, M.~Orhant, Fr{\'e}d{\'e}ric Turquier, and Thierry Hoc.
\newblock Contribution of collagen and elastin fibers to the mechanical
  behavior of an abdominal connective tissue.
\newblock \emph{Journal of the Mechanical Behavior of Biomedical Materials},
  61:\penalty0 308--317, August 2016.
\newblock ISSN 1751-6161.
\newblock \doi{10.1016/j.jmbbm.2016.04.006}.

\bibitem[Kirilova et~al.(2011)Kirilova, Stoytchev, Pashkouleva, and
  Kavardzhikov]{kirilovaExperimentalStudyMechanical2011}
Miglena Kirilova, Stoyan Stoytchev, Dessislava Pashkouleva, and Vasil
  Kavardzhikov.
\newblock Experimental study of the mechanical properties of human abdominal
  fascia.
\newblock \emph{Medical Engineering \& Physics}, 33\penalty0 (1):\penalty0
  1--6, January 2011.
\newblock ISSN 1350-4533.
\newblock \doi{10.1016/j.medengphy.2010.07.017}.

\bibitem[Martins et~al.(2012)Martins, Pe{\~n}a, Jorge, Santos, Santos,
  Mascarenhas, and Calvo]{martinsMechanicalCharacterizationConstitutive2012}
Pedro Martins, Estefan{\'i}a Pe{\~n}a, R.~M.~Natal Jorge, A.~Santos, L.~Santos,
  T.~Mascarenhas, and Bego{\~n}a Calvo.
\newblock Mechanical characterization and constitutive modelling of the damage
  process in rectus sheath.
\newblock \emph{Journal of the Mechanical Behavior of Biomedical Materials},
  8:\penalty0 111--122, April 2012.
\newblock ISSN 1751-6161.
\newblock \doi{10.1016/j.jmbbm.2011.12.005}.

\bibitem[Lubowiecka et~al.(2022)Lubowiecka, Szepietowska, Tomaszewska, Bielski,
  Chmielewski, {Lichodziejewska-Niemierko}, and
  Szymczak]{lubowieckaNovelVivoApproach2022}
Izabela Lubowiecka, Katarzyna Szepietowska, Agnieszka Tomaszewska, Pawel~Michal
  Bielski, Michal Chmielewski, Monika {Lichodziejewska-Niemierko}, and Czeslaw
  Szymczak.
\newblock A novel in vivo approach to assess strains of the human abdominal
  wall under known intraabdominal pressure.
\newblock \emph{JOURNAL OF THE MECHANICAL BEHAVIOR OF BIOMEDICAL MATERIALS},
  125, January 2022.
\newblock ISSN 1751-6161.
\newblock \doi{10.1016/j.jmbbm.2021.104902}.

\bibitem[Kriener et~al.(2023)Kriener, Lala, Homes, Finley, Sinclair, Williams,
  and Midwinter]{krienerMechanicalCharacterizationHuman2023}
Kyleigh Kriener, Raushan Lala, Ryan Anthony~Peter Homes, Hayley Finley, Kate
  Sinclair, Mason~Kelley Williams, and Mark~John Midwinter.
\newblock Mechanical {{Characterization}} of the {{Human Abdominal Wall Using
  Uniaxial Tensile Testing}}.
\newblock \emph{Bioengineering}, 10\penalty0 (10):\penalty0 1213, October 2023.
\newblock ISSN 2306-5354.
\newblock \doi{10.3390/bioengineering10101213}.

\bibitem[McDougall et~al.(1994)McDougall, Figenshau, Clayman, Monk, and
  Smith]{mcdougallLaparoscopicPneumoperitoneumImpact1994}
Elspeth~M. McDougall, Robert~S. Figenshau, Ralph~V. Clayman, Terri~G. Monk, and
  Deborah~S. Smith.
\newblock Laparoscopic {{Pneumoperitoneum}}: {{Impact}} of {{Body Habitus}}.
\newblock \emph{Journal of Laparoendoscopic Surgery}, 4\penalty0 (6):\penalty0
  385--391, December 1994.
\newblock ISSN 1052-3901.
\newblock \doi{10.1089/lps.1994.4.385}.

\bibitem[Ott(2019)]{ottAbdominalComplianceLaparoscopy2019}
Douglas~E. Ott.
\newblock Abdominal {{Compliance}} and {{Laparoscopy}}: {{A Review}}.
\newblock \emph{JSLS : Journal of the Society of Laparoendoscopic Surgeons},
  23\penalty0 (1):\penalty0 e2018.00080, 2019.
\newblock ISSN 1086-8089.
\newblock \doi{10.4293/JSLS.2018.00080}.

\bibitem[Becker et~al.(2017)Becker, Plymale, Wennergren, Totten, Stigall, and
  Roth]{beckerComplianceAbdominalWall2017}
Chuck Becker, Margaret~A. Plymale, John Wennergren, Crystal Totten, Kyle
  Stigall, and J.~Scott Roth.
\newblock Compliance of the abdominal wall during laparoscopic insufflation.
\newblock \emph{Surgical Endoscopy}, 31\penalty0 (4):\penalty0 1947--1951,
  April 2017.
\newblock ISSN 0930-2794, 1432-2218.
\newblock \doi{10.1007/s00464-016-5201-6}.

\bibitem[Casati et~al.(1997)Casati, Valentini, Ferrari, Senatore, Zangrillo,
  and Torri]{casatiCardiorespiratoryChangesGynaecological1997}
Andrea Casati, Giovanna Valentini, S~Ferrari, R~Senatore, Alberto Zangrillo,
  and Giorgio Torri.
\newblock Cardiorespiratory changes during gynaecological laparoscopy by
  abdominal wall elevation: Comparison with carbon dioxide pneumoperitoneum.
\newblock \emph{British Journal of Anaesthesia}, 78\penalty0 (1):\penalty0
  51--54, January 1997.
\newblock ISSN 0007-0912.
\newblock \doi{10.1093/bja/78.1.51}.

\bibitem[Kamine et~al.(2014)Kamine, Papavassiliou, and
  Schneider]{kamineEffectAbdominalInsufflation2014}
Tovy~Haber Kamine, Efstathios Papavassiliou, and Benjamin~E. Schneider.
\newblock Effect of {{Abdominal Insufflation}} for {{Laparoscopy}} on
  {{Intracranial Pressure}}.
\newblock \emph{JAMA Surgery}, 149\penalty0 (4):\penalty0 380--382, April 2014.
\newblock ISSN 2168-6254.
\newblock \doi{10.1001/jamasurg.2013.3024}.

\bibitem[Holzapfel(2001)]{holzapfelBiomechanicsSoftTissue2001}
Gerhard~A Holzapfel.
\newblock Biomechanics of {{Soft Tissue}}.
\newblock 2001.

\bibitem[Cooney et~al.(2016)Cooney, Lake, Thompson, Castile, Winter, and
  Simms]{cooneyUniaxialBiaxialTensile2016}
Gerard~M. Cooney, Spencer~P. Lake, Dominic~M. Thompson, Ryan~M. Castile, Des~C.
  Winter, and Ciaran~K. Simms.
\newblock Uniaxial and biaxial tensile stress--stretch response of human linea
  alba.
\newblock \emph{Journal of the Mechanical Behavior of Biomedical Materials},
  63:\penalty0 134--140, October 2016.
\newblock ISSN 17516161.
\newblock \doi{10.1016/j.jmbbm.2016.06.015}.

\bibitem[Anurov et~al.(2012)Anurov, Titkova, and
  Oettinger]{anurovBiomechanicalCompatibilitySurgical2012}
Mikhail Anurov, S.~M. Titkova, and Alexander Oettinger.
\newblock Biomechanical compatibility of surgical mesh and fascia being
  reinforced: Dependence of experimental hernia defect repair results on
  anisotropic surgical mesh positioning.
\newblock \emph{Hernia}, 16\penalty0 (2):\penalty0 199--210, April 2012.
\newblock ISSN 1248-9204.
\newblock \doi{10.1007/s10029-011-0877-y}.

\bibitem[Kureshi et~al.(2008)Kureshi, Vaiude, Nazhat, Petrie, and
  Brown]{kureshiMatrixMechanicalProperties2008}
Alvena Kureshi, Partha Vaiude, Showan~N. Nazhat, Aviva Petrie, and Robert~A.
  Brown.
\newblock Matrix mechanical properties of transversalis fascia in inguinal
  herniation as a model for tissue expansion.
\newblock \emph{Journal of Biomechanics}, 41\penalty0 (16):\penalty0
  3462--3468, December 2008.
\newblock ISSN 0021-9290.
\newblock \doi{10.1016/j.jbiomech.2008.08.018}.

\bibitem[Yang et~al.(2018)Yang, Wang, Yan, Xiang, Tang, Zhang, Liu, and
  Qiu]{yangDeterminationNormalSkin2018}
Yujia Yang, Liyun Wang, Feng Yan, Xi~Xiang, Yuanjiao Tang, Lingyan Zhang, Jibin
  Liu, and Li~Qiu.
\newblock Determination of {{Normal Skin Elasticity}} by {{Using Real-time
  Shear Wave Elastography}}.
\newblock \emph{Journal of Ultrasound in Medicine}, 37\penalty0 (11):\penalty0
  2507--2516, 2018.
\newblock ISSN 1550-9613.
\newblock \doi{10.1002/jum.14608}.

\bibitem[Song et~al.(2006{\natexlab{b}})Song, Alijani, Frank, Hanna, and
  Cuschieri]{songMechanicalPropertiesHuman2006}
Chengli Song, Afshin Alijani, Tim Frank, George Hanna, and Alfred Cuschieri.
\newblock Mechanical properties of the human abdominal wall measured in vivo
  during insufflation for laparoscopic surgery.
\newblock \emph{Surgical Endoscopy And Other Interventional Techniques},
  20\penalty0 (6):\penalty0 987--990, June 2006{\natexlab{b}}.
\newblock ISSN 1432-2218.
\newblock \doi{10.1007/s00464-005-0676-6}.

\bibitem[Le~Ruyet et~al.(2020)Le~Ruyet, Yurtkap, den Hartog, Vegleur, Turquier,
  Lange, and Kleinrensink]{leruyetDifferencesBiomechanicsAbdominal2020}
Anicet Le~Ruyet, Y.~Yurtkap, F.~P.~J. den Hartog, A.~Vegleur, F.~Turquier,
  Johan~F. Lange, and Gert-Jan Kleinrensink.
\newblock Differences in biomechanics of abdominal wall closure with and
  without mesh reinforcement: {{A}} study in post mortem human specimens.
\newblock \emph{Journal of the Mechanical Behavior of Biomedical Materials},
  105:\penalty0 103683, May 2020.
\newblock ISSN 1751-6161.
\newblock \doi{10.1016/j.jmbbm.2020.103683}.

\bibitem[Siassi et~al.(2014)Siassi, Mahn, Baumann, Vollmer, Huber, Morlock, and
  Kallinowski]{siassiDevelopmentDynamicModel2014}
Michael Siassi, A.~Mahn, E.~Baumann, M.~Vollmer, G.~Huber, Michael Morlock, and
  Friedrich Kallinowski.
\newblock Development of a dynamic model for ventral hernia mesh repair.
\newblock \emph{Langenbeck's Archives of Surgery}, 399\penalty0 (7):\penalty0
  857--862, October 2014.
\newblock ISSN 1435-2451.
\newblock \doi{10.1007/s00423-014-1239-x}.

\bibitem[Lyons et~al.(2015)Lyons, Mohan, Winter, and
  Simms]{lyonsBiomechanicalAbdominalWall2015}
Mathew Lyons, H~Mohan, D~C Winter, and Ciaran~K. Simms.
\newblock Biomechanical abdominal wall model applied to hernia repair.
\newblock \emph{British Journal of Surgery}, 102\penalty0 (2):\penalty0
  e133--e139, January 2015.
\newblock ISSN 0007-1323.
\newblock \doi{10.1002/bjs.9687}.

\bibitem[Kroese et~al.(2017)Kroese, Harlaar, Ordrenneau, Verhelst, Gu{\'e}rin,
  Turquier, Goossens, Kleinrensink, Jeekel, and
  Lange]{kroeseAbdoMANArtificialAbdominal2017}
Leonard~F. Kroese, J.~J. Harlaar, C.~Ordrenneau, J.~Verhelst, Gaetan
  Gu{\'e}rin, Fr{\'e}d{\'e}ric Turquier, R.~H.~M. Goossens, G.-J. Kleinrensink,
  J.~Jeekel, and Johan~F. Lange.
\newblock The `{{AbdoMAN}}': An artificial abdominal wall simulator for
  biomechanical studies on laparotomy closure techniques.
\newblock \emph{Hernia}, 21\penalty0 (5):\penalty0 783--791, October 2017.
\newblock ISSN 1248-9204.
\newblock \doi{10.1007/s10029-017-1615-x}.

\bibitem[Roy et~al.(2023)Roy, Khan, Krishna, Bhatia, Prakash, and
  Bansal]{royComparativeStudyEvaluate2023}
Nilanjan~Barman Roy, Washim~Firoz Khan, Asuri Krishna, Renu Bhatia, Om~Prakash,
  and Virinder~Kumar Bansal.
\newblock A comparative study to evaluate abdominal wall dynamics in patients
  with incisional hernia compared to healthy controls.
\newblock \emph{Surgical Endoscopy}, 37\penalty0 (12):\penalty0 9414--9419,
  December 2023.
\newblock ISSN 1432-2218.
\newblock \doi{10.1007/s00464-023-10408-z}.

\bibitem[Strig{\aa}rd et~al.(2016)Strig{\aa}rd, Clay, Stark, Gunnarsson, and
  Falk]{strigardGiantVentralHernia2016}
Karin Strig{\aa}rd, L.~Clay, B.~Stark, U.~Gunnarsson, and Peter Falk.
\newblock Giant ventral hernia---relationship between abdominal wall muscle
  strength and hernia area.
\newblock \emph{BMC Surgery}, 16\penalty0 (1):\penalty0 1--6, December 2016.
\newblock ISSN 1471-2482.
\newblock \doi{10.1186/s12893-016-0166-x}.

\bibitem[Criss et~al.(2014)Criss, Petro, Krpata, Seafler, Lai, Fiutem,
  Novitsky, and Rosen]{crissFunctionalAbdominalWall2014}
Cory~N. Criss, Clayton~C. Petro, David~M. Krpata, Christina~M. Seafler, Nicola
  Lai, Justin Fiutem, Yuri~W. Novitsky, and Michael~J. Rosen.
\newblock Functional abdominal wall reconstruction improves core physiology and
  quality-of-life.
\newblock \emph{Surgery}, 156\penalty0 (1):\penalty0 176--182, July 2014.
\newblock ISSN 0039-6060.
\newblock \doi{10.1016/j.surg.2014.04.010}.

\bibitem[Ahmed et~al.(2018)Ahmed, Mohamed, {ABOU EL-NAGA}, and
  {El-Kablawy}]{ahmedEffectPreoperativeAbdominal2018}
Mohamed~G. Ahmed, Salah~A. Mohamed, Walid~A. {ABOU EL-NAGA}, and Mohamed~M.
  {El-Kablawy}.
\newblock Effect of {{Preoperative Abdominal Training}} on {{Abdominal Muscles
  Strength Outcomes}} after {{Ventral Hernia Repair}}.
\newblock \emph{The Medical Journal of Cairo University}, 86\penalty0
  (12):\penalty0 4495--4501, December 2018.
\newblock ISSN 2536-9806.
\newblock \doi{10.21608/mjcu.2018.63151}.

\bibitem[S{\'a}nchez~Arteaga et~al.(2024)S{\'a}nchez~Arteaga, Gil~Delgado,
  Feria~Madue{\~n}o, Tall{\'o}n~Aguilar, Sa{\~n}udo, and
  Padillo~Ruiz]{sanchezarteagaImpactIncisionalHernia2024}
Alejandro S{\'a}nchez~Arteaga, Jos{\'e}~Luis Gil~Delgado, Adri{\'a}n
  Feria~Madue{\~n}o, Luis Tall{\'o}n~Aguilar, Borja Sa{\~n}udo, and Javier
  Padillo~Ruiz.
\newblock Impact of incisional hernia on abdominal wall strength.
\newblock \emph{BJS Open}, 8\penalty0 (3):\penalty0 zrae045, June 2024.
\newblock ISSN 2474-9842.
\newblock \doi{10.1093/bjsopen/zrae045}.

\bibitem[Garc{\'i}a~Moriana et~al.(2023)Garc{\'i}a~Moriana,
  S{\'a}nchez~Arteaga, Gil~Delgado, Maroto~S{\'a}nchez, Feria~Madue{\~n}o,
  Tall{\'o}n~Aguilar, Padillo~Ruiz, and
  Sa{\~n}udo]{garciamorianaEvaluationRectusAbdominis2023}
Antonio~Jes{\'u}s Garc{\'i}a~Moriana, Alejandros S{\'a}nchez~Arteaga, J.~L.
  Gil~Delgado, R.~Maroto~S{\'a}nchez, A.~Feria~Madue{\~n}o,
  L.~Tall{\'o}n~Aguilar, J.~Padillo~Ruiz, and Borja Sa{\~n}udo.
\newblock Evaluation of rectus abdominis muscle strength and width of hernia
  defect in patients undergoing incisional hernia surgery.
\newblock \emph{Hernia}, 27\penalty0 (4):\penalty0 919--926, August 2023.
\newblock ISSN 1248-9204.
\newblock \doi{10.1007/s10029-023-02834-8}.

\bibitem[Cobb et~al.(2005)Cobb, Burns, Kercher, Matthews, Norton, and
  Heniford]{cobbNormalIntraabdominalPressure2005}
William~S. Cobb, Justin~M. Burns, Kent~W. Kercher, Brent~D. Matthews, H.~James
  Norton, and B.~Todd Heniford.
\newblock Normal {{Intraabdominal Pressure}} in {{Healthy Adults}}.
\newblock \emph{Journal of Surgical Research}, 129\penalty0 (2):\penalty0
  231--235, December 2005.
\newblock ISSN 0022-4804, 1095-8673.
\newblock \doi{10.1016/j.jss.2005.06.015}.

\bibitem[Liao et~al.(2020)Liao, Cheng, Chen, Jow, Chen, Lai, Chen, and
  Ho]{liaoIngestibleElectronicsContinuous2020}
Chien-Hung Liao, Chi-Tung Cheng, Chih-Chi Chen, Uei-Ming Jow, Chun-Hung Chen,
  Yen-Liang Lai, Ya-Chuan Chen, and Dong-Ru Ho.
\newblock An {{Ingestible Electronics}} for {{Continuous}} and {{Real-Time
  Intraabdominal Pressure Monitoring}}.
\newblock \emph{Journal of Personalized Medicine}, 11\penalty0 (1):\penalty0
  12, December 2020.
\newblock ISSN 2075-4426.
\newblock \doi{10.3390/jpm11010012}.

\bibitem[Tayebi et~al.(2021)Tayebi, Gutierrez, Mohout, Smets, Wise, Stiens, and
  Malbrain]{tayebiConciseOverviewNoninvasive2021}
Salar Tayebi, Adrian Gutierrez, Ikram Mohout, Evelien Smets, Robert Wise, Johan
  Stiens, and Manu L. N.~G. Malbrain.
\newblock A concise overview of non-invasive intra-abdominal pressure
  measurement techniques: From bench to bedside.
\newblock \emph{Journal of Clinical Monitoring and Computing}, 35\penalty0
  (1):\penalty0 51--70, February 2021.
\newblock ISSN 1387-1307, 1573-2614.
\newblock \doi{10.1007/s10877-020-00561-4}.

\bibitem[Vincent et~al.(2023)Vincent, Mietzsch, Braun, Trochimiuk, Reinshagen,
  and Boettcher]{vincentAbdominalWallMovements2023}
Deirdre Vincent, Stefan Mietzsch, Wolfgang Braun, Magdalena Trochimiuk, Konrad
  Reinshagen, and Michael Boettcher.
\newblock Abdominal {{Wall Movements Predict Intra-Abdominal Pressure Changes}}
  in {{Rats}}: {{A Novel Non-Invasive Intra-Abdominal Pressure Detection
  Method}}.
\newblock \emph{Children}, 10\penalty0 (8):\penalty0 1422, August 2023.
\newblock ISSN 2227-9067.
\newblock \doi{10.3390/children10081422}.

\bibitem[Li et~al.(2024)Li, Wang, and
  Lu]{liDevelopmentFeasibilityCredibility2024}
ZhiRu Li, HuaFen Wang, and FangYan Lu.
\newblock The development, feasibility and credibility of intra-abdominal
  pressure measurement techniques: {{A}} scoping review.
\newblock \emph{PLOS ONE}, 19\penalty0 (3):\penalty0 e0297982, March 2024.
\newblock ISSN 1932-6203.
\newblock \doi{10.1371/journal.pone.0297982}.

\bibitem[Szepietowska et~al.(2023)Szepietowska, Troka,
  {Lichodziejewska-Niemierko}, Chmielewski, and
  Lubowiecka]{szepietowskaFullfieldVivoExperimental2023}
Katarzyna Szepietowska, Mateusz Troka, Monika {Lichodziejewska-Niemierko},
  Micha{\l} Chmielewski, and Izabela Lubowiecka.
\newblock Full-field in vivo experimental study of the strains of a breathing
  human abdominal wall with intra-abdominal pressure variation.
\newblock \emph{Journal of the Mechanical Behavior of Biomedical Materials},
  147:\penalty0 106148, November 2023.
\newblock ISSN 1751-6161.
\newblock \doi{10.1016/j.jmbbm.2023.106148}.

\bibitem[Hope et~al.(2018)Hope, Williams, Rawles, Hooks, Clancy, and
  Eckhauser]{hopeRationaleTechniqueMeasuring2018}
William~W. Hope, Zachary~F. Williams, James~W. Rawles, W.~Borden Hooks,
  Thomas~V. Clancy, and Frederic~E. Eckhauser.
\newblock Rationale and {{Technique}} for {{Measuring Abdominal Wall Tension}}
  in {{Hernia Repair}}.
\newblock \emph{The American Surgeon}, 84\penalty0 (9):\penalty0 1446--1449,
  September 2018.
\newblock ISSN 0003-1348.
\newblock \doi{10.1177/000313481808400947}.

\bibitem[Tenzel et~al.(2019)Tenzel, Bilezikian, Eckhauser, and
  Hope]{tenzelTensionMeasurementsAbdominal2019}
Paul~L. Tenzel, Jordan~A. Bilezikian, Frederic~E. Eckhauser, and William~W.
  Hope.
\newblock Tension measurements in abdominal wall hernia repair: {{Concept}} and
  clinical applications.
\newblock \emph{International Journal of Abdominal Wall and Hernia Surgery},
  2\penalty0 (4):\penalty0 119, October 2019.
\newblock ISSN 2589-8736.
\newblock \doi{10.4103/ijawhs.ijawhs_37_19}.

\bibitem[Miller et~al.(2023)Miller, Ellis, Walsh, Joyce, Simon, Almassi, Lee,
  DeBernardo, Steele, Haywood, Beffa, Tu, and
  Rosen]{millerPhysiologicTensionAbdominal2023}
Benjamin~T. Miller, Ryan~C. Ellis, R.~Matthew Walsh, Daniel Joyce, Robert
  Simon, Nima Almassi, Byron Lee, Robert DeBernardo, Scott Steele, Samuel
  Haywood, Lindsey Beffa, Chao Tu, and Michael~J. Rosen.
\newblock Physiologic tension of the abdominal wall.
\newblock \emph{Surgical Endoscopy}, 37\penalty0 (12):\penalty0 9347--9350,
  December 2023.
\newblock ISSN 1432-2218.
\newblock \doi{10.1007/s00464-023-10346-w}.

\bibitem[Dudley(1970)]{dudleyLayeredMassClosure1970}
H~A~F Dudley.
\newblock Layered and mass closure of the abdominal wall: {{A}} theoretical and
  experimental analysis.
\newblock \emph{British Journal of Surgery}, 57\penalty0 (9):\penalty0
  664--667, September 1970.
\newblock ISSN 0007-1323.
\newblock \doi{10.1002/bjs.1800570908}.

\bibitem[Zienkiewicz et~al.(2025)Zienkiewicz, Taylor, and
  Govindjee]{zienkiewicz1StandardDiscrete2025}
Olgierd Zienkiewicz, R.~L. Taylor, and S.~Govindjee.
\newblock 1 - {{The}} standard discrete system and origins of the finite
  element method.
\newblock In O.~C. Zienkiewicz, R.~L. Taylor, and Sanjay Govindjee, editors,
  \emph{The {{Finite Element Method}} ({{Eighth Edition}})}, pages 1--20.
  Butterworth-Heinemann, January 2025.
\newblock ISBN 978-0-443-16044-8.
\newblock \doi{10.1016/B978-0-44-316044-8.00010-7}.

\bibitem[{Hern{\'a}ndez-Gasc{\'o}n} et~al.(2014){Hern{\'a}ndez-Gasc{\'o}n},
  Esp{\'e}s, Pe{\~n}a, Pascual, Bell{\'o}n, and
  Calvo]{hernandez-gasconComputationalFrameworkModel2014}
Bel{\'e}n {Hern{\'a}ndez-Gasc{\'o}n}, N.~Esp{\'e}s, E.~Pe{\~n}a, G.~Pascual,
  J.M. Bell{\'o}n, and B.~Calvo.
\newblock Computational framework to model and design surgical meshes for
  hernia repair.
\newblock \emph{Computer Methods in Biomechanics and Biomedical Engineering},
  17\penalty0 (10):\penalty0 1071--1085, July 2014.
\newblock ISSN 1025-5842.
\newblock \doi{10.1080/10255842.2012.736967}.

\bibitem[Lohr et~al.(2022)Lohr, Sugerman, Kakaletsis, Lejeune, and
  Rausch]{lohrIntroductionOgdenModel2022}
Matthew~J. Lohr, Gabriella~P. Sugerman, Sotirios Kakaletsis, Emma Lejeune, and
  Manuel~K. Rausch.
\newblock An introduction to the {{Ogden}} model in biomechanics: Benefits,
  implementation tools and limitations.
\newblock \emph{Philosophical Transactions of the Royal Society A:
  Mathematical, Physical and Engineering Sciences}, 380\penalty0
  (2234):\penalty0 20210365, August 2022.
\newblock \doi{10.1098/rsta.2021.0365}.

\bibitem[Grasa et~al.(2016)Grasa, Sierra, Lauzeral, Mu{\~n}oz, {Miana-Mena},
  and Calvo]{grasaActiveBehaviorAbdominal2016}
Jorge Grasa, M.~Sierra, N.~Lauzeral, M.~J. Mu{\~n}oz, F.~J. {Miana-Mena}, and
  Bego{\~n}a Calvo.
\newblock Active behavior of abdominal wall muscles: {{Experimental}} results
  and numerical model formulation.
\newblock \emph{Journal of the Mechanical Behavior of Biomedical Materials},
  61:\penalty0 444--454, August 2016.
\newblock ISSN 1751-6161.
\newblock \doi{10.1016/j.jmbbm.2016.04.013}.

\bibitem[Kauer et~al.(2002)Kauer, Vuskovic, Dual, Szekely, and
  Bajka]{kauerInverseFiniteElement2002}
Matthias Kauer, V~Vuskovic, J~Dual, G~Szekely, and Michael Bajka.
\newblock Inverse finite element characterization of soft tissues.
\newblock \emph{Medical Image Analysis}, 6\penalty0 (3):\penalty0 275--287,
  September 2002.
\newblock ISSN 1361-8415.
\newblock \doi{10.1016/S1361-8415(02)00085-3}.

\bibitem[Remus et~al.(2024)Remus, Sure, Selkmann, Uttich, and
  Bender]{remusSoftTissueMaterial2024}
Robin Remus, Christian Sure, Sascha Selkmann, Eike Uttich, and Beate Bender.
\newblock Soft tissue material properties based on human abdominal in vivo
  macro-indenter measurements.
\newblock \emph{Frontiers in Bioengineering and Biotechnology}, 12, May 2024.
\newblock ISSN 2296-4185.
\newblock \doi{10.3389/fbioe.2024.1384062}.

\bibitem[{Sim{\'o}n-Allu{\'e}} et~al.(2017){Sim{\'o}n-Allu{\'e}}, Calvo,
  Oberai, and Barbone]{simon-allueMechanicalCharacterizationAbdominal2017}
Raquel {Sim{\'o}n-Allu{\'e}}, Begona Calvo, Assad~Anshuman Oberai, and Paul~E
  Barbone.
\newblock Towards the mechanical characterization of abdominal wall by inverse
  analysis.
\newblock \emph{Journal of the Mechanical Behavior of Biomedical Materials},
  66:\penalty0 127--137, February 2017.
\newblock ISSN 1751-6161.
\newblock \doi{10.1016/j.jmbbm.2016.11.007}.

\bibitem[Cooney et~al.(2015)Cooney, Moerman, Takaza, Winter, and
  Simms]{cooneyUniaxialBiaxialMechanical2015}
Gerard~M. Cooney, Kevin~M. Moerman, Michael Takaza, Des~C. Winter, and
  Ciaran~K. Simms.
\newblock Uniaxial and biaxial mechanical properties of porcine linea alba.
\newblock \emph{Journal of the Mechanical Behavior of Biomedical Materials},
  41:\penalty0 68--82, January 2015.
\newblock ISSN 1751-6161.
\newblock \doi{10.1016/j.jmbbm.2014.09.026}.

\bibitem[Pachera et~al.(2016)Pachera, Pavan, Todros, Cavinato, Fontanella, and
  Natali]{pacheraNumericalInvestigationHealthy2016}
Paola Pachera, Piero~G. Pavan, Silvia Todros, C.~Cavinato, C.~G. Fontanella,
  and Arturo~N. Natali.
\newblock A numerical investigation of the healthy abdominal wall structures.
\newblock \emph{Journal of Biomechanics}, 49\penalty0 (9):\penalty0 1818--1823,
  June 2016.
\newblock ISSN 0021-9290.
\newblock \doi{10.1016/j.jbiomech.2016.04.019}.

\bibitem[Jourdan et~al.(2024)Jourdan, Dhume, Gu{\'e}rin, Siegel, Le~Ruyet, and
  Palmer]{jourdanNumericalInvestigationFinite2024}
Arthur Jourdan, Rohit Dhume, Elisabeth Gu{\'e}rin, Alice Siegel, Anicet
  Le~Ruyet, and Mark Palmer.
\newblock Numerical investigation of a finite element abdominal wall model
  during breathing and muscular contraction.
\newblock \emph{Computer Methods and Programs in Biomedicine}, 244:\penalty0
  107985, February 2024.
\newblock ISSN 0169-2607.
\newblock \doi{10.1016/j.cmpb.2023.107985}.

\bibitem[Zamkowski et~al.(2023)Zamkowski, Tomaszewska, Lubowiecka, and
  {\'S}mieta{\'n}ski]{zamkowskiBiomechanicalCausesFailure2023}
Mateusz Zamkowski, Agnieszka Tomaszewska, Izabela Lubowiecka, and Maciej
  {\'S}mieta{\'n}ski.
\newblock Biomechanical causes for failure of the
  {{Physiomesh}}/{{Securestrap}} system.
\newblock \emph{Scientific Reports}, 13\penalty0 (1):\penalty0 17504, October
  2023.
\newblock ISSN 2045-2322.
\newblock \doi{10.1038/s41598-023-44940-8}.

\bibitem[{Hern{\'a}ndez-Gasc{\'o}n}
  et~al.(2013{\natexlab{b}}){Hern{\'a}ndez-Gasc{\'o}n}, Pe{\~n}a, Grasa,
  Pascual, Bell{\'o}n, and
  Calvo]{hernandez-gasconMechanicalResponseHerniated2013}
Bel{\'e}n {Hern{\'a}ndez-Gasc{\'o}n}, Estefan{\'i}a Pe{\~n}a, Jorge Grasa,
  Gemma Pascual, Juan~M. Bell{\'o}n, and Bego{\~n}a Calvo.
\newblock Mechanical {{Response}} of the {{Herniated Human Abdomen}} to the
  {{Placement}} of {{Different Prostheses}}.
\newblock \emph{Journal of Biomechanical Engineering}, 135\penalty0 (051004),
  April 2013{\natexlab{b}}.
\newblock ISSN 0148-0731.
\newblock \doi{10.1115/1.4023703}.

\bibitem[Gu{\'e}rin and Turquier(2013)]{guerinImpactDefectSize2013}
Gaetan Gu{\'e}rin and Fr{\'e}d{\'e}ric Turquier.
\newblock Impact of the defect size, the mesh overlap and the fixation depth on
  ventral hernia repairs: A combined experimental and numerical approach.
\newblock \emph{Hernia: The Journal of Hernias and Abdominal Wall Surgery},
  17\penalty0 (5):\penalty0 647--655, October 2013.
\newblock ISSN 1248-9204.
\newblock \doi{10.1007/s10029-013-1050-6}.

\bibitem[Todros(2020)]{todrosBiomechanicsSurgicalMesh2020}
Silvia Todros.
\newblock Biomechanics in {{Surgical Mesh Fixation}} for {{Abdominal Wall
  Repair}}.
\newblock \emph{Journal of Investigative Surgery: The Official Journal of the
  Academy of Surgical Research}, pages 1--2, December 2020.
\newblock ISSN 1521-0553.
\newblock \doi{10.1080/08941939.2020.1849468}.

\bibitem[Karrech et~al.(2023)Karrech, Ahmad, and
  Hamdorf]{karrechBiomechanicalStabilityHerniadamaged2023}
Ali Karrech, Hairul Ahmad, and Jeffrey~M. Hamdorf.
\newblock Biomechanical stability of hernia-damaged abdominal walls.
\newblock \emph{Scientific Reports}, 13\penalty0 (1):\penalty0 4936, March
  2023.
\newblock ISSN 2045-2322.
\newblock \doi{10.1038/s41598-023-31674-w}.

\bibitem[{Hern{\'a}ndez-Gasc{\'o}n} et~al.(2011){Hern{\'a}ndez-Gasc{\'o}n},
  Pe{\~n}a, Melero, Pascual, Doblar{\'e}, Ginebra, Bell{\'o}n, and
  Calvo]{hernandez-gasconMechanicalBehaviourSynthetic2011}
Bel{\'e}n {Hern{\'a}ndez-Gasc{\'o}n}, E.~Pe{\~n}a, H.~Melero, G.~Pascual,
  M.~Doblar{\'e}, M.~P. Ginebra, J.~M. Bell{\'o}n, and B.~Calvo.
\newblock Mechanical behaviour of synthetic surgical meshes: {{Finite}} element
  simulation of the herniated abdominal wall.
\newblock \emph{Acta Biomaterialia}, 7\penalty0 (11):\penalty0 3905--3913,
  November 2011.
\newblock ISSN 1742-7061.
\newblock \doi{10.1016/j.actbio.2011.06.033}.

\bibitem[He et~al.(2020)He, Liu, Wu, Liao, Cao, Fan, and
  Li]{heNumericalMethodGuiding2020}
Wei He, Xiaoyu Liu, Shuai Wu, Jie Liao, Guangxiu Cao, Yubo Fan, and Xiaoming
  Li.
\newblock A numerical method for guiding the design of surgical meshes with
  suitable mechanical properties for specific abdominal hernias.
\newblock \emph{COMPUTERS IN BIOLOGY AND MEDICINE}, 116, January 2020.
\newblock ISSN 0010-4825.
\newblock \doi{10.1016/j.compbiomed.2019.103531}.

\bibitem[VV4()]{VV40AssessingCredibilityComputational}
{{VV40-Assessing Credibility}} of {{Computational Modeling}} through
  {{Verification}} and {{Validation Application}} to {{Medical Devices}} -
  {{ASME}}.
\newblock
  https://www.asme.org/codes-standards/find-codes-standards/assessing-credibility-of-computational-modeling-through-verification-and-validation-application-to-medical-devices.

\bibitem[{Kychot}(2009)]{kychotEnglishNeedleEMG2009}
{Kychot}.
\newblock English: {{The}} needle {{EMG}} of paravertebral muscles, May 2009.

\bibitem[H{\"a}ggstr{\"o}m(2018)]{haggstromUltrasonographyDiastasisRecti2018}
Mikael H{\"a}ggstr{\"o}m.
\newblock Ultrasonography of diastasis recti, January 2018.

\bibitem[{MRelasto}(2023)]{mrelastoMRElastographyKidney2023}
{MRelasto}.
\newblock {{MR}} elastography of (a) the kidney, (b) the prostate,, February
  2023.

\bibitem[Urschel et~al.(1988{\natexlab{a}})Urschel, Scott, and
  Williams]{urschelEtiologyLateDeveloping1988}
John~D. Urschel, Paul~G. Scott, and Henry Thomas~G. Williams.
\newblock Etiology of late developing incisional hernias --- the possible role
  of mechanical stress.
\newblock \emph{Medical Hypotheses}, 25\penalty0 (1):\penalty0 31--34, January
  1988{\natexlab{a}}.
\newblock ISSN 0306-9877.
\newblock \doi{10.1016/0306-9877(88)90043-6}.

\bibitem[Kroese et~al.(2018)Kroese, Gillion, Jeekel, Kleinrensink, and
  Lange]{kroesePrimaryIncisionalVentral2018}
Leonard~F. Kroese, Jean-Francois Gillion, Johannes Jeekel, Gert-Jan
  Kleinrensink, and Johan~F. Lange.
\newblock Primary and incisional ventral hernias are different in terms of
  patient characteristics and postoperative complications - {{A}} prospective
  cohort study of 4,565 patients.
\newblock \emph{International Journal of Surgery}, 51:\penalty0 114--119, March
  2018.
\newblock ISSN 1743-9191.
\newblock \doi{10.1016/j.ijsu.2018.01.010}.

\bibitem[Parker et~al.(2021)Parker, Mallett, Quinn, Wood, Boulton, Jamshaid,
  .~Erotocritou, .~Gowda, .~Collier, Plumb, Windsor, Archer, and
  Halligan]{parkerIdentifyingPredictorsVentral2021}
Sam~G Parker, S~Mallett, L~Quinn, C~P~J Wood, R~W Boulton, S~Jamshaid,
  M~.~Erotocritou, S~.~Gowda, W~.~Collier, Andrew A~O Plumb, Alastair C~J
  Windsor, L~Archer, and S~Halligan.
\newblock Identifying predictors of ventral hernia recurrence: Systematic
  review and meta-analysis.
\newblock \emph{BJS Open}, 5\penalty0 (2):\penalty0 zraa071, March 2021.
\newblock ISSN 2474-9842.
\newblock \doi{10.1093/bjsopen/zraa071}.

\bibitem[{Al-Mansour} et~al.(2024){Al-Mansour}, Ding, Yergin, Tamer, and
  Huang]{al-mansourAssociationHerniaspecificProcedural2024}
Mazen~R. {Al-Mansour}, Delaney~D. Ding, Celeste~G. Yergin, Robert Tamer, and
  Li-Ching Huang.
\newblock The association of hernia-specific and procedural risk factors with
  early complications in ventral hernia repair: {{ACHQC}} analysis.
\newblock \emph{The American Journal of Surgery}, 233:\penalty0 100--107, July
  2024.
\newblock ISSN 0002-9610.
\newblock \doi{10.1016/j.amjsurg.2024.02.028}.

\bibitem[Franz(2006)]{franzBiologyHerniasAbdominal2006}
Michael~G. Franz.
\newblock The biology of hernias and the abdominal wall.
\newblock \emph{Hernia}, 10\penalty0 (6):\penalty0 462--471, December 2006.
\newblock ISSN 1248-9204.
\newblock \doi{10.1007/s10029-006-0144-9}.

\bibitem[{van Ramshorst} et~al.(2011){van Ramshorst}, Salih, Hop, {van Waes},
  Kleinrensink, Goossens, and
  Lange]{vanramshorstNoninvasiveAssessmentIntraabdominal2011}
Gabrielle~H. {van Ramshorst}, Mahdi Salih, Wim C.~J. Hop, Oscar J.~F. {van
  Waes}, Gert-Jan Kleinrensink, Richard H.~M. Goossens, and Johan~F. Lange.
\newblock Noninvasive assessment of intra-abdominal pressure by measurement of
  abdominal wall tension.
\newblock \emph{The Journal of Surgical Research}, 171\penalty0 (1):\penalty0
  240--244, November 2011.
\newblock ISSN 1095-8673.
\newblock \doi{10.1016/j.jss.2010.02.007}.

\bibitem[Werner and Dayan(2019)]{wernerDiastasisRectiAbdominisdiagnosis2019}
Laura~Anne Werner and Marcy Dayan.
\newblock Diastasis {{Recti Abdominis-diagnosis}}, {{Risk Factors}}, {{Effect}}
  on {{Musculoskeletal Function}}, {{Framework}} for {{Treatment}} and
  {{Implications}} for the {{Pelvic Floor}}.
\newblock \emph{Current Women`s Health Reviews}, 15\penalty0 (2):\penalty0
  86--101, 2019.
\newblock \doi{10.2174/1573404814666180222152952}.

\bibitem[Gasser and
  Holzapfel(2002)]{gasserRateindependentElastoplasticConstitutive2002}
Thomas~Christian Gasser and Gerhard~A Holzapfel.
\newblock A rate-independent elastoplastic constitutive model for biological
  fiber-reinforced composites at finite strains: Continuum basis, algorithmic
  formulation and finite element implementation.
\newblock \emph{Computational Mechanics}, 29\penalty0 (4):\penalty0 340--360,
  October 2002.
\newblock ISSN 1432-0924.
\newblock \doi{10.1007/s00466-002-0347-6}.

\bibitem[Martin and Sun(2015)]{martinFatigueDamageCollagenous2015}
Caitlin Martin and Wei Sun.
\newblock Fatigue {{Damage}} of {{Collagenous Tissues}}: {{Experiment}},
  {{Modeling}} and {{Simulation Studies}}.
\newblock \emph{Journal of Long-Term Effects of Medical Implants}, 25\penalty0
  (1-2), 2015.
\newblock ISSN 1050-6934, 1940-4379.
\newblock \doi{10.1615/JLongTermEffMedImplants.2015011749}.

\bibitem[Mota et~al.(2015)Mota, Pascoal, and
  Bo]{motaDiastasisRectiAbdominis2015}
Patricia Mota, Augusto~Gil Pascoal, and Kari Bo.
\newblock Diastasis {{Recti Abdominis}} in {{Pregnancy}} and {{Postpartum
  Period}}. {{Risk Factors}}, {{Functional Implications}} and {{Resolution}}.
\newblock \emph{Current Women`s Health Reviews}, 11\penalty0 (1):\penalty0
  59--67, 2015.
\newblock \doi{10.2174/157340481101150914201735}.

\bibitem[Cavalli et~al.(2021)Cavalli, Aiolfi, Bruni, Manfredini, Lombardo,
  Bonfanti, Bona, and Campanelli]{cavalliPrevalenceRiskFactors2021}
Marta Cavalli, A.~Aiolfi, P.~G. Bruni, L.~Manfredini, F.~Lombardo, M.~T.
  Bonfanti, D.~Bona, and Giampiero Campanelli.
\newblock Prevalence and risk factors for diastasis recti abdominis: A review
  and proposal of a new anatomical variation.
\newblock \emph{Hernia}, 25\penalty0 (4):\penalty0 883--890, August 2021.
\newblock ISSN 1248-9204.
\newblock \doi{10.1007/s10029-021-02468-8}.

\bibitem[Gueroult et~al.(2024)Gueroult, Joppin, Chaumoitre, Di~Bisceglie,
  Masson, and Bege]{gueroultLineaAlba3D2024}
Pierre Gueroult, Victoria Joppin, Kathia Chaumoitre, Mathieu Di~Bisceglie,
  Catherine Masson, and Thierry Bege.
\newblock Linea alba {{3D}} morphometric variability by {{CT}} scan
  exploration.
\newblock \emph{Hernia}, 28\penalty0 (2):\penalty0 485--494, April 2024.
\newblock ISSN 1248-9204.
\newblock \doi{10.1007/s10029-023-02939-0}.

\bibitem[Verstoep et~al.(2021)Verstoep, {de Smet}, Sneiders, Kroese,
  Kleinrensink, Lange, Gillion, and {The Hernia-Club
  Members}]{verstoepHerniaWidthExplains2021}
Laura Verstoep, G.~H.~J. {de Smet}, D.~Sneiders, Leonard~F. Kroese, G.-J.
  Kleinrensink, Johan~F. Lange, Jean-Francois Gillion, and {The Hernia-Club
  Members}.
\newblock Hernia width explains differences in outcomes between primary and
  incisional hernias: A prospective cohort study of 9159 patients.
\newblock \emph{Hernia}, 25\penalty0 (2):\penalty0 463--469, April 2021.
\newblock ISSN 1248-9204.
\newblock \doi{10.1007/s10029-020-02340-1}.

\bibitem[{Sim{\'o}n-Allu{\'e}} et~al.(2018){Sim{\'o}n-Allu{\'e}}, Ortill{\'e}s,
  and Calvo]{simon-allueMechanicalBehaviorSurgical2018}
Raquel {Sim{\'o}n-Allu{\'e}}, A.~Ortill{\'e}s, and Bego{\~n}a Calvo.
\newblock Mechanical behavior of surgical meshes for abdominal wall repair:
  {{In}} vivo versus biaxial characterization.
\newblock \emph{Journal of the Mechanical Behavior of Biomedical Materials},
  82:\penalty0 102--111, June 2018.
\newblock ISSN 1751-6161.
\newblock \doi{10.1016/j.jmbbm.2018.03.011}.

\bibitem[Romain et~al.(2020)Romain, Renard, Binquet, Poghosyan, Moszkowicz,
  Gillion, {Ortega-Deballon}, Gillion, {Ortega-Deballon}, Gadiri, Mesli,
  Poghosyan, Moszkowicz, Bouillot, Mariette, Chau, Arvieux, Abet, Renard,
  Marion, Dubuisson, David, Mercoli, Manfredelli, Glehen, Passot, Lamblin,
  Arnalsteen, Constantin, Vauchaussade, El~Nakadi, Demian, R{\'e}gimbeau,
  Demartines, Romain, Chollet, Binot, Massalou, Benizri, Pichot, Blanc,
  Baraket, Jurczak, Rouqui{\'e}, Mzoughi, Soler, Putinier, Ain, Bellouard,
  Mathonnet, Najim, Vinatier, Lep{\`e}re, Cas, Cossa, Frileux, Tzanis,
  Hennequin, Demaret, Merabet, Bilem, Boukortt, Blazquez, Magne, Khalil,
  Largenton, Lavy, Isambert, Br{\'e}hant, Odet, Firtion, Manouvrier, Soufron,
  and Letoux]{romainRecurrenceElectiveIncisional2020}
Beno{\^i}t Romain, Yohann Renard, Christine Binquet, Tigran Poghosyan, David
  Moszkowicz, Jean-Fran{\c c}ois Gillion, Pablo {Ortega-Deballon}, Jean-Fran{\c
  c}ois Gillion, Pablo {Ortega-Deballon}, Naziha Gadiri, Smain Mesli, Tigran
  Poghosyan, David Moszkowicz, Jean-Luc Bouillot, Christophe Mariette,
  Am{\'e}lie Chau, Catherine Arvieux, Emeric Abet, Yohann Renard, Yohann
  Marion, Vincent Dubuisson, Anaelle David, Henry-Alexis Mercoli, Simone
  Manfredelli, Olivier Glehen, Guillaume Passot, Antoine Lamblin, Laurent
  Arnalsteen, Maita Constantin, Arthus Vauchaussade, Issam El~Nakadi, Hassan
  Demian, Jean-Marc R{\'e}gimbeau, Nicolas Demartines, Benoit Romain,
  Jean-Michel Chollet, Daniel Binot, Damien Massalou, Emmanuel Benizri,
  Virginie Pichot, Benjamin Blanc, Oussama Baraket, Florent Jurczak, Delphine
  Rouqui{\'e}, Zeineb Mzoughi, Marc Soler, Jean-Baptiste Putinier, Jean-Fran{\c
  c}ois Ain, Arnauld Bellouard, Muriel Mathonnet, Mohammed Najim, Edouard
  Vinatier, Marc Lep{\`e}re, Olivier Cas, Jean-Pierre Cossa, Pascal Frileux,
  Dimitri Tzanis, Sandra Hennequin, Sebastien Demaret, Mustapha Merabet,
  Djaouad Bilem, Tayb Boukortt, Denis Blazquez, Eric Magne, Haitham Khalil,
  Claude Largenton, Marianne Lavy, Mil{\`e}ne Isambert, Olivier Br{\'e}hant,
  Emmanuel Odet, Olivier Firtion, Jean-Luc Manouvrier, Jacques Soufron, and
  Nathalie Letoux.
\newblock Recurrence after elective incisional hernia repair is more frequent
  than you think: {{An}} international prospective cohort from the {{French
  Society}} of {{Surgery}}.
\newblock \emph{Surgery}, 168\penalty0 (1):\penalty0 125--134, July 2020.
\newblock ISSN 0039-6060.
\newblock \doi{10.1016/j.surg.2020.02.016}.

\bibitem[Burger et~al.(2005)Burger, Lange, Halm, Kleinrensink, and
  Jeekel]{burgerIncisionalHerniaEarly2005}
Jacobus~W.A. Burger, Johan~F. Lange, Jens~A. Halm, Gert-Jan Kleinrensink, and
  Hans Jeekel.
\newblock Incisional {{Hernia}}: {{Early Complication}} of {{Abdominal
  Surgery}}.
\newblock \emph{World Journal of Surgery}, 29\penalty0 (12):\penalty0
  1608--1613, December 2005.
\newblock ISSN 1432-2323.
\newblock \doi{10.1007/s00268-005-7929-3}.

\bibitem[Bhardwaj et~al.(2024)Bhardwaj, Huayllani, Olson, and
  Janis]{bhardwajYearOverYearVentralHernia2024}
Priya Bhardwaj, Maria~T. Huayllani, Molly~A. Olson, and Jeffrey~E. Janis.
\newblock Year-{{Over-Year Ventral Hernia Recurrence Rates}} and {{Risk
  Factors}}.
\newblock \emph{JAMA Surgery}, 159\penalty0 (6):\penalty0 651--658, June 2024.
\newblock ISSN 2168-6254.
\newblock \doi{10.1001/jamasurg.2024.0233}.

\bibitem[Fink et~al.(2014)Fink, Baumann, Wente, Knebel, Bruckner, Ulrich,
  Werner, B{\"u}chler, and Diener]{finkIncisionalHerniaRate2014}
Christine Fink, Petra Baumann, M~N Wente, P~Knebel, T~Bruckner, A~Ulrich,
  J~Werner, Markus~W. B{\"u}chler, and Markus~K. Diener.
\newblock Incisional hernia rate 3 years after midline laparotomy.
\newblock \emph{British Journal of Surgery}, 101\penalty0 (2):\penalty0 51--54,
  January 2014.
\newblock ISSN 0007-1323.
\newblock \doi{10.1002/bjs.9364}.

\bibitem[Stirler et~al.(2014)Stirler, Schoenmaeckers, {de Haas}, Raymakers, and
  Rakic]{stirlerLaparoscopicRepairPrimary2014}
Vincent M.~A. Stirler, Ernst J.~P. Schoenmaeckers, Robbert~J. {de Haas}, Johan
  T. F.~J. Raymakers, and Srdjan Rakic.
\newblock Laparoscopic repair of primary and incisional ventral hernias: The
  differences must be acknowledged.
\newblock \emph{Surgical Endoscopy}, 28\penalty0 (3):\penalty0 891--895, March
  2014.
\newblock ISSN 1432-2218.
\newblock \doi{10.1007/s00464-013-3243-6}.

\bibitem[Kurian et~al.(2010)Kurian, Gallagher, Cheeyandira, and
  Josloff]{kurianLaparoscopicRepairPrimary2010}
Ashwin Kurian, S.~Gallagher, A.~Cheeyandira, and Robert Josloff.
\newblock Laparoscopic repair of primary versus incisional ventral hernias:
  Time to recognize the differences?
\newblock \emph{Hernia}, 14\penalty0 (4):\penalty0 383--387, August 2010.
\newblock ISSN 1248-9204.
\newblock \doi{10.1007/s10029-010-0649-0}.

\bibitem[K{\"o}ckerling et~al.(2015)K{\"o}ckerling, {Schug-Pa{\ss}}, Adolf,
  Reinpold, and Stechemesser]{kockerlingPooledDataAnalysis2015}
Ferdinand K{\"o}ckerling, Christine {Schug-Pa{\ss}}, Daniela Adolf, Wolfgang
  Reinpold, and Bernd Stechemesser.
\newblock Is {{Pooled Data Analysis}} of {{Ventral}} and {{Incisional Hernia
  Repair Acceptable}}?
\newblock \emph{Frontiers in Surgery}, 2, May 2015.
\newblock ISSN 2296-875X.
\newblock \doi{10.3389/fsurg.2015.00015}.

\bibitem[Muysoms et~al.(2009)Muysoms, Miserez, Berrevoet, Campanelli,
  Champault, Chelala, Dietz, Eker, El~Nakadi, Hauters, Hidalgo~Pascual,
  Hoeferlin, Klinge, Montgomery, Simmermacher, Simons, {\'S}mieta{\'n}ski,
  Sommeling, Tollens, Vierendeels, and
  Kingsnorth]{muysomsClassificationPrimaryIncisional2009}
Filip~E. Muysoms, M.~Miserez, F.~Berrevoet, G.~Campanelli, G.~G. Champault,
  E.~Chelala, U.~A. Dietz, H.~H. Eker, I.~El~Nakadi, P.~Hauters,
  M.~Hidalgo~Pascual, A.~Hoeferlin, U.~Klinge, A.~Montgomery, R.~K.~J.
  Simmermacher, M.~P. Simons, M.~{\'S}mieta{\'n}ski, C.~Sommeling, T.~Tollens,
  T.~Vierendeels, and Andrew~N. Kingsnorth.
\newblock Classification of primary and incisional abdominal wall hernias.
\newblock \emph{Hernia}, 13\penalty0 (4):\penalty0 407--414, August 2009.
\newblock ISSN 1248-9204.
\newblock \doi{10.1007/s10029-009-0518-x}.

\bibitem[Juvany et~al.(2022)Juvany, Guillaumes, Hoyuela, Bachero, Trias, Ardid,
  and Martrat]{juvanyResultsProspectiveCohort2022}
Montserrat Juvany, Salvador Guillaumes, Carlos Hoyuela, Irene Bachero, Miguel
  Trias, Jordi Ardid, and Antoni Martrat.
\newblock Results of a {{Prospective Cohort Study}} on {{Open Rives Technique}}
  of the {{Midline Incisional Hernia}}: {{Midline Closure}} and {{Mesh
  Overlap}}.
\newblock \emph{Surgical Innovation}, 29\penalty0 (3):\penalty0 321--328, June
  2022.
\newblock ISSN 1553-3506.
\newblock \doi{10.1177/15533506211033137}.

\bibitem[Helgstrand et~al.(2013)Helgstrand, Rosenberg, Kehlet, Jorgensen, and
  Bisgaard]{helgstrandNationwideProspectiveStudy2013}
Frederik Helgstrand, Jacob Rosenberg, Henrik Kehlet, Lars~N. Jorgensen, and
  Thue Bisgaard.
\newblock Nationwide {{Prospective Study}} of {{Outcomes}} after {{Elective
  Incisional Hernia Repair}}.
\newblock \emph{Journal of the American College of Surgeons}, 216\penalty0
  (2):\penalty0 217--228, February 2013.
\newblock ISSN 1879-1190.
\newblock \doi{10.1016/j.jamcollsurg.2012.10.013}.

\bibitem[Lin et~al.(2024)Lin, Shi, Yang, Li, Xu, Yang, Song, and
  Li]{linRiskFactorsRecurrence2024}
Yiming Lin, Hekai Shi, Rongduo Yang, Shaochun Li, Zijin Xu, Dongchao Yang,
  Zhicheng Song, and Shaojie Li.
\newblock Risk factors of recurrence after incisional hernia preperitoneal
  repair: A long-term retrospective single-center cohort study.
\newblock \emph{Langenbeck's Archives of Surgery}, 409\penalty0 (1):\penalty0
  164, May 2024.
\newblock ISSN 1435-2451.
\newblock \doi{10.1007/s00423-024-03352-6}.

\bibitem[{Sim{\'o}n-Allu{\'e}} et~al.(2016){Sim{\'o}n-Allu{\'e}},
  {Hern{\'a}ndez-Gasc{\'o}n}, L{\`e}oty, Bell{\'o}n, Pe{\~n}a, and
  Calvo]{simon-allueProsthesesSizeDependency2016}
Raquel {Sim{\'o}n-Allu{\'e}}, Bel{\'e}n {Hern{\'a}ndez-Gasc{\'o}n},
  L.~L{\`e}oty, J.~M. Bell{\'o}n, Estefan{\'i}a Pe{\~n}a, and Bego{\~n}a Calvo.
\newblock Prostheses size dependency of the mechanical response of the
  herniated human abdomen.
\newblock \emph{Hernia: The Journal of Hernias and Abdominal Wall Surgery},
  20\penalty0 (6):\penalty0 839--848, December 2016.
\newblock ISSN 1248-9204.
\newblock \doi{10.1007/s10029-016-1525-3}.

\bibitem[Luijendijk et~al.(2000)Luijendijk, Hop, {van den Tol}, {de Lange},
  Braaksma, IJzermans, Boelhouwer, {de Vries}, Salu, Wereldsma, Bruijninckx,
  and Jeekel]{luijendijkComparisonSutureRepair2000}
Roland~W. Luijendijk, Wim~C.J. Hop, M.~Petrousjka {van den Tol}, Diederik~C.D.
  {de Lange}, Marijel~M.J. Braaksma, Jan~N.M. IJzermans, Roelof~U. Boelhouwer,
  Bas~C. {de Vries}, Marc~K.M. Salu, Jack~C.J. Wereldsma, Cornelis~M.A.
  Bruijninckx, and Johannes Jeekel.
\newblock A {{Comparison}} of {{Suture Repair}} with {{Mesh Repair}} for
  {{Incisional Hernia}}.
\newblock \emph{New England Journal of Medicine}, 343\penalty0 (6):\penalty0
  392--398, August 2000.
\newblock ISSN 0028-4793.
\newblock \doi{10.1056/NEJM200008103430603}.

\bibitem[Poruk et~al.(2016)Poruk, Farrow, Azar, Burce, Hicks, Azoury, Cornell,
  Cooney, and Eckhauser]{porukEffectHerniaSize2016}
Katherine~E. Poruk, N.~Farrow, F.~Azar, K.~K. Burce, C.~W. Hicks, S.~C. Azoury,
  P.~Cornell, Gerard~M. Cooney, and Frederic~E. Eckhauser.
\newblock Effect of hernia size on operative repair and post-operative outcomes
  after open ventral hernia repair.
\newblock \emph{Hernia}, 20\penalty0 (6):\penalty0 805--810, December 2016.
\newblock ISSN 1248-9204.
\newblock \doi{10.1007/s10029-016-1542-2}.

\bibitem[Hesselink et~al.(1993)Hesselink, Luijendijk, {de Wilt}, Heide, and
  Jeekel]{hesselinkEvaluationRiskFactors1993}
Vincent~J. Hesselink, R.~W. Luijendijk, J.~H. {de Wilt}, R.~Heide, and Johannes
  Jeekel.
\newblock An evaluation of risk factors in incisional hernia recurrence.
\newblock \emph{Surgery, Gynecology \& Obstetrics}, 176\penalty0 (3):\penalty0
  228--234, March 1993.
\newblock ISSN 0039-6087.

\bibitem[{Guti{\'e}rrez de la Pe{\~n}a} et~al.(2001){Guti{\'e}rrez de la
  Pe{\~n}a}, Vargas~Romero, and
  Di{\'e}guez~Garc{\'i}a]{gutierrezdelapenaValueCTDiagnosis2001}
Carlos {Guti{\'e}rrez de la Pe{\~n}a}, J.~Vargas~Romero, and Juan~Antonio
  Di{\'e}guez~Garc{\'i}a.
\newblock The value of {{CT}} diagnosis of hernia recurrence after prosthetic
  repair of ventral incisional hernias.
\newblock \emph{European Radiology}, 11\penalty0 (7):\penalty0 1161--1164, July
  2001.
\newblock ISSN 1432-1084.
\newblock \doi{10.1007/s003300000743}.

\bibitem[Baucom et~al.(2014)Baucom, Beck, Holzman, Sharp, Nealon, and
  Poulose]{baucomProspectiveEvaluationSurgeon2014}
Rebeccah~B. Baucom, William~C. Beck, Michael~D. Holzman, Kenneth~W. Sharp,
  William~H. Nealon, and Benjamin~K. Poulose.
\newblock Prospective {{Evaluation}} of {{Surgeon Physical Examination}} for
  {{Detection}} of {{Incisional Hernias}}.
\newblock \emph{Journal of the American College of Surgeons}, 218\penalty0
  (3):\penalty0 363--366, March 2014.
\newblock ISSN 1072-7515.
\newblock \doi{10.1016/j.jamcollsurg.2013.12.007}.

\bibitem[Muysoms et~al.(2015)Muysoms, Antoniou, Bury, Campanelli, Conze,
  Cuccurullo, {de Beaux}, Deerenberg, East, Fortelny, Gillion, Henriksen,
  Israelsson, Jairam, J{\"a}nes, Jeekel, {L{\'o}pez-Cano}, Miserez,
  {Morales-Conde}, Sanders, Simons, {\'S}mieta{\'n}ski, Venclauskas, and
  Berrevoet]{muysomsEuropeanHerniaSociety2015}
Filip~E. Muysoms, S.~A. Antoniou, Kamil Bury, G.~Campanelli, Joachim Conze,
  D.~Cuccurullo, A.~C. {de Beaux}, Eva~B Deerenberg, B.~East, R.~H. Fortelny,
  Jean-Fran{\c c}ois Gillion, Nadia Henriksen, L.~Israelsson, A.~Jairam,
  A.~J{\"a}nes, J.~Jeekel, M.~{L{\'o}pez-Cano}, M.~Miserez, S.~{Morales-Conde},
  D.~L. Sanders, M.~P. Simons, M.~{\'S}mieta{\'n}ski, L.~Venclauskas, and
  Frederik Berrevoet.
\newblock European {{Hernia Society}} guidelines on the closure of abdominal
  wall incisions.
\newblock \emph{Hernia}, 19\penalty0 (1):\penalty0 1--24, February 2015.
\newblock ISSN 1248-9204.
\newblock \doi{10.1007/s10029-014-1342-5}.

\bibitem[Gossios et~al.(2003)Gossios, Zikou, Vazakas, Passas, Glantzouni,
  Glantzounis, Kontogiannis, and Tsimoyiannis]{gossiosValueCTLaparoscopic2003}
Konstantinos Gossios, A.~Zikou, P.~Vazakas, G.~Passas, A.~Glantzouni,
  G.~Glantzounis, D.~Kontogiannis, and Evangelos Tsimoyiannis.
\newblock Value of {{CT}} after laparoscopic repair of postsurgical ventral
  hernia.
\newblock \emph{Abdominal Imaging}, 28\penalty0 (1):\penalty0 0099--0102,
  January 2003.
\newblock ISSN 1432-0509.
\newblock \doi{10.1007/s00261-001-0156-y}.

\bibitem[Du et~al.(2023)Du, Jin, Yan, Sun, Shen, Pan, and
  Jiang]{duCTmeasuredHerniaParameters2023}
Xiangying Du, C.~Jin, Y.~Yan, P.~Sun, Y.~Shen, Z.~Pan, and Tao Jiang.
\newblock {{CT-measured}} hernia parameters can predict component separation: A
  cross-sectional study from {{China}}.
\newblock \emph{Hernia}, 27\penalty0 (4):\penalty0 979--986, August 2023.
\newblock ISSN 1248-9204.
\newblock \doi{10.1007/s10029-023-02761-8}.

\bibitem[Winters et~al.(2019)Winters, Knaapen, Buyne, Hummelink, Ulrich, {van
  Goor}, {van Geffen}, and Slater]{wintersPreoperativeCTScan2019}
Harm Winters, L.~Knaapen, O.~R. Buyne, S.~Hummelink, D.~J.~O. Ulrich, H.~{van
  Goor}, E.~{van Geffen}, and Nicholas~J. Slater.
\newblock Pre-operative {{CT}} scan measurements for predicting complications
  in patients undergoing complex ventral hernia repair using the component
  separation technique.
\newblock \emph{Hernia}, 23\penalty0 (2):\penalty0 347--354, April 2019.
\newblock ISSN 1248-9204.
\newblock \doi{10.1007/s10029-019-01899-8}.

\bibitem[Bellio et~al.(2019)Bellio, Cipolat~Mis, Del~Giudice, and
  Munegato]{bellioPreoperativeAbdominalComputed2019}
Gabriele Bellio, Tommaso Cipolat~Mis, Roberto Del~Giudice, and Gabriele
  Munegato.
\newblock Preoperative {{Abdominal Computed Tomography}} at {{Rest}} and
  {{During Valsalva}}'s {{Maneuver}} to {{Evaluate Incisional Hernias}}.
\newblock \emph{Surgical Innovation}, 26\penalty0 (5):\penalty0 519--527,
  October 2019.
\newblock ISSN 1553-3506.
\newblock \doi{10.1177/1553350619849986}.

\bibitem[Jourdan et~al.(2020)Jourdan, Soucasse, Scemama, Gillion, Chaumoitre,
  Masson, and Bege]{jourdanAbdominalWallMorphometric2020}
Arthur Jourdan, Andrea Soucasse, Ugo Scemama, Jean~F. Gillion, Kathia
  Chaumoitre, Catherine Masson, and Thierry Bege.
\newblock Abdominal wall morphometric variability based on computed tomography:
  {{Influence}} of age, gender, and body mass index.
\newblock \emph{Clinical Anatomy}, 33\penalty0 (8):\penalty0 1110--1119, 2020.
\newblock ISSN 1098-2353.
\newblock \doi{10.1002/ca.23548}.

\bibitem[Kushner et~al.(2021)Kushner, Starnes, Sehnert, Holden, and
  Blatnik]{kushnerIdentifyingCriticalComputed2021}
Bradley Kushner, Carter Starnes, Maggie Sehnert, Sara Holden, and Jeffrey
  Blatnik.
\newblock Identifying critical computed tomography ({{CT}}) imaging findings
  for the preoperative planning of ventral hernia repairs.
\newblock \emph{Hernia}, 25\penalty0 (4):\penalty0 963--969, August 2021.
\newblock ISSN 1248-9204.
\newblock \doi{10.1007/s10029-020-02314-3}.

\bibitem[Holihan et~al.(2016)Holihan, Karanjawala, Ko, Askenasy, Matta,
  Gharbaoui, Hasapes, Tammisetti, Thupili, Alawadi, Bondre, {Flores-Gonzalez},
  Kao, and Liang]{holihanUseComputedTomography2016}
Julie~L. Holihan, Burzeen Karanjawala, Annie Ko, Erik~P. Askenasy, Eduardo~J.
  Matta, Latifa Gharbaoui, Joseph~P. Hasapes, Varaha~S. Tammisetti,
  Chakradhar~R. Thupili, Zeinab~M. Alawadi, Ioana Bondre, Juan~R.
  {Flores-Gonzalez}, Lillian~S. Kao, and Mike~K. Liang.
\newblock Use of {{Computed Tomography}} in {{Diagnosing Ventral Hernia
  Recurrence}}: {{A Blinded}}, {{Prospective}}, {{Multispecialty Evaluation}}.
\newblock \emph{JAMA Surgery}, 151\penalty0 (1):\penalty0 7--13, January 2016.
\newblock ISSN 2168-6254.
\newblock \doi{10.1001/jamasurg.2015.2580}.

\bibitem[Messer et~al.(2024)Messer, Melland, Miller, Krpata, Beffa, Chao,
  Petro, Maskal, Ellis, Rosen, and Prabhu]{messerCanSurgeonsAccurately2024}
Nir Messer, M.~S. Melland, B.~T. Miller, D.~M. Krpata, L.~R.~A. Beffa, T.~Chao,
  C.~C. Petro, S.~M. Maskal, R.~C. Ellis, M.~J. Rosen, and Ajita~S. Prabhu.
\newblock Can surgeons accurately identify mesh type when interpreting computed
  tomography scans after ventral hernia repair?
\newblock \emph{Hernia}, 28\penalty0 (4):\penalty0 1275--1281, August 2024.
\newblock ISSN 1248-9204.
\newblock \doi{10.1007/s10029-024-03024-w}.

\bibitem[Read et~al.(2022)Read, Ibrahim, Jacombs, Elstner, Saunders, and
  {Rodriguez-Acevedo}]{readImagingInsightsAbdominal2022}
John~W. Read, Nabeel Ibrahim, Anita S.~W. Jacombs, Kristen~E. Elstner, Jeni
  Saunders, and Omar {Rodriguez-Acevedo}.
\newblock Imaging {{Insights Into Abdominal Wall Function}}.
\newblock \emph{Frontiers in Surgery}, 9, 2022.
\newblock ISSN 2296-875X.

\bibitem[Plumb et~al.(2021)Plumb, Windsor, and
  Ross]{plumbContemporaryImagingRectus2021}
Andrew~A. Plumb, Alastair C.~J. Windsor, and David Ross.
\newblock Contemporary imaging of rectus diastasis and the abdominal wall.
\newblock \emph{Hernia}, 25\penalty0 (4):\penalty0 921--927, August 2021.
\newblock ISSN 1248-9204.
\newblock \doi{10.1007/s10029-021-02463-z}.

\bibitem[Vo{\ss} et~al.(2020)Vo{\ss}, L{\"o}sel, Heuveline, Saalfeld, Berg, and
  Kallinowski]{vossAutomatedIncisionalHernia2020}
Samuel Vo{\ss}, Philipp~D. L{\"o}sel, Vincent Heuveline, Sylvia Saalfeld,
  Philipp Berg, and Friedrich Kallinowski.
\newblock Automated incisional hernia characterization by non-rigid
  registration of {{CT}} images -- a pilot study.
\newblock \emph{Current Directions in Biomedical Engineering}, 6\penalty0
  (3):\penalty0 91--94, September 2020.
\newblock ISSN 2364-5504.
\newblock \doi{10.1515/cdbme-2020-3024}.

\bibitem[Jaffe et~al.(2005)Jaffe, O'Connell, Harris, Paulson, and
  DeLong]{jaffeMDCTAbdominalWall2005}
Tracy~A. Jaffe, Martin~J. O'Connell, John~P. Harris, Erik~K. Paulson, and
  David~M. DeLong.
\newblock {{MDCT}} of {{Abdominal Wall Hernias}}: {{Is There}} a {{Role}} for
  {{Valsalva}}'s {{Maneuver}}?
\newblock \emph{American Journal of Roentgenology}, 184\penalty0 (3):\penalty0
  847--851, March 2005.
\newblock ISSN 0361-803X.
\newblock \doi{10.2214/ajr.184.3.01840847}.

\bibitem[Kallinowski et~al.(2019)Kallinowski, Nessel, G{\"o}rich, Grimm, and
  L{\"o}ffler]{kallinowskiCTAbdomenValsalva2019}
Friedrich Kallinowski, Regine Nessel, Johannes G{\"o}rich, Annette Grimm, and
  Thorsten L{\"o}ffler.
\newblock {{CT Abdomen}} with {{Valsalva}}'s {{Maneuver Facilitates Grip-Based
  Incisional Hernia Repair}}.
\newblock \emph{SSRN Electronic Journal}, 2019.
\newblock ISSN 1556-5068.
\newblock \doi{10.2139/ssrn.3482813}.

\bibitem[Dallaudiere et~al.(2024)Dallaudiere, Sans, Reboul, Dallet, Reau, Bise,
  Bouguennec, Pesquer, Dallaudiere, Sans, Sr, Dallet, Reau, Bise, Bouguennec,
  and Pesquer]{dallaudiereDynamicMagneticResonance2024}
Benjamin Dallaudiere, Hugo Sans, Gilles Reboul, Laurence Dallet, Patricia Reau,
  Sylvain Bise, Nicolas Bouguennec, Lionel Pesquer, Benjamin Dallaudiere, Hugo
  Sans, Gilles~Reboul Sr, Laurence Dallet, Patricia Reau, Sylvain Bise, Nicolas
  Bouguennec, and Lionel Pesquer.
\newblock Dynamic {{Magnetic Resonance Imaging}} ({{MRI}}) in
  {{Inguinal-Related Chronic Groin Pain}} ({{CGP}}): {{Comparison With
  Systematic Surgical Assessment}}.
\newblock \emph{Cureus}, 16\penalty0 (3), March 2024.
\newblock ISSN 2168-8184.
\newblock \doi{10.7759/cureus.55947}.

\bibitem[Qiu et~al.(2023)Qiu, Li, Tang, Fang, Pang, and
  Chen]{qiuMeasurementsAbdominalWall2023}
Zhiying Qiu, Shaojie Li, JianXiong Tang, Liang Fang, Yun Pang, and Lin Chen.
\newblock Measurements of abdominal wall defect and hernia sac volume for the
  treatment of incisional hernia: {{Application}} of the ultrasonic volume
  auto-scan in 50 cases.
\newblock \emph{Asian Journal of Surgery}, 46\penalty0 (9):\penalty0
  3601--3606, September 2023.
\newblock ISSN 1015-9584.
\newblock \doi{10.1016/j.asjsur.2023.04.034}.

\bibitem[Harlaar et~al.(2017)Harlaar, Deerenberg, Dwarkasing, Kamperman,
  Kleinrensink, Jeekel, and Lange]{harlaarDevelopmentIncisionalHerniation2017}
Joris~J. Harlaar, Eva~B Deerenberg, R.~S. Dwarkasing, A.~M. Kamperman, G.~J.
  Kleinrensink, J.~Jeekel, and Johan~F. Lange.
\newblock Development of incisional herniation after midline laparotomy.
\newblock \emph{BJS open}, 1\penalty0 (1):\penalty0 18--23, February 2017.
\newblock ISSN 2474-9842.
\newblock \doi{10.1002/bjs5.3}.

\bibitem[Wang et~al.(2023)Wang, Mao, Yan, Ling, and
  Cai]{wangDeepLearningbasedApproach2023}
Fei Wang, Rongsong Mao, Laifa Yan, Shan Ling, and Zhenyu Cai.
\newblock A deep learning-based approach for rectus abdominis segmentation and
  distance measurement in ultrasonography.
\newblock \emph{Frontiers in Physiology}, 14, 2023.
\newblock ISSN 1664-042X.

\bibitem[Vergari et~al.(2023)Vergari, Persohn, and
  Rohan]{vergariEffectBreathingVivo2023}
Claudio Vergari, Sylvain Persohn, and Pierre-Yves Rohan.
\newblock The effect of breathing on the {\emph{in vivo}} mechanical
  characterization of {\emph{linea alba}} by ultrasound shearwave elastography.
\newblock \emph{Computers in Biology and Medicine}, 167:\penalty0 107637,
  December 2023.
\newblock ISSN 0010-4825.
\newblock \doi{10.1016/j.compbiomed.2023.107637}.

\bibitem[Wang et~al.(2020)Wang, He, Zhu, Fu, Huang, Ding, Yao, and
  Chen]{wangUseShearWave2020}
Xiaohong Wang, Kai He, Yulan Zhu, Xiaojian Fu, Zhifang Huang, Rui Ding, Qiyuan
  Yao, and Hao Chen.
\newblock Use of {{Shear Wave Elastography}} to {{Quantify Abdominal Wall
  Muscular Properties}} in {{Patients With Incisional Hernia}}.
\newblock \emph{Ultrasound in Medicine \& Biology}, 46\penalty0 (7):\penalty0
  1651--1657, July 2020.
\newblock ISSN 0301-5629.
\newblock \doi{10.1016/j.ultrasmedbio.2020.03.027}.

\bibitem[Beamish et~al.(2019)Beamish, Green, Nieuwold, and
  McLean]{beamishDifferencesLineaAlba2019}
Nicole Beamish, Natasha Green, Elyse Nieuwold, and Linda McLean.
\newblock Differences in {{Linea Alba Stiffness}} and {{Linea Alba Distortion
  Between Women With}} and {{Without Diastasis Recti Abdominis}}: {{The
  Impact}} of {{Measurement Site}} and {{Task}}.
\newblock \emph{Journal of Orthopaedic \& Sports Physical Therapy}, 49\penalty0
  (9):\penalty0 656--665, September 2019.
\newblock ISSN 0190-6011.
\newblock \doi{10.2519/jospt.2019.8543}.

\bibitem[Chaudhry et~al.(2017)Chaudhry, {Fernandez-Moure}, Shajudeen, Eps,
  Cabrera, Weiner, Dunkin, Tasciotti, and
  Righetti]{chaudhryCharacterizationVentralIncisional2017}
Anuj Chaudhry, Joseph~S. {Fernandez-Moure}, Peer~Shafeeq Shajudeen, Jeffrey
  L.~Van Eps, Fernando~J. Cabrera, Bradley~K. Weiner, Brian~J. Dunkin, Ennio
  Tasciotti, and Raffaella Righetti.
\newblock Characterization of ventral incisional hernia and repair using shear
  wave elastography.
\newblock \emph{Journal of Surgical Research}, 210:\penalty0 244--252, April
  2017.
\newblock ISSN 0022-4804, 1095-8673.
\newblock \doi{10.1016/j.jss.2016.11.041}.

\bibitem[{Rodriguez-Acevedo} et~al.(2020){Rodriguez-Acevedo}, Elstner, Read,
  Jacombs, and Ibrahim]{rodriguez-acevedoFunctional3DVRImaging2020}
Omar {Rodriguez-Acevedo}, Kristen Elstner, John~W Read, Anita Jacombs, and
  Nabeel Ibrahim.
\newblock Functional {{3DVR}} imaging of abdominal wall hernias.
\newblock \emph{Journal of Medical Imaging and Radiation Oncology}, 64\penalty0
  (5):\penalty0 663--667, 2020.
\newblock ISSN 1754-9485.
\newblock \doi{10.1111/1754-9485.13085}.

\bibitem[Randall et~al.(2017)Randall, Joosten, Ten~Broek, Gillott, Bardhan,
  Strik, Prins, {van Goor}, and Fenner]{randallNovelDiagnosticAid2017}
David Randall, Frank Joosten, Richard~Peter Ten~Broek, Richard Gillott,
  Karna~Dev Bardhan, Chema Strik, Wiesje Prins, Harry {van Goor}, and
  John~Wesley Fenner.
\newblock A novel diagnostic aid for intra-abdominal adhesion detection in
  cine-{{MRI}}: Pilot study and initial diagnostic impressions.
\newblock \emph{British Journal of Radiology}, 90\penalty0 (1077):\penalty0
  20170158, September 2017.
\newblock ISSN 0007-1285.
\newblock \doi{10.1259/bjr.20170158}.

\bibitem[Blanchard et~al.(2016)Blanchard, Smith, and
  Grenier]{blanchardDynamicLiftingTask2016}
Trevor~W. Blanchard, Camille Smith, and Sylvain~G. Grenier.
\newblock In a dynamic lifting task, the relationship between cross-sectional
  abdominal muscle thickness and the corresponding muscle activity is affected
  by the combined use of a weightlifting belt and the {{Valsalva}} maneuver.
\newblock \emph{Journal of Electromyography and Kinesiology}, 28:\penalty0
  99--103, June 2016.
\newblock ISSN 1050-6411.
\newblock \doi{10.1016/j.jelekin.2016.03.006}.

\bibitem[Yasemin et~al.(2020)Yasemin, Mehmet, and
  Omer]{yaseminAssessmentDiagnosticEfficacy2020}
Alt{\i}ntas Yasemin, Bayrak Mehmet, and Alabaz Omer.
\newblock Assessment of the diagnostic efficacy of abdominal ultrasonography
  and cine magnetic resonance imaging in detecting abdominal adhesions: {{A}}
  double-blind research study.
\newblock \emph{European Journal of Radiology}, 126:\penalty0 108922, May 2020.
\newblock ISSN 0720-048X.
\newblock \doi{10.1016/j.ejrad.2020.108922}.

\bibitem[Miyamoto et~al.(2002)Miyamoto, Shimizu, and
  Masuda]{miyamotoFastMRIUsed2002}
Kei Miyamoto, Katsuji Shimizu, and Kazuaki Masuda.
\newblock Fast {{MRI Used}} to {{Evaluate}} the {{Effect}} of {{Abdominal Belts
  During Contraction}} of {{Trunk Muscles}}.
\newblock \emph{Spine}, 27\penalty0 (16):\penalty0 1749, August 2002.
\newblock ISSN 0362-2436.

\bibitem[Fischer et~al.(2007)Fischer, Ladurner, Gangkofer, Mussack, Reiser, and
  Lienemann]{fischerFunctionalCineMRI2007}
Tanja Fischer, Roland Ladurner, Alexander Gangkofer, Thomas Mussack, Maximilian
  Reiser, and Andreas Lienemann.
\newblock Functional cine {{MRI}} of the abdomen for the assessment of
  implanted synthetic mesh in patients after incisional hernia repair: Initial
  results.
\newblock \emph{European Radiology}, 17\penalty0 (12):\penalty0 3123--3129,
  December 2007.
\newblock ISSN 1432-1084.
\newblock \doi{10.1007/s00330-007-0678-y}.

\bibitem[Ciritsis et~al.(2014)Ciritsis, Hansen, Barabasch, Kuehnert, Otto,
  Conze, Klinge, Kuhl, and Kraemer]{ciritsisTimeDependentChangesMagnetic2014}
Alexander Ciritsis, Nienke~Lynn Hansen, Alexandra Barabasch, Nicolas Kuehnert,
  Jens Otto, Joachim Conze, Uwe Klinge, Christiane~K. Kuhl, and Nils~Andreas
  Kraemer.
\newblock Time-{{Dependent Changes}} of {{Magnetic Resonance
  Imaging}}--{{Visible Mesh Implants}} in {{Patients}}.
\newblock \emph{Investigative Radiology}, 49\penalty0 (7):\penalty0 439--444,
  July 2014.
\newblock ISSN 0020-9996.
\newblock \doi{10.1097/RLI.0000000000000051}.

\bibitem[Weston et~al.(2019)Weston, Korfiatis, Kline, Philbrick, Kostandy,
  Sakinis, Sugimoto, Takahashi, and
  Erickson]{westonAutomatedAbdominalSegmentation2019}
Alexander~D. Weston, Panagiotis Korfiatis, Timothy~L. Kline, Kenneth~A.
  Philbrick, Petro Kostandy, Tomas Sakinis, Motokazu Sugimoto, Naoki Takahashi,
  and Bradley~J. Erickson.
\newblock Automated {{Abdominal Segmentation}} of {{CT Scans}} for {{Body
  Composition Analysis Using Deep Learning}}.
\newblock \emph{Radiology}, 290\penalty0 (3):\penalty0 669--679, March 2019.
\newblock ISSN 0033-8419, 1527-1315.
\newblock \doi{10.1148/radiol.2018181432}.

\bibitem[Park et~al.(2020)Park, Shin, Park, Kim, Lee, Seo, Huh, Lee, Park, Lee,
  and Kim]{parkDevelopmentValidationDeep2020}
Hyo~Jung Park, Yongbin Shin, Jisuk Park, Hyosang Kim, In~Seob Lee, Dong-Woo
  Seo, Jimi Huh, Tae~Young Lee, TaeYong Park, Jeongjin Lee, and Kyung~Won Kim.
\newblock Development and {{Validation}} of a {{Deep Learning System}} for
  {{Segmentation}} of {{Abdominal Muscle}} and {{Fat}} on {{Computed
  Tomography}}.
\newblock \emph{Korean Journal of Radiology}, 21\penalty0 (1):\penalty0
  88--100, January 2020.
\newblock ISSN 1229-6929.
\newblock \doi{10.3348/kjr.2019.0470}.

\bibitem[Kway et~al.(2021)Kway, Thirumurugan, Tint, Michael, Shek, Yap, Tan,
  Godfrey, Chong, Fortier, Marx, Eriksson, Lee, Velan, Feng, and
  Sadananthan]{kwayAutomatedSegmentationVisceral2021}
Yeshe~Manuel Kway, Kashthuri Thirumurugan, Mya~Thway Tint, Navin Michael,
  Lynette Pei-Chi Shek, Fabian Kok~Peng Yap, Kok~Hian Tan, Keith~M. Godfrey,
  Yap~Seng Chong, Marielle~Valerie Fortier, Ute~C. Marx, Johan~G. Eriksson,
  Yung~Seng Lee, S.~Sendhil Velan, Mengling Feng, and Suresh~Anand Sadananthan.
\newblock Automated {{Segmentation}} of {{Visceral}}, {{Deep Subcutaneous}},
  and {{Superficial Subcutaneous Adipose Tissue Volumes}} in {{MRI}} of
  {{Neonates}} and {{Young}} {{Children}}.
\newblock \emph{Radiology: Artificial Intelligence}, 3\penalty0 (5):\penalty0
  e200304, September 2021.
\newblock \doi{10.1148/ryai.2021200304}.

\bibitem[Joppin et~al.(2024)Joppin, Belton, Hostin, Bellemare, Lawlor, Curran,
  B{\`e}ge, Masson, and Bendahan]{joppinAutomaticMuscleSegmentation2024}
Victoria Joppin, Niamh Belton, Marc~Adrien Hostin, Marc-Emmanuel Bellemare,
  Aonghus Lawlor, Kathleen~M. Curran, Thierry B{\`e}ge, Catherine Masson, and
  David Bendahan.
\newblock Automatic muscle segmentation on healthy abdominal {{MRI}} using
  {{nnUNet}}.
\newblock In \emph{Medical {{Imaging}} with {{Deep Learning}}}, April 2024.

\bibitem[Lima et~al.(2024)Lima, Kasakewitch, Nguyen, Nogueira, Cavazzola,
  Heniford, and Malcher]{limaMachineLearningDeep2024}
Diego~L. Lima, J.~Kasakewitch, D.~Q. Nguyen, R.~Nogueira, L.~T. Cavazzola,
  B.~T. Heniford, and Flavio Malcher.
\newblock Machine learning, deep learning and hernia surgery. {{Are}} we
  pushing the limits of abdominal core health? {{A}} qualitative systematic
  review.
\newblock \emph{Hernia}, 28\penalty0 (4):\penalty0 1405--1412, August 2024.
\newblock ISSN 1248-9204.
\newblock \doi{10.1007/s10029-024-03069-x}.

\bibitem[Borga(2018)]{borgaMRIAdiposeTissue2018}
Magnus Borga.
\newblock {{MRI}} adipose tissue and muscle composition analysis---a review of
  automation techniques.
\newblock \emph{British Journal of Radiology}, 91\penalty0 (1089):\penalty0
  20180252, September 2018.
\newblock ISSN 0007-1285.
\newblock \doi{10.1259/bjr.20180252}.

\bibitem[Todros et~al.(2018)Todros, Pachera, Baldan, Pavan, Pianigiani,
  Merigliano, and Natali]{todrosComputationalModelingAbdominal2018}
Silvia Todros, Paola Pachera, Nicola Baldan, Piero~G. Pavan, Silvia Pianigiani,
  Stefano Merigliano, and Arturo~N. Natali.
\newblock Computational modeling of abdominal hernia laparoscopic repair with a
  surgical mesh.
\newblock \emph{International Journal of Computer Assisted Radiology and
  Surgery}, 13\penalty0 (1):\penalty0 73--81, January 2018.
\newblock ISSN 1861-6429.
\newblock \doi{10.1007/s11548-017-1681-7}.

\bibitem[Podwojewski et~al.(2013)Podwojewski, Ott{\'e}nio, Beillas, Gu{\'e}rin,
  Turquier, and Mitton]{podwojewskiMechanicalResponseAnimal2013}
Florence Podwojewski, M{\'e}lanie Ott{\'e}nio, Philippe Beillas, Gaetan
  Gu{\'e}rin, Fr{\'e}d{\'e}ric Turquier, and David Mitton.
\newblock Mechanical response of animal abdominal walls in vitro:
  {{Evaluation}} of the influence of a hernia defect and a repair with a mesh
  implanted intraperitoneally.
\newblock \emph{Journal of Biomechanics}, 46\penalty0 (3):\penalty0 561--566,
  February 2013.
\newblock ISSN 0021-9290.
\newblock \doi{10.1016/j.jbiomech.2012.09.014}.

\bibitem[Podwojewski et~al.(2014)Podwojewski, Ott{\'e}nio, Beillas, Gu{\'e}rin,
  Turquier, and Mitton]{podwojewskiMechanicalResponseHuman2014}
Florence Podwojewski, M{\'e}lanie Ott{\'e}nio, Philippe Beillas, Gaetan
  Gu{\'e}rin, Fr{\'e}d{\'e}ric Turquier, and David Mitton.
\newblock Mechanical response of human abdominal walls ex vivo: {{Effect}} of
  an incisional hernia and a mesh repair.
\newblock \emph{Journal of the Mechanical Behavior of Biomedical Materials},
  38:\penalty0 126--133, October 2014.
\newblock ISSN 1751-6161.
\newblock \doi{10.1016/j.jmbbm.2014.07.002}.

\bibitem[Kallinowski et~al.(2021{\natexlab{a}})Kallinowski, Ludwig, Gutjahr,
  Gerhard, {Schulte-H{\"o}rmann}, Krimmel, Lesch, Uhr, L{\"o}sel, Vo{\ss},
  Heuveline, Vollmer, G{\"o}rich, and
  Nessel]{kallinowskiBiomechanicalInfluencesMeshRelated2021}
Friedrich Kallinowski, Yannique Ludwig, Dominik Gutjahr, Christian Gerhard,
  Hannah {Schulte-H{\"o}rmann}, Lena Krimmel, Carolin Lesch, Katharina Uhr,
  Philipp L{\"o}sel, Samuel Vo{\ss}, Vincent Heuveline, Matthias Vollmer,
  Johannes G{\"o}rich, and Regine Nessel.
\newblock Biomechanical {{Influences}} on {{Mesh-Related Complications}} in
  {{Incisional Hernia Repair}}.
\newblock \emph{Frontiers in Surgery}, 8:\penalty0 518, 2021{\natexlab{a}}.
\newblock ISSN 2296-875X.
\newblock \doi{10.3389/fsurg.2021.763957}.

\bibitem[DuBay et~al.(2007)DuBay, Choi, Urbanchek, Wang, Adamson, Dennis,
  Kuzon, and Franz]{dubayIncisionalHerniationInduces2007}
Derek~A. DuBay, Winston Choi, Melanie~G. Urbanchek, Xue Wang, Belinda Adamson,
  Robert~G. Dennis, William~M. Kuzon, and Michael~G. Franz.
\newblock Incisional herniation induces decreased abdominal wall compliance via
  oblique muscle atrophy and fibrosis.
\newblock \emph{Annals of Surgery}, 245\penalty0 (1):\penalty0 140--146,
  January 2007.
\newblock ISSN 0003-4932.
\newblock \doi{10.1097/01.sla.0000251267.11012.85}.

\bibitem[Qandeel and O'Dwyer(2016)]{qandeelRelationshipVentralHernia2016}
Haitham Qandeel and Patrick~J. O'Dwyer.
\newblock Relationship between ventral hernia defect area and intra-abdominal
  pressure: Dynamic in vivo measurement.
\newblock \emph{Surgical Endoscopy}, 30\penalty0 (4):\penalty0 1480--1484,
  April 2016.
\newblock ISSN 1432-2218.
\newblock \doi{10.1007/s00464-015-4356-x}.

\bibitem[Shafik et~al.(1997)Shafik, {El-Sharkawy}, and
  Sharaf]{shafikDirectMeasurementIntraabdominal1997}
Ahmed Shafik, A.~{El-Sharkawy}, and Walid~Mohamed Sharaf.
\newblock Direct measurement of intra-abdominal pressure in various conditions.
\newblock \emph{The European Journal of Surgery = Acta Chirurgica},
  163\penalty0 (12):\penalty0 883--887, December 1997.
\newblock ISSN 1102-4151.

\bibitem[Smith et~al.(2022)Smith, Mierzwinski, Chitsabesan, and
  Chintapatla]{smithHealthrelatedQualityLife2022}
Olivia Smith, M.~F. Mierzwinski, P.~Chitsabesan, and Srinivas Chintapatla.
\newblock Health-related quality of life in abdominal wall hernia: Let's ask
  patients what matters to them?
\newblock \emph{Hernia}, 26\penalty0 (3):\penalty0 795--808, June 2022.
\newblock ISSN 1248-9204.
\newblock \doi{10.1007/s10029-022-02599-6}.

\bibitem[Lesch et~al.(2024)Lesch, Nessel, Adolf, Hukauf, K{\"o}ckerling,
  Kallinowski, Willms, Schwab, Zarras, and {For the
  STRONGHOLD/Herniamed-Collaborators
  GROUP}]{leschSTRONGHOLDFirstyearResults2024}
Carolin Lesch, Regine Nessel, D.~Adolf, M.~Hukauf, F.~K{\"o}ckerling, Friedrich
  Kallinowski, A.~Willms, R.~Schwab, Konstantinos Zarras, and {For the
  STRONGHOLD/Herniamed-Collaborators GROUP}.
\newblock {{STRONGHOLD}} first-year results of biomechanically calculated
  abdominal wall repair: A propensity score matching.
\newblock \emph{Hernia}, 28\penalty0 (1):\penalty0 63--73, February 2024.
\newblock ISSN 1248-9204.
\newblock \doi{10.1007/s10029-023-02897-7}.

\bibitem[Nessel et~al.(2024)Nessel, L{\"o}ffler, Rinn, and
  Kallinowski]{nesselThreeyearFollowupGrip2024}
Regine Nessel, T.~L{\"o}ffler, J.~Rinn, and Friedrich Kallinowski.
\newblock Three-year follow-up of the grip concept: An open, prospective,
  observational registry study on biomechanically calculated abdominal wall
  repair for complex incisional hernias.
\newblock \emph{Hernia}, 28\penalty0 (3):\penalty0 913--924, June 2024.
\newblock ISSN 1248-9204.
\newblock \doi{10.1007/s10029-024-03064-2}.

\bibitem[Kallinowski et~al.(2021{\natexlab{b}})Kallinowski, Ludwig,
  L{\"o}ffler, Vollmer, L{\"o}sel, Vo{\ss}, G{\"o}rich, Heuveline, and
  Nessel]{kallinowskiBiomechanicsAppliedIncisional2021}
Friedrich Kallinowski, Y.~Ludwig, T.~L{\"o}ffler, M.~Vollmer, P.~D. L{\"o}sel,
  S.~Vo{\ss}, J.~G{\"o}rich, V.~Heuveline, and R.~Nessel.
\newblock Biomechanics applied to incisional hernia repair -- {{Considering}}
  the critical and the gained resistance towards impacts related to pressure.
\newblock \emph{Clinical Biomechanics}, 82:\penalty0 105253, February
  2021{\natexlab{b}}.
\newblock ISSN 0268-0033.
\newblock \doi{10.1016/j.clinbiomech.2020.105253}.

\bibitem[Diener et~al.(2010)Diener, Voss, Jensen, B{\"u}chler, and
  Seiler]{dienerElectiveMidlineLaparotomy2010}
Markus~K. Diener, Sabine Voss, Katrin Jensen, Markus~W. B{\"u}chler, and
  Christoph~M. Seiler.
\newblock Elective {{Midline Laparotomy Closure}}: {{The INLINE Systematic
  Review}} and {{Meta-Analysis}}.
\newblock \emph{Annals of Surgery}, 251\penalty0 (5):\penalty0 843--856, May
  2010.
\newblock ISSN 0003-4932.
\newblock \doi{10.1097/SLA.0b013e3181d973e4}.

\bibitem[Millbourn et~al.(2009)Millbourn, Cengiz, and
  Israelsson]{millbournEffectStitchLength2009}
Daniel Millbourn, Yucel Cengiz, and Leif~A. Israelsson.
\newblock Effect of {{Stitch Length}} on {{Wound Complications After Closure}}
  of {{Midline Incisions}}: {{A Randomized Controlled Trial}}.
\newblock \emph{Archives of Surgery}, 144\penalty0 (11):\penalty0 1056--1059,
  November 2009.
\newblock ISSN 0004-0010.
\newblock \doi{10.1001/archsurg.2009.189}.

\bibitem[Deerenberg et~al.(2015)Deerenberg, Harlaar, Steyerberg, Lont, van
  Doorn, Heisterkamp, Wijnhoven, Schouten, Cense, Stockmann, Berends,
  Dijkhuizen, Dwarkasing, Jairam, van Ramshorst, Kleinrensink, Jeekel, and
  Lange]{deerenbergSmallBitesLarge2015}
Eva~B. Deerenberg, Joris~J. Harlaar, Ewout~W. Steyerberg, Harold~E. Lont,
  Helena~C. van Doorn, Joos Heisterkamp, Bas~PL Wijnhoven, Willem~R. Schouten,
  Huib~A. Cense, Hein~BAC Stockmann, Frits~J. Berends, F.~Paul~HLJ Dijkhuizen,
  Roy~S. Dwarkasing, An~P. Jairam, Gabrielle~H. van Ramshorst, Gert-Jan
  Kleinrensink, Johannes Jeekel, and Johan~F. Lange.
\newblock Small bites versus large bites for closure of abdominal midline
  incisions ({{STITCH}}): A double-blind, multicentre, randomised controlled
  trial.
\newblock \emph{The Lancet}, 386\penalty0 (10000):\penalty0 1254--1260,
  September 2015.
\newblock ISSN 0140-6736, 1474-547X.
\newblock \doi{10.1016/S0140-6736(15)60459-7}.

\bibitem[Deerenberg et~al.(2022)Deerenberg, Henriksen, Antoniou, Antoniou,
  Bramer, Fischer, Fortelny, G{\"o}k, Harris, Hope, Horne, Jensen,
  K{\"o}ckerling, Kretschmer, {L{\'o}pez-Cano}, Malcher, Shao, Slieker, {de
  Smet}, Stabilini, Torkington, and
  Muysoms]{deerenbergUpdatedGuidelineClosure2022}
Eva~B Deerenberg, Nadia~A Henriksen, George~A Antoniou, Stavros~A Antoniou,
  Wichor~M Bramer, John~P Fischer, Rene~H Fortelny, Hakan G{\"o}k, Hobart~W
  Harris, William Hope, Charlotte~M Horne, Thomas~K Jensen, Ferdinand
  K{\"o}ckerling, Alexander Kretschmer, Manuel {L{\'o}pez-Cano}, Flavio
  Malcher, Jenny~M Shao, Juliette~C Slieker, Gijs H~J {de Smet}, Cesare
  Stabilini, Jared Torkington, and Filip~E Muysoms.
\newblock Updated guideline for closure of abdominal wall incisions from the
  {{European}} and {{American Hernia Societies}}.
\newblock \emph{British Journal of Surgery}, 109\penalty0 (12):\penalty0
  1239--1250, December 2022.
\newblock ISSN 0007-1323.
\newblock \doi{10.1093/bjs/znac302}.

\bibitem[Varshney et~al.(1999)Varshney, Manek, and
  Johnson]{varshneySixfoldSutureWound1999}
Shweta Varshney, P.~Manek, and Colin~D. Johnson.
\newblock Six-fold suture:wound length ratio for abdominal closure.
\newblock \emph{Annals of The Royal College of Surgeons of England},
  81\penalty0 (5):\penalty0 333--336, September 1999.
\newblock ISSN 0035-8843.

\bibitem[H{\"o}er et~al.(2001)H{\"o}er, Klinge, Schachtrupp, T{\"o}ns, and
  Schumpelick]{hoerInfluenceSutureTechnique2001}
J{\"o}rg H{\"o}er, Uwe Klinge, Alexander Schachtrupp, Christian T{\"o}ns, and
  Volker Schumpelick.
\newblock Influence of suture technique on laparotomy wound healing: An
  experimental study in the rat.
\newblock \emph{Langenbeck's Archives of Surgery}, 386\penalty0 (3):\penalty0
  218--223, April 2001.
\newblock ISSN 1435-2451.
\newblock \doi{10.1007/s004230000196}.

\bibitem[Meijer et~al.(2016)Meijer, Timmermans, Jeekel, Lange, and
  Muysoms]{meijerPrinciplesAbdominalWound2016}
Evert-Jan Meijer, L.~Timmermans, J.~Jeekel, J.F. Lange, and Filip~E Muysoms.
\newblock The {{Principles}} of {{Abdominal Wound Closure}}.
\newblock \emph{Acta Chirurgica Belgica}, 113\penalty0 (4):\penalty0 239--244,
  2016.
\newblock ISSN 0001-5458.
\newblock \doi{10.1080/00015458.2013.11680920}.

\bibitem[Harlaar et~al.(2009)Harlaar, {van Ramshorst}, Nieuwenhuizen, {ten
  Brinke}, Hop, Kleinrensink, Jeekel, and Lange]{harlaarSmallStitchesSmall2009}
Joris~J. Harlaar, Gabrielle~H. {van Ramshorst}, Jeroen Nieuwenhuizen, Joost~G.
  {ten Brinke}, Wim C.~J. Hop, Gert-Jan Kleinrensink, Hans Jeekel, and Johan~F.
  Lange.
\newblock Small stitches with small suture distances increase laparotomy
  closure strength.
\newblock \emph{The American Journal of Surgery}, 198\penalty0 (3):\penalty0
  392--395, September 2009.
\newblock ISSN 0002-9610.
\newblock \doi{10.1016/j.amjsurg.2008.10.018}.

\bibitem[Cooney et~al.(2018)Cooney, Kiernan, Winter, and
  Simms]{cooneyOptimizedWoundClosure2018}
Gerard~M. Cooney, A.~Kiernan, Des~C. Winter, and Ciaran~K. Simms.
\newblock Optimized wound closure using a biomechanical abdominal model.
\newblock \emph{British Journal of Surgery}, 105\penalty0 (4):\penalty0
  395--400, March 2018.
\newblock ISSN 0007-1323.
\newblock \doi{10.1002/bjs.10753}.

\bibitem[Henriksen et~al.(2020)Henriksen, Montgomery, Kaufmann, Berrevoet,
  East, Fischer, Hope, Klassen, Lorenz, Renard, Garcia~Urena, and
  Simons]{henriksenGuidelinesTreatmentUmbilical2020}
Nadia Henriksen, A~Montgomery, R~Kaufmann, F~Berrevoet, B~East, J~Fischer,
  W~Hope, D~Klassen, R~Lorenz, Y~Renard, M~A Garcia~Urena, and Maarten~P
  Simons.
\newblock Guidelines for treatment of umbilical and epigastric hernias from the
  {{European Hernia Society}} and {{Americas Hernia Society}}.
\newblock \emph{British Journal of Surgery}, 107\penalty0 (3):\penalty0
  171--190, February 2020.
\newblock ISSN 0007-1323.
\newblock \doi{10.1002/bjs.11489}.

\bibitem[Rath and Chevrel(1998)]{rathHealingLaparotomiesReview1998}
Ana~M. Rath and Jean~Paul Chevrel.
\newblock The healing of laparotomies: Review of the literature.
\newblock \emph{Hernia}, 2\penalty0 (3):\penalty0 145--149, September 1998.
\newblock ISSN 1248-9204.
\newblock \doi{10.1007/BF01250034}.

\bibitem[Vasanthan et~al.(2009)Vasanthan, Satheesh, Hoopes, Lucaci, Williams,
  and Rapley]{vasanthanComparingSutureStrengths2009}
Asvin Vasanthan, Keerthana Satheesh, Wyeth Hoopes, Patrick Lucaci, Karen
  Williams, and John Rapley.
\newblock Comparing {{Suture Strengths}} for {{Clinical Applications}}: {{A
  Novel In Vitro Study}}.
\newblock \emph{Journal of Periodontology}, 80\penalty0 (4):\penalty0 618--624,
  2009.
\newblock ISSN 1943-3670.
\newblock \doi{10.1902/jop.2009.080490}.

\bibitem[Christoffersen et~al.(2013)Christoffersen, Helgstrand, Rosenberg,
  Kehlet, and Bisgaard]{christoffersenLowerReoperationRate2013}
Mette~Willaume Christoffersen, Frederik Helgstrand, J.~Rosenberg, H.~Kehlet,
  and Thue Bisgaard.
\newblock Lower {{Reoperation Rate}} for {{Recurrence}} after {{Mesh}} versus
  {{Sutured Elective Repair}} in {{Small Umbilical}} and {{Epigastric
  Hernias}}. {{A Nationwide Register Study}}.
\newblock \emph{World Journal of Surgery}, 37\penalty0 (11):\penalty0 1, 2013.
\newblock ISSN 1432-2323.
\newblock \doi{10.1007/s00268-013-2160-0}.

\bibitem[Jeroukhimov et~al.(2014)Jeroukhimov, Wiser, Karasic, Nesterenko,
  Poluksht, Lavy, and Halevy]{jeroukhimovReducedPostoperativeChronic2014}
Igor Jeroukhimov, Itay Wiser, Evgeny Karasic, Vladimir Nesterenko, Natan
  Poluksht, Ron Lavy, and Ariel Halevy.
\newblock Reduced {{Postoperative Chronic Pain}} after {{Tension-Free Inguinal
  Hernia Repair Using Absorbable Sutures}}: {{A Single-Blind Randomized
  Clinical Trial}}.
\newblock \emph{Journal of the American College of Surgeons}, 218\penalty0
  (1):\penalty0 102, January 2014.
\newblock ISSN 1879-1190.
\newblock \doi{10.1016/j.jamcollsurg.2013.09.010}.

\bibitem[Patel et~al.(2017)Patel, Paskar, Nelson, Vedula, and
  Steele]{patelClosureMethodsLaparotomy2017}
Sunil~V. Patel, David~D. Paskar, Richard~L. Nelson, Satyanarayana~S. Vedula,
  and Scott~R. Steele.
\newblock Closure methods for laparotomy incisions for preventing incisional
  hernias and other wound complications.
\newblock \emph{Cochrane Database of Systematic Reviews}, \penalty0 (11), 2017.
\newblock ISSN 1465-1858.
\newblock \doi{10.1002/14651858.CD005661.pub2}.

\bibitem[Greenall et~al.(1980)Greenall, Evans, and
  Pollock]{greenallMidlineTransverseLaparotomy1980}
Michael~J. Greenall, Mary Evans, and Alan~V. Pollock.
\newblock Midline or transverse laparotomy? {{A}} random controlled clinical
  trial. {{Part I}}: {{Influence}} on healing.
\newblock \emph{British Journal of Surgery}, 67\penalty0 (3):\penalty0
  188--190, March 1980.
\newblock ISSN 0007-1323.
\newblock \doi{10.1002/bjs.1800670308}.

\bibitem[Le~Huu~Nho et~al.(2012)Le~Huu~Nho, Mege, Oua{\"i}ssi, Sielezneff, and
  Sastre]{lehuunhoIncidencePreventionVentral2012}
Remy Le~Huu~Nho, D.~Mege, M.~Oua{\"i}ssi, I.~Sielezneff, and Bernard Sastre.
\newblock Incidence and prevention of ventral incisional hernia.
\newblock \emph{Journal of Visceral Surgery}, 149\penalty0 (5,
  Supplement):\penalty0 e3--e14, October 2012.
\newblock ISSN 1878-7886.
\newblock \doi{10.1016/j.jviscsurg.2012.05.004}.

\bibitem[Lee et~al.(2018)Lee, Mata, Droeser, Kaneva, Liberman, Charlebois,
  Stein, Fried, and Feldman]{leeIncisionalHerniaMidline2018}
Lawrence Lee, Juan Mata, Raoul~A. Droeser, Pepa Kaneva, Sender Liberman,
  Patrick Charlebois, Barry Stein, Gerald~M. Fried, and Liane~S. Feldman.
\newblock Incisional {{Hernia After Midline Versus Transverse Specimen
  Extraction Incision}}: {{A Randomized Trial}} in {{Patients Undergoing
  Laparoscopic Colectomy}}.
\newblock \emph{Annals of Surgery}, 268\penalty0 (1):\penalty0 41, July 2018.
\newblock ISSN 0003-4932.
\newblock \doi{10.1097/SLA.0000000000002615}.

\bibitem[Brown and Tiernan(2005)]{brownTransverseVersesMidline2005}
Steven~R. Brown and Jim Tiernan.
\newblock Transverse verses midline incisions for abdominal surgery.
\newblock \emph{Cochrane Database of Systematic Reviews}, \penalty0 (4), 2005.
\newblock ISSN 1465-1858.
\newblock \doi{10.1002/14651858.CD005199.pub2}.

\bibitem[Grantcharov and
  Rosenberg(2001)]{grantcharovVerticalComparedTransverse2001}
Teodor~P Grantcharov and Jacob Rosenberg.
\newblock Vertical compared with transverse incisions in abdominal surgery.
\newblock \emph{European Journal of Surgery}, 167\penalty0 (4):\penalty0
  260--267, April 2001.
\newblock ISSN 1741-9271.
\newblock \doi{10.1080/110241501300091408}.

\bibitem[Jenkins(1976)]{jenkinsBurstAbdominalWound1976}
T~P~N Jenkins.
\newblock The burst abdominal wound: {{A}} mechanical approach.
\newblock \emph{British Journal of Surgery}, 63\penalty0 (11):\penalty0
  873--876, November 1976.
\newblock ISSN 0007-1323.
\newblock \doi{10.1002/bjs.1800631110}.

\bibitem[Burger et~al.(2004)Burger, Luijendijk, Hop, Halm, Verdaasdonk, and
  Jeekel]{burgerLongtermFollowupRandomized2004}
Jacobus W.~A. Burger, Roland~W. Luijendijk, Wim C.~J. Hop, Jens~A. Halm, Emiel
  G.~G. Verdaasdonk, and Johannes Jeekel.
\newblock Long-term {{Follow-up}} of a {{Randomized Controlled Trial}} of
  {{Suture Versus Mesh Repair}} of {{Incisional Hernia}}.
\newblock \emph{Annals of Surgery}, 240\penalty0 (4):\penalty0 578, October
  2004.
\newblock ISSN 0003-4932.
\newblock \doi{10.1097/01.sla.0000141193.08524.e7}.

\bibitem[Liang et~al.(2017)Liang, Holihan, Itani, Alawadi, Gonzalez, Askenasy,
  Ballecer, Chong, Goldblatt, Greenberg, Harvin, Keith, Martindale, Orenstein,
  Richmond, Roth, Szotek, Towfigh, Tsuda, Vaziri, and
  Berger]{liangVentralHerniaManagement2017}
Mike~K. Liang, Julie~L. Holihan, Kamal Itani, Zeinab~M. Alawadi, Juan R.~Flores
  Gonzalez, Erik~P. Askenasy, Conrad Ballecer, Hui~Sen Chong, Matthew~I.
  Goldblatt, Jacob~A. Greenberg, John~A. Harvin, Jerrod~N. Keith, Robert~G.
  Martindale, Sean Orenstein, Bryan Richmond, John~Scott Roth, Paul Szotek,
  Shirin Towfigh, Shawn Tsuda, Khashayar Vaziri, and David~H. Berger.
\newblock Ventral {{Hernia Management}}: {{Expert Consensus Guided}} by
  {{Systematic Review}}.
\newblock \emph{Annals of Surgery}, 265\penalty0 (1):\penalty0 80--89, January
  2017.
\newblock ISSN 0003-4932.
\newblock \doi{10.1097/SLA.0000000000001701}.

\bibitem[Jinka and Janis(2024)]{jinkaClinicallyAppliedBiomechanics2024}
Sanjay K.~A. Jinka and Jeffrey~E. Janis.
\newblock Clinically {{Applied Biomechanics}} of {{Mesh-reinforced Ventral
  Hernia Repair}}: {{A Practical Review}}.
\newblock \emph{Plastic and Reconstructive Surgery -- Global Open}, 12\penalty0
  (11):\penalty0 e6294, November 2024.
\newblock \doi{10.1097/GOX.0000000000006294}.

\bibitem[{Pereira-Rodr{\'i}guez} et~al.(2023){Pereira-Rodr{\'i}guez},
  {Bravo-Salva}, {Argudo-Aguirre}, {Amador-Gil}, and
  {Pera-Rom{\'a}n}]{pereira-rodriguezDefiningHighRiskPatients2023}
Jose~Antonio {Pereira-Rodr{\'i}guez}, Alejandro {Bravo-Salva}, N{\'u}ria
  {Argudo-Aguirre}, Sara {Amador-Gil}, and Miguel {Pera-Rom{\'a}n}.
\newblock Defining {{High-Risk Patients Suitable}} for {{Incisional Hernia
  Prevention}}.
\newblock \emph{Journal of Abdominal Wall Surgery}, 2:\penalty0 10899, February
  2023.
\newblock ISSN 2813-2092.
\newblock \doi{10.3389/jaws.2023.10899}.

\bibitem[Sadava et~al.(2022)Sadava, Bras~Harriott, Angeramo, and
  Schlottmann]{sadavaSyntheticMeshContaminated2022}
Emmanuel~E. Sadava, Camila Bras~Harriott, Cristian~A. Angeramo, and Francisco
  Schlottmann.
\newblock Synthetic {{Mesh}} in {{Contaminated Abdominal Wall Surgery}}:
  {{Friend}} or {{Foe}}? {{A Literature Review}}.
\newblock \emph{Journal of Gastrointestinal Surgery}, 26\penalty0 (1):\penalty0
  235--244, January 2022.
\newblock ISSN 1091-255X.
\newblock \doi{10.1007/s11605-021-05155-2}.

\bibitem[Zogbi et~al.(2010)Zogbi, Portella, Trindade, and
  Trindade]{zogbiRetractionFibroplasiaPolypropylene2010}
Luciano Zogbi, A.~O.~V. Portella, Manoel~R.M. Trindade, and Eduardo~Neubarth
  Trindade.
\newblock Retraction and fibroplasia in a polypropylene prosthesis:
  Experimental study in rats.
\newblock \emph{Hernia}, 14\penalty0 (3):\penalty0 291--298, June 2010.
\newblock ISSN 1248-9204.
\newblock \doi{10.1007/s10029-009-0607-x}.

\bibitem[Idrees et~al.(2018)Idrees, Jindal, Gupta, and
  Sarangi]{idreesSurgicalMeshesSearch2018}
Sarrah Idrees, Sanam Jindal, Manish Gupta, and Rathindra Sarangi.
\newblock Surgical meshes -- {{The}} search continues.
\newblock \emph{Current Medicine Research and Practice}, 8\penalty0
  (5):\penalty0 177--182, September 2018.
\newblock ISSN 2352-0817.
\newblock \doi{10.1016/j.cmrp.2018.08.005}.

\bibitem[Rossi et~al.(2017)Rossi, Trindade, D`acampora, and
  Meurer]{rossiPeritonealAdhesionsType2017}
Lucas~F{\'e}lix Rossi, Manoel Roberto~Maciel Trindade, Armando~Jos{\'e}
  D`acampora, and Luise Meurer.
\newblock Peritoneal adhesions {{Type I}}, {{III}} and total collagen on
  polypropylene and coated polypropylene meshes experimental study in rats.
\newblock \emph{ABCD. Arquivos Brasileiros de Cirurgia Digestiva (S{\~a}o
  Paulo)}, 30:\penalty0 77--82, 2017.
\newblock ISSN 0102-6720, 2317-6326.
\newblock \doi{10.1590/0102-6720201700020001}.

\bibitem[Woloson and Greisler(2001)]{wolosonBiochemistryImmunologyTissue2001}
Susanne~K. Woloson and Howard~P. Greisler.
\newblock Biochemistry, {{Immunology}}, and {{Tissue Response}} to {{Prosthetic
  Material}}.
\newblock In Robert Bendavid, Jack Abrahamson, Maurice~E. Arregui, Jean~Bernard
  Flament, and Edward~H. Phillips, editors, \emph{Abdominal {{Wall Hernias}}:
  {{Principles}} and {{Management}}}, pages 201--207. Springer, New York, NY,
  2001.
\newblock ISBN 978-1-4419-8574-3.
\newblock \doi{10.1007/978-1-4419-8574-3_26}.

\bibitem[Earle and Mark(2008)]{earleProstheticMaterialInguinal2008}
David~B. Earle and Lisa~A. Mark.
\newblock Prosthetic {{Material}} in {{Inguinal Hernia Repair}}: {{How Do I
  Choose}}?
\newblock \emph{Surgical Clinics of North America}, 88\penalty0 (1):\penalty0
  179--201, February 2008.
\newblock ISSN 0039-6109.
\newblock \doi{10.1016/j.suc.2007.11.002}.

\bibitem[DuBay et~al.(2006)DuBay, Wang, Adamson, Kuzon, Dennis, and
  Franz]{dubayMeshIncisionalHerniorrhaphy2006}
Derek~A. DuBay, Xue Wang, Belinda Adamson, William~M. Kuzon, Robert~G. Dennis,
  and Michael~G. Franz.
\newblock Mesh incisional herniorrhaphy increases abdominal wall elastic
  properties: A mechanism for decreased hernia recurrences in comparison with
  suture repair.
\newblock \emph{Surgery}, 140\penalty0 (1):\penalty0 14--24, July 2006.
\newblock ISSN 0039-6060.
\newblock \doi{10.1016/j.surg.2006.01.007}.

\bibitem[Kallinowski et~al.(2018)Kallinowski, Harder, Gutjahr, Raschidi, Silva,
  Vollmer, and Nessel]{kallinowskiAssessingGRIPVentral2018}
Friedrich Kallinowski, F.~Harder, D.~Gutjahr, R.~Raschidi, T.~G. Silva,
  M.~Vollmer, and Regine Nessel.
\newblock Assessing the {{GRIP}} of {{Ventral Hernia Repair}}: {{How}} to
  {{Securely Fasten DIS Classified Meshes}}.
\newblock \emph{Frontiers in Surgery}, 4, January 2018.
\newblock ISSN 2296-875X.
\newblock \doi{10.3389/fsurg.2017.00078}.

\bibitem[Brown and Finch(2010)]{brownWhichMeshHernia2010}
{\relax CN}.~Brown and {\relax JG}.~Finch.
\newblock Which mesh for hernia repair?
\newblock \emph{Annals of The Royal College of Surgeons of England},
  92\penalty0 (4):\penalty0 272--278, May 2010.
\newblock ISSN 0035-8843.
\newblock \doi{10.1308/003588410X12664192076296}.

\bibitem[Conze and Klinge(1999)]{conzeBiocompatibilityBiomaterialsClinical1999}
Joachim Conze and Uwe Klinge.
\newblock Biocompatibility of {{Biomaterials}} - {{Clinical}} and {{Mechanical
  Aspects}}.
\newblock In Volker Schumpelick and Andrew~N. Kingsnorth, editors,
  \emph{Incisional {{Hernia}}}, pages 169--177, Berlin, Heidelberg, 1999.
  Springer.
\newblock ISBN 978-3-642-60123-1.
\newblock \doi{10.1007/978-3-642-60123-1_14}.

\bibitem[Todros et~al.(2017{\natexlab{a}})Todros, Pavan, Pachera, and
  Natali]{todrosSyntheticSurgicalMeshes2017a}
{\relax Silvia}.~Todros, Piero~G. Pavan, Paola Pachera, and Arturo~N. Natali.
\newblock Synthetic surgical meshes used in abdominal wall surgery: {{Part
  II}}---{{Biomechanical}} aspects.
\newblock \emph{Journal of Biomedical Materials Research Part B: Applied
  Biomaterials}, 105\penalty0 (4):\penalty0 892--903, 2017{\natexlab{a}}.
\newblock ISSN 1552-4981.
\newblock \doi{10.1002/jbm.b.33584}.

\bibitem[Deeken et~al.(2011)Deeken, Abdo, Frisella, and
  Matthews]{deekenPhysicomechanicalEvaluationAbsorbable2011}
Corey~R. Deeken, Michael~S. Abdo, Margaret~M. Frisella, and Brent~D. Matthews.
\newblock Physicomechanical evaluation of absorbable and nonabsorbable barrier
  composite meshes for laparoscopic ventral hernia repair.
\newblock \emph{Surgical Endoscopy}, 25\penalty0 (5):\penalty0 1541--1552, May
  2011.
\newblock ISSN 1432-2218.
\newblock \doi{10.1007/s00464-010-1432-0}.

\bibitem[Lubowiecka et~al.(2020)Lubowiecka, Tomaszewska, Szepietowska,
  Szymczak, and
  {\'S}mieta{\'n}ski]{lubowieckaVivoPerformanceIntraperitoneal2020}
Izabela Lubowiecka, Agnieszka Tomaszewska, Katarzyna Szepietowska, Czes{\l}aw
  Szymczak, and Maciej {\'S}mieta{\'n}ski.
\newblock In vivo performance of intraperitoneal onlay mesh after ventral
  hernia repair.
\newblock \emph{Clinical Biomechanics}, 78:\penalty0 105076, August 2020.
\newblock ISSN 0268-0033.
\newblock \doi{10.1016/j.clinbiomech.2020.105076}.

\bibitem[Szymczak et~al.(2010)Szymczak, Lubowiecka, Tomaszewska, and
  {\'S}mieta{\'n}ski]{szymczakModelingFasciameshSystem2010}
Czes{\l}aw Szymczak, Izabela Lubowiecka, Agnieszka Tomaszewska, and Maciej
  {\'S}mieta{\'n}ski.
\newblock Modeling of the fascia-mesh system and sensitivity analysis of a
  junction force after a laparoscopic ventral hernia repair.
\newblock \emph{undefined}, 2010.

\bibitem[Tomaszewska et~al.(2013)Tomaszewska, Lubowiecka, Szymczak,
  {\'S}mieta{\'n}ski, Meronk, K{\l}osowski, and
  Bury]{tomaszewskaPhysicalMathematicalModelling2013}
Agnieszka Tomaszewska, Izabela Lubowiecka, Czes{\l}aw Szymczak, Maciej
  {\'S}mieta{\'n}ski, B.~Meronk, P.~K{\l}osowski, and Kamil Bury.
\newblock Physical and mathematical modelling of implant--fascia system in
  order to improve laparoscopic repair of ventral hernia.
\newblock \emph{Clinical Biomechanics}, 28\penalty0 (7):\penalty0 743--751,
  August 2013.
\newblock ISSN 0268-0033.
\newblock \doi{10.1016/j.clinbiomech.2013.06.009}.

\bibitem[Chanda et~al.(2018)Chanda, Ruchti, and
  Upchurch]{chandaBiomechanicalModelingProsthetic2018}
Arnab Chanda, Tysum Ruchti, and Weston Upchurch.
\newblock Biomechanical {{Modeling}} of {{Prosthetic Mesh}} and {{Human Tissue
  Surrogate Interaction}}.
\newblock \emph{Biomimetics}, 3\penalty0 (3):\penalty0 27, September 2018.
\newblock ISSN 2313-7673.
\newblock \doi{10.3390/biomimetics3030027}.

\bibitem[Bl{\'a}zquez~Hernando et~al.(2015)Bl{\'a}zquez~Hernando,
  Garc{\'i}a~Ure{\~n}a, L{\'o}pez~Moncl{\'u}s, {Robin del Valle Lersundi},
  Melero~Montes, Cruz~Cidoncha, Jim{\'e}nez~Ceinos, and
  Castell{\'o}n~Pav{\'o}n]{blazquezhernandoRoturasMallaCausa2015}
Luis~Alberto Bl{\'a}zquez~Hernando, Miguel~{\'A}ngel Garc{\'i}a~Ure{\~n}a,
  Javier L{\'o}pez~Moncl{\'u}s, {\'A}lvaro {Robin del Valle Lersundi}, Daniel
  Melero~Montes, Arturo Cruz~Cidoncha, Carmen Jim{\'e}nez~Ceinos, and Camilo
  Castell{\'o}n~Pav{\'o}n.
\newblock {Roturas de malla: una causa poco frecuente de recidiva herniaria}.
\newblock \emph{Revista Hispanoamericana de Hernia}, 3\penalty0 (4):\penalty0
  155--159, October 2015.
\newblock ISSN 2255-2677.
\newblock \doi{10.1016/j.rehah.2015.02.006}.

\bibitem[Deerenberg et~al.(2016)Deerenberg, Verhelst, Hovius, and
  Lange]{deerenbergMeshExpansionCause2016}
Eva~B Deerenberg, J.~Verhelst, S.~E.~R. Hovius, and Johan~F. Lange.
\newblock Mesh expansion as the cause of bulging after abdominal wall hernia
  repair.
\newblock \emph{International Journal of Surgery Case Reports}, 28:\penalty0
  200--203, 2016.
\newblock ISSN 2210-2612.
\newblock \doi{10.1016/j.ijscr.2016.09.051}.

\bibitem[{Hern{\'a}ndez-Gasc{\'o}n} et~al.(2012){Hern{\'a}ndez-Gasc{\'o}n},
  Pe{\~n}a, Pascual, Rodr{\'i}guez, Bell{\'o}n, and
  Calvo]{hernandez-gasconLongtermAnisotropicMechanical2012}
Bel{\'e}n {Hern{\'a}ndez-Gasc{\'o}n}, Estefan{\'i}a Pe{\~n}a, G.~Pascual,
  M.~Rodr{\'i}guez, J.~M. Bell{\'o}n, and Bego{\~n}a Calvo.
\newblock Long-term anisotropic mechanical response of surgical meshes used to
  repair abdominal wall defects.
\newblock \emph{Journal of the Mechanical Behavior of Biomedical Materials},
  5\penalty0 (1):\penalty0 257--271, January 2012.
\newblock ISSN 1751-6161.
\newblock \doi{10.1016/j.jmbbm.2011.09.005}.

\bibitem[Todros et~al.(2017{\natexlab{b}})Todros, Pavan, and
  Natali]{todrosSyntheticSurgicalMeshes2017}
Silvia Todros, Piero~G. Pavan, and Arturo~N. Natali.
\newblock Synthetic surgical meshes used in abdominal wall surgery: {{Part
  I}}---materials and structural conformation.
\newblock \emph{Journal of Biomedical Materials Research Part B: Applied
  Biomaterials}, 105\penalty0 (3):\penalty0 689--699, 2017{\natexlab{b}}.
\newblock ISSN 1552-4981.
\newblock \doi{10.1002/jbm.b.33586}.

\bibitem[Li et~al.(2014)Li, Kruger, Jor, Wong, Dietz, Nash, and
  Nielsen]{liCharacterizingExVivo2014}
Xinxin Li, Jennifer~A. Kruger, Jessica W.~Y. Jor, Vivien Wong, Hans~P. Dietz,
  Martyn~P. Nash, and Poul M.~F. Nielsen.
\newblock Characterizing the ex vivo mechanical properties of synthetic
  polypropylene surgical mesh.
\newblock \emph{Journal of the Mechanical Behavior of Biomedical Materials},
  37:\penalty0 48--55, September 2014.
\newblock ISSN 1751-6161.
\newblock \doi{10.1016/j.jmbbm.2014.05.005}.

\bibitem[Velayudhan et~al.(2009)Velayudhan, Martin, and
  {Cooper-White}]{velayudhanEvaluationDynamicCreep2009}
Shiny Velayudhan, Darren Martin, and Justin {Cooper-White}.
\newblock Evaluation of dynamic creep properties of surgical mesh
  prostheses---{{Uniaxial}} fatigue.
\newblock \emph{Journal of Biomedical Materials Research Part B: Applied
  Biomaterials}, 91B\penalty0 (1):\penalty0 287--296, 2009.
\newblock ISSN 1552-4981.
\newblock \doi{10.1002/jbm.b.31401}.

\bibitem[Tomaszewska et~al.(2018)Tomaszewska, Lubowiecka, and
  Szymczak]{tomaszewskaMechanicsMeshImplanted2018}
Agnieszka Tomaszewska, Izabela Lubowiecka, and Czes{\l}aw Szymczak.
\newblock Mechanics of mesh implanted into abdominal wall under repetitive
  load. {{Experimental}} and numerical study.
\newblock \emph{Journal of Biomedical Materials Research Part B: Applied
  Biomaterials}, 107\penalty0 (5):\penalty0 1400--1409, 2018.
\newblock ISSN 1552-4981.
\newblock \doi{10.1002/jbm.b.34232}.

\bibitem[Eliason et~al.(2011)Eliason, Frisella, Matthews, and
  Deeken]{eliasonEffectRepetitiveLoading2011}
Braden~J. Eliason, Margaret~M. Frisella, Brent~D. Matthews, and Corey~R.
  Deeken.
\newblock Effect of {{Repetitive Loading}} on the {{Mechanical Properties}} of
  {{Synthetic Hernia Repair Materials}}.
\newblock \emph{Journal of the American College of Surgeons}, 213\penalty0
  (3):\penalty0 430--435, September 2011.
\newblock ISSN 1072-7515.
\newblock \doi{10.1016/j.jamcollsurg.2011.05.018}.

\bibitem[Kallinowski et~al.(2021{\natexlab{c}})Kallinowski, Gutjahr, Harder,
  Sabagh, Ludwig, Lozanovski, L{\"o}ffler, Rinn, G{\"o}rich, Grimm, Vollmer,
  and Nessel]{kallinowskiGripConceptIncisional2021}
Friedrich Kallinowski, Dominik Gutjahr, Felix Harder, Mohammad Sabagh, Yannique
  Ludwig, Vladimir~J. Lozanovski, Thorsten L{\"o}ffler, Johannes Rinn, Johannes
  G{\"o}rich, Annette Grimm, Matthias Vollmer, and Regine Nessel.
\newblock The {{Grip Concept}} of {{Incisional Hernia Repair}}---{{Dynamic
  Bench Test}}, {{CT Abdomen With Valsalva}} and 1-{{Year Clinical Results}}.
\newblock \emph{Frontiers in Surgery}, 8, April 2021{\natexlab{c}}.
\newblock ISSN 2296-875X.
\newblock \doi{10.3389/fsurg.2021.602181}.

\bibitem[Wang~See et~al.(2020)Wang~See, Kim, and
  Zhu]{wangseeHerniaMeshHernia2020}
Carmine Wang~See, Tiffany Kim, and Donghui Zhu.
\newblock Hernia {{Mesh}} and {{Hernia Repair}}: {{A Review}}.
\newblock \emph{Engineered Regeneration}, 1:\penalty0 19--33, January 2020.
\newblock ISSN 2666-1381.
\newblock \doi{10.1016/j.engreg.2020.05.002}.

\bibitem[Kalaba et~al.(2016)Kalaba, Gerhard, Winder, Pauli, Haluck, and
  Yang]{kalabaDesignStrategiesApplications2016}
Surge Kalaba, Ethan Gerhard, Joshua~S. Winder, Eric~M. Pauli, Randy~S. Haluck,
  and Jian Yang.
\newblock Design strategies and applications of biomaterials and devices for
  {{Hernia}} repair.
\newblock \emph{Bioactive Materials}, 1\penalty0 (1):\penalty0 2--17, September
  2016.
\newblock ISSN 2452-199X.
\newblock \doi{10.1016/j.bioactmat.2016.05.002}.

\bibitem[Schneeberger et~al.(2019)Schneeberger, Phillips, Huang, Pierce,
  Etemad, and Poulose]{schneebergerCostUtilityAnalysisBiologic2019}
Steven Schneeberger, Sharon Phillips, Li-Ching Huang, Richard~A. Pierce,
  Shervin~A. Etemad, and Benjamin~K. Poulose.
\newblock Cost-{{Utility Analysis}} of {{Biologic}} and {{Biosynthetic Mesh}}
  in {{Ventral Hernia Repair}}: {{When Are They Worth It}}?
\newblock \emph{Journal of the American College of Surgeons}, 228\penalty0
  (1):\penalty0 66--71, January 2019.
\newblock ISSN 1072-7515.
\newblock \doi{10.1016/j.jamcollsurg.2018.10.009}.

\bibitem[{Mazzola Poli de Figueiredo} et~al.(2023){Mazzola Poli de Figueiredo},
  Tastaldi, Mao, Lima, Huang, and
  Lu]{mazzolapolidefigueiredoBiologicSyntheticMesh2023}
Sergio {Mazzola Poli de Figueiredo}, Luciano Tastaldi, Rui-Min~Diana Mao,
  Diego~Laurentino Lima, Li-Ching Huang, and Richard Lu.
\newblock Biologic versus synthetic mesh in open ventral hernia repair: {{A}}
  systematic review and meta-analysis of randomized controlled trials.
\newblock \emph{Surgery}, 173\penalty0 (4):\penalty0 1001--1007, April 2023.
\newblock ISSN 0039-6060.
\newblock \doi{10.1016/j.surg.2022.12.002}.

\bibitem[Bittner et~al.(2019)Bittner, Bain, Bansal, Berrevoet,
  {Bingener-Casey}, Chen, Chen, Chowbey, Dietz, {de Beaux}, Ferzli, Fortelny,
  Hoffmann, Iskander, Ji, Jorgensen, Khullar, Kirchhoff, K{\"o}ckerling,
  Kukleta, LeBlanc, Li, Lomanto, Mayer, Meytes, Misra, {Morales-Conde},
  Niebuhr, Radvinsky, Ramshaw, Ranev, Reinpold, Sharma, Schrittwieser,
  Stechemesser, Sutedja, Tang, Warren, Weyhe, Wiegering, Woeste, and
  Yao]{bittnerUpdateGuidelinesLaparoscopic2019}
Reinhard Bittner, K.~Bain, V.~K. Bansal, F.~Berrevoet, J.~{Bingener-Casey},
  D.~Chen, J.~Chen, P.~Chowbey, U.~A. Dietz, A.~{de Beaux}, G.~Ferzli,
  R.~Fortelny, H.~Hoffmann, M.~Iskander, Z.~Ji, L.~N. Jorgensen, R.~Khullar,
  P.~Kirchhoff, F.~K{\"o}ckerling, J.~Kukleta, K.~LeBlanc, J.~Li, D.~Lomanto,
  F.~Mayer, V.~Meytes, M.~Misra, S.~{Morales-Conde}, H.~Niebuhr, D.~Radvinsky,
  B.~Ramshaw, D.~Ranev, W.~Reinpold, A.~Sharma, R.~Schrittwieser,
  B.~Stechemesser, B.~Sutedja, J.~Tang, J.~Warren, D.~Weyhe, A.~Wiegering,
  G.~Woeste, and Qiyuan Yao.
\newblock Update of {{Guidelines}} for laparoscopic treatment of ventral and
  incisional abdominal wall hernias ({{International Endohernia Society}}
  ({{IEHS}}))---{{Part A}}.
\newblock \emph{Surgical Endoscopy}, 33\penalty0 (10):\penalty0 3069--3139,
  October 2019.
\newblock ISSN 1432-2218.
\newblock \doi{10.1007/s00464-019-06907-7}.

\bibitem[Coda et~al.(2012)Coda, Lamberti, and
  Martorana]{codaClassificationProstheticsUsed2012}
Andrea Coda, Roberta Lamberti, and Selanna Martorana.
\newblock Classification of prosthetics used in hernia repair based on weight
  and biomaterial.
\newblock \emph{Hernia}, 16\penalty0 (1):\penalty0 9--20, February 2012.
\newblock ISSN 1248-9204.
\newblock \doi{10.1007/s10029-011-0868-z}.

\bibitem[Klosterhalfen et~al.(2005)Klosterhalfen, Junge, and
  Klinge]{klosterhalfenLightweightLargePorous2005}
Bernd Klosterhalfen, Karsten Junge, and Uwe Klinge.
\newblock The lightweight and large porous mesh concept for hernia repair.
\newblock \emph{Expert Review of Medical Devices}, 2\penalty0 (1):\penalty0
  103--117, January 2005.
\newblock ISSN 1743-4440.
\newblock \doi{10.1586/17434440.2.1.103}.

\bibitem[Yu and Ma(2021)]{yuMechanicalPropertiesWarpknitted2021}
Shuang Yu and Pibo Ma.
\newblock Mechanical properties of warp-knitted hernia repair mesh with various
  boundary conditions.
\newblock \emph{Journal of the Mechanical Behavior of Biomedical Materials},
  114:\penalty0 104192, February 2021.
\newblock ISSN 1751-6161.
\newblock \doi{10.1016/j.jmbbm.2020.104192}.

\bibitem[Petro et~al.(2015)Petro, Nahabet, Criss, Orenstein, {von Recum},
  Novitsky, and Rosen]{petroCentralFailuresLightweight2015}
Clayton~C Petro, E.~H. Nahabet, C.~N. Criss, S.~B. Orenstein, H.~A. {von
  Recum}, Y.~W. Novitsky, and Michael~J Rosen.
\newblock Central failures of lightweight monofilament polyester mesh causing
  hernia recurrence: A cautionary note.
\newblock \emph{Hernia}, 19\penalty0 (1):\penalty0 155--159, February 2015.
\newblock ISSN 1248-9204.
\newblock \doi{10.1007/s10029-014-1237-5}.

\bibitem[Warren et~al.(2017)Warren, McGrath, Hale, Ewing, CarbonellII, and
  CobbIV]{warrenPatternsRecurrenceMechanisms2017}
Jeremy~A. Warren, Sean~P. McGrath, Allyson~L. Hale, Joseph~A. Ewing, Alfredo~M.
  CarbonellII, and William~S. CobbIV.
\newblock Patterns of {{Recurrence}} and {{Mechanisms}} of {{Failure}} after
  {{Open Ventral Hernia Repair}} with {{Mesh}}.
\newblock \emph{The American Surgeon™}, 83\penalty0 (11):\penalty0
  1275--1282, November 2017.
\newblock ISSN 0003-1348.
\newblock \doi{10.1177/000313481708301131}.

\bibitem[Hauters et~al.(2017)Hauters, Desmet, Gherardi, Dewaele, Poilvache, and
  Malvaux]{hautersAssessmentPredictiveFactors2017}
Philippe Hauters, J.~Desmet, D.~Gherardi, S.~Dewaele, H.~Poilvache, and
  Philippe Malvaux.
\newblock Assessment of predictive factors for recurrence in laparoscopic
  ventral hernia repair using a bridging technique.
\newblock \emph{Surgical Endoscopy}, 31\penalty0 (9):\penalty0 3656--3663,
  September 2017.
\newblock ISSN 1432-2218.
\newblock \doi{10.1007/s00464-016-5401-0}.

\bibitem[Parker et~al.(2020)Parker, Halligan, Liang, Muysoms, Adrales, Boutall,
  Beaux, Dietz, Divino, Hawn, Heniford, Hong, Ibrahim, Itani, Jorgensen,
  Montgomery, {Morales-Conde}, Renard, Sanders, Smart, Torkington, and
  Windsor]{parkerInternationalClassificationAbdominal2020}
Sam~G Parker, S~Halligan, M~K Liang, F~E Muysoms, G~L Adrales, A~Boutall, A~C
  Beaux, U~A Dietz, C~M Divino, M~T Hawn, T~B Heniford, J~P Hong, N~Ibrahim,
  K~M~F Itani, L~N Jorgensen, A~Montgomery, S~{Morales-Conde}, Y~Renard, D~L
  Sanders, N~J Smart, J~J Torkington, and Alastair C~J Windsor.
\newblock International classification of abdominal wall planes ({{ICAP}}) to
  describe mesh insertion for ventral hernia repair.
\newblock \emph{British Journal of Surgery}, 107\penalty0 (3):\penalty0
  209--217, February 2020.
\newblock ISSN 0007-1323.
\newblock \doi{10.1002/bjs.11400}.

\bibitem[Albino et~al.(2013)Albino, Patel, Nahabedian, Sosin, Attinger, and
  Bhanot]{albinoDoesMeshLocation2013}
Frank~P. Albino, Ketan~M. Patel, Maurice~Y. Nahabedian, Michael Sosin,
  Christopher~E. Attinger, and Parag Bhanot.
\newblock Does {{Mesh Location Matter}} in {{Abdominal Wall Reconstruction}}?
  {{A Systematic Review}} of the {{Literature}} and a {{Summary}} of
  {{Recommendations}}.
\newblock \emph{Plastic and Reconstructive Surgery}, 132\penalty0 (5):\penalty0
  1295, November 2013.
\newblock ISSN 0032-1052.
\newblock \doi{10.1097/PRS.0b013e3182a4c393}.

\bibitem[Holihan et~al.(2017)Holihan, Hannon, Goodenough, {Flores-Gonzalez},
  Itani, Olavarria, Mo, Ko, Kao, and Liang]{holihanVentralHerniaRepair2017}
Julie~L. Holihan, Craig Hannon, Christopher Goodenough, Juan~R.
  {Flores-Gonzalez}, Kamal~M. Itani, Oscar Olavarria, Jiandi Mo, Tien~C. Ko,
  Lillian~S. Kao, and Mike~K. Liang.
\newblock Ventral {{Hernia Repair}}: {{A Meta-Analysis}} of {{Randomized
  Controlled Trials}}.
\newblock \emph{Surgical Infections}, 18\penalty0 (6):\penalty0 647--658,
  August 2017.
\newblock ISSN 1096-2964.
\newblock \doi{10.1089/sur.2017.029}.

\bibitem[Kallinowski et~al.(2015)Kallinowski, Baumann, Harder, Siassi, Mahn,
  Vollmer, and Morlock]{kallinowskiDynamicIntermittentStrain2015}
Friedrich Kallinowski, Elena Baumann, Felix Harder, Michael Siassi, Axel Mahn,
  Matthias Vollmer, and Michael~M. Morlock.
\newblock Dynamic intermittent strain can rapidly impair ventral hernia repair.
\newblock \emph{Journal of Biomechanics}, 48\penalty0 (15):\penalty0
  4026--4036, November 2015.
\newblock ISSN 0021-9290.
\newblock \doi{10.1016/j.jbiomech.2015.09.045}.

\bibitem[Sosin et~al.(2018)Sosin, Nahabedian, and
  Bhanot]{sosinPerfectPlaneSystematic2018}
Michael Sosin, Maurice~Y. Nahabedian, and Parag Bhanot.
\newblock The {{Perfect Plane}}: {{A Systematic Review}} of {{Mesh Location}}
  and {{Outcomes}}, {{Update}} 2018.
\newblock \emph{Plastic and Reconstructive Surgery}, 142\penalty0
  (3S):\penalty0 107S, September 2018.
\newblock ISSN 0032-1052.
\newblock \doi{10.1097/PRS.0000000000004864}.

\bibitem[Roca et~al.(2018)Roca, Nogu{\'e}s, Villalobos, M{\'i}as, Comellas,
  Gas, and Olsina]{rocaSurgicalDynamometerSimultaneously2018}
Joan Roca, Miquel Nogu{\'e}s, Rafael Villalobos, Mar{\'i}a~Carmen M{\'i}as,
  Mart{\'i} Comellas, Cristina Gas, and Jorge~Juan Olsina.
\newblock Surgical {{Dynamometer}} to {{Simultaneously Measure}} the {{Tension
  Forces}} and the {{Distance}} between {{Wound Edges}} during the {{Closure}}
  of a {{Laparotomy}}.
\newblock \emph{Sensors}, 18\penalty0 (1):\penalty0 189, January 2018.
\newblock ISSN 1424-8220.
\newblock \doi{10.3390/s18010189}.

\bibitem[Schachtrupp et~al.(2016)Schachtrupp, Wetter, and
  H{\"o}er]{schachtruppImplantableSensorDevice2016}
Alexander Schachtrupp, Olivier Wetter, and J{\"o}rg H{\"o}er.
\newblock An implantable sensor device measuring suture tension dynamics:
  Results of developmental and experimental work.
\newblock \emph{Hernia}, 20\penalty0 (4):\penalty0 601--606, August 2016.
\newblock ISSN 1248-9204.
\newblock \doi{10.1007/s10029-015-1433-y}.

\bibitem[Pereira et~al.(2024)Pereira, De~Oliveira, Tagerman, {Romero-Velez},
  Liu, and Malcher]{pereiraHowItUsing2024}
Xavier Pereira, Pedro De~Oliveira, Daniel Tagerman, Gustavo {Romero-Velez},
  Rockson Liu, and Flavio Malcher.
\newblock How {{I}} do it: Using physics and progressive defect tensioning to
  close large hernia defects during {{MIS}} ventral hernia repair.
\newblock \emph{Hernia}, 29\penalty0 (1):\penalty0 55, December 2024.
\newblock ISSN 1248-9204.
\newblock \doi{10.1007/s10029-024-03230-6}.

\bibitem[Reza~Zahiri et~al.(2018)Reza~Zahiri, Belyansky, and
  Park]{rezazahiriAbdominalWallHernia2018}
Hamid Reza~Zahiri, Igor Belyansky, and Adrian Park.
\newblock Abdominal {{Wall Hernia}}.
\newblock \emph{Current Problems in Surgery}, 55\penalty0 (8):\penalty0
  286--317, August 2018.
\newblock ISSN 0011-3840.
\newblock \doi{10.1067/j.cpsurg.2018.08.005}.

\bibitem[Muysoms et~al.(2013)Muysoms, Vander~Mijnsbrugge, Pletinckx, Boldo,
  Jacobs, Michiels, and Ceulemans]{muysomsRandomizedClinicalTrial2013}
Filip~E Muysoms, G.~Vander~Mijnsbrugge, P.~Pletinckx, E.~Boldo, I.~Jacobs,
  M.~Michiels, and R.~Ceulemans.
\newblock Randomized clinical trial of mesh fixation with ``double crown''
  versus ``sutures and tackers'' in laparoscopic ventral hernia repair.
\newblock \emph{Hernia}, 17\penalty0 (5):\penalty0 603--612, October 2013.
\newblock ISSN 1248-9204.
\newblock \doi{10.1007/s10029-013-1084-9}.

\bibitem[Khan et~al.(2018)Khan, Bughio, Ali, Hajibandeh, and
  Hajibandeh]{khanAbsorbableNonabsorbableTacks2018}
Rao Muhammad~Asaf Khan, Mumtaz Bughio, Baqar Ali, Shahin Hajibandeh, and Shahab
  Hajibandeh.
\newblock Absorbable versus non-absorbable tacks for mesh fixation in
  laparoscopic ventral hernia repair: {{A}} systematic review and
  meta-analysis.
\newblock \emph{International Journal of Surgery}, 53:\penalty0 184--192, May
  2018.
\newblock ISSN 1743-9191.
\newblock \doi{10.1016/j.ijsu.2018.03.042}.

\bibitem[Harsl{\o}f et~al.(2018)Harsl{\o}f, {Krum-M{\o}ller}, Sommer, Zinther,
  Wara, and {Friis-Andersen}]{harslofEffectFixationDevices2018}
Sanne Harsl{\o}f, Pia {Krum-M{\o}ller}, Thorbj{\o}rn Sommer, Nellie Zinther,
  P{\aa}l Wara, and Hans {Friis-Andersen}.
\newblock Effect of fixation devices on postoperative pain after laparoscopic
  ventral hernia repair: A randomized clinical trial of permanent tacks,
  absorbable tacks, and synthetic glue.
\newblock \emph{Langenbeck's Archives of Surgery}, 403\penalty0 (4):\penalty0
  529--537, June 2018.
\newblock ISSN 1435-2451.
\newblock \doi{10.1007/s00423-018-1676-z}.

\bibitem[Calpin et~al.(2024)Calpin, Davey, Whooley, Ryan, Ryan, Ponten, Weiss,
  Conneely, Robb, and Donlon]{calpinEvaluatingMeshFixation2024}
Gavin~G. Calpin, Matthew~G. Davey, Jack Whooley, Eanna~J. Ryan, Odhran~K. Ryan,
  Jeroen E.~H. Ponten, Andreas Weiss, John~B. Conneely, William~B. Robb, and
  Noel~E. Donlon.
\newblock Evaluating mesh fixation techniques for ventral hernia repair: {{A}}
  systematic review and network meta-analysis of randomised control trials.
\newblock \emph{The American Journal of Surgery}, 228:\penalty0 62--69,
  February 2024.
\newblock ISSN 0002-9610.
\newblock \doi{10.1016/j.amjsurg.2023.09.015}.

\bibitem[Afifi et~al.(2017)Afifi, Hartmann, Talaat, Alfotooh, Omar, Mareei,
  Sanchez, and Kempton]{afifiQuantitativeAssessmentTension2017}
Ahmed~M. Afifi, Emily Hartmann, Ahmed Talaat, Ashraf~Abo Alfotooh, Omar~S.
  Omar, Sayed Mareei, Ruston Sanchez, and Steve~J. Kempton.
\newblock Quantitative {{Assessment}} of {{Tension Reduction}} at the {{Midline
  Closure During Abdominal Component Separation}}.
\newblock \emph{Journal of the American College of Surgeons}, 224\penalty0
  (5):\penalty0 954--961, May 2017.
\newblock ISSN 1879-1190.
\newblock \doi{10.1016/j.jamcollsurg.2016.12.052}.

\bibitem[Lisiecki et~al.(2015)Lisiecki, Kozlow, Agarwal, Ranganathan,
  Terjimanian, Rinkinen, Brownley, Enchakalody, Wang, and
  Levi]{lisieckiAbdominalWallDynamics2015}
Jeffrey Lisiecki, Jeffrey~H. Kozlow, Shailesh Agarwal, Kavitha Ranganathan,
  Michael~N. Terjimanian, Jacob Rinkinen, R.~Cameron Brownley, Binu
  Enchakalody, Stewart~C. Wang, and Benjamin Levi.
\newblock Abdominal wall dynamics after component separation hernia repair.
\newblock \emph{Journal of Surgical Research}, 193\penalty0 (1):\penalty0
  497--503, January 2015.
\newblock ISSN 0022-4804.
\newblock \doi{10.1016/j.jss.2014.08.008}.

\bibitem[Daes et~al.(2022)Daes, Oma, and
  Jorgensen]{daesChangesAbdominalWall2022}
Jorge Daes, E.~Oma, and Lars~Nannestad Jorgensen.
\newblock Changes in the abdominal wall after anterior, posterior, and combined
  component separation.
\newblock \emph{Hernia}, 26\penalty0 (1):\penalty0 17--27, February 2022.
\newblock ISSN 1248-9204.
\newblock \doi{10.1007/s10029-021-02535-0}.

\bibitem[Love et~al.(2021)Love, Warren, Davis, Ewing, Hall, Cobb, and
  Carbonell]{loveComputedTomographyImaging2021}
Michael~Wes Love, J.~A. Warren, S.~Davis, J.~A. Ewing, A.~M. Hall, W.~S. Cobb,
  and Alfredo~Maximiliano Carbonell.
\newblock Computed tomography imaging in ventral hernia repair: Can we predict
  the need for myofascial release?
\newblock \emph{Hernia}, 25\penalty0 (2):\penalty0 471--477, April 2021.
\newblock ISSN 1248-9204.
\newblock \doi{10.1007/s10029-020-02181-y}.

\bibitem[Fas()]{Fasciotens}
Fasciotens.
\newblock https://www.fasciotens.com.

\bibitem[Hees and Willeke(2020)]{heesPreventionFascialRetraction2020}
Anita Hees and Frank Willeke.
\newblock Prevention of {{Fascial Retraction}} in the {{Open Abdomen}} with a
  {{Novel Device}}.
\newblock \emph{Case Reports in Surgery}, 2020\penalty0 (1):\penalty0 8254804,
  2020.
\newblock ISSN 2090-6919.
\newblock \doi{10.1155/2020/8254804}.

\bibitem[Niebuhr et~al.(2021)Niebuhr, Aufenberg, Dag, Reinpold, Peiper,
  Schardey, Renter, Aly, Eucker, K{\"o}ckerling, and
  Eichelter]{niebuhrIntraoperativeFasciaTension2021}
Henning Niebuhr, Thomas Aufenberg, Halil Dag, Wolfgang Reinpold, Christian
  Peiper, Hans~Martin Schardey, Marc~Alexander Renter, Mohamed Aly, Dietmar
  Eucker, Ferdinand K{\"o}ckerling, and Jakob Eichelter.
\newblock Intraoperative {{Fascia Tension}} as an {{Alternative}} to
  {{Component Separation}}. {{A Prospective Observational Study}}.
\newblock \emph{Frontiers in Surgery}, 7, February 2021.
\newblock ISSN 2296-875X.
\newblock \doi{10.3389/fsurg.2020.616669}.

\bibitem[Liu et~al.(2021)Liu, Xie, Zheng, He, Qiao, and
  Meng]{liuRegulatoryScienceHernia2021}
Wenbo Liu, Yajie Xie, Yudong Zheng, Wei He, Kun Qiao, and Haoye Meng.
\newblock Regulatory science for hernia mesh: {{Current}} status and future
  perspectives.
\newblock \emph{BIOACTIVE MATERIALS}, 6\penalty0 (2):\penalty0 420--432,
  February 2021.
\newblock \doi{10.1016/j.bioactmat.2020.08.021}.

\bibitem[Welty et~al.(2001)Welty, Klinge, Klosterhalfen, Kasperk, and
  Schumpelick]{weltyFunctionalImpairmentComplaints2001}
Georg Welty, Uwe Klinge, Bernd Klosterhalfen, R.~Kasperk, and Volker
  Schumpelick.
\newblock Functional impairment and complaints following incisional hernia
  repair with different polypropylene meshes.
\newblock \emph{Hernia}, 5\penalty0 (3):\penalty0 142--147, September 2001.
\newblock ISSN 1248-9204.
\newblock \doi{10.1007/s100290100017}.

\bibitem[Crocetti et~al.(2020)Crocetti, Carbotta, Cantelli, Iorio, Gurrado,
  Sibio, Brauneis, and Cavallaro]{crocettiDietaryProteinSupplementation2020}
Daniele Crocetti, Giovanni Carbotta, Fabio Cantelli, Olga Iorio, Angela
  Gurrado, Simone Sibio, Stefano Brauneis, and Giuseppe Cavallaro.
\newblock Dietary {{Protein Supplementation Helps}} in {{Muscle Thickness
  Regain}} after {{Abdominal Wall Reconstruction}} for {{Incisional Hernia}}.
\newblock \emph{The American Surgeon™}, 86\penalty0 (3):\penalty0 232--236,
  March 2020.
\newblock ISSN 0003-1348.
\newblock \doi{10.1177/000313482008600333}.

\bibitem[Liang et~al.(2018)Liang, Bernardi, Holihan, Cherla, Escamilla, Lew,
  Berger, Ko, and Kao]{liangModifyingRisksVentral2018}
Mike~K. Liang, Karla Bernardi, Julie~L. Holihan, Deepa~V. Cherla, Richard
  Escamilla, Debbie~F. Lew, David~H. Berger, Tien~C. Ko, and Lillian~S. Kao.
\newblock Modifying {{Risks}} in {{Ventral Hernia Patients With
  Prehabilitation}}: {{A Randomized Controlled Trial}}.
\newblock \emph{Annals of Surgery}, 268\penalty0 (4):\penalty0 674, October
  2018.
\newblock ISSN 0003-4932.
\newblock \doi{10.1097/SLA.0000000000002961}.

\bibitem[Pezeshk et~al.(2015)Pezeshk, Pulikkottil, Mapula, Schaffer, Yap,
  Scott, Gordon, and Hoxworth]{pezeshkComplexAbdominalWall2015}
Ronnie~A. Pezeshk, Benson~J. Pulikkottil, Steven Mapula, Nathaniel~E. Schaffer,
  Lori Yap, Kelly Scott, Patricia Gordon, and Ronald~E. Hoxworth.
\newblock Complex {{Abdominal Wall Reconstruction}}: {{A Novel Approach}} to
  {{Postoperative Care Using Physical Medicine}} and {{Rehabilitation}}.
\newblock \emph{Plastic and Reconstructive Surgery}, 136\penalty0 (3):\penalty0
  362e, September 2015.
\newblock ISSN 0032-1052.
\newblock \doi{10.1097/PRS.0000000000001532}.

\bibitem[Angelici et~al.(2016)Angelici, Perotti, Dezzi, Amatucci, Mancuso,
  Caronna, and Palumbo]{angeliciMeasurementIntraabdominalPressure2016}
Alberto~M. Angelici, B.~Perotti, C.~Dezzi, C.~Amatucci, G.~Mancuso, R.~Caronna,
  and Piergaspare Palumbo.
\newblock Measurement of intra-abdominal pressure in large incisional hernia
  repair to prevent abdominal compartmental syndrome.
\newblock \emph{Il Giornale Di Chirurgia}, 37\penalty0 (1):\penalty0 31--36,
  2016.
\newblock ISSN 0391-9005.
\newblock \doi{10.11138/gchir/2016.37.1.031}.

\bibitem[{Espinosa-de-los-Monteros} et~al.(2022){Espinosa-de-los-Monteros},
  {Dominguez-Arellano}, {Vazquez-Guadalupe}, {de-la-Garza-Elizondo}, and
  {Caralampio-Castro}]{espinosa-de-los-monterosImmediateChangesIntraabdominal2022}
Antonio {Espinosa-de-los-Monteros}, S.~{Dominguez-Arellano},
  J.~{Vazquez-Guadalupe}, C.~{de-la-Garza-Elizondo}, and Ali
  {Caralampio-Castro}.
\newblock Immediate changes in intra-abdominal pressure and lung indicators in
  patients undergoing complex ventral hernia repair with the transversus
  abdominis muscle release, with and without preoperative botulinum toxin.
\newblock \emph{Hernia}, March 2022.
\newblock ISSN 1248-9204.
\newblock \doi{10.1007/s10029-022-02601-1}.

\bibitem[Luj{\'a}n et~al.(2004)Luj{\'a}n, Frutos, Hern{\'a}ndez, Liron, Cuenca,
  Valero, and Parrilla]{lujanLaparoscopicOpenGastric2004}
Juan~A. Luj{\'a}n, M.~Dolores Frutos, Quiteria Hern{\'a}ndez, Ram{\'o}n Liron,
  Jose~R. Cuenca, Graciela Valero, and Pascual Parrilla.
\newblock Laparoscopic {{Versus Open Gastric Bypass}} in the {{Treatment}} of
  {{Morbid Obesity}}: {{A Randomized Prospective Study}}.
\newblock \emph{Annals of Surgery}, 239\penalty0 (4):\penalty0 433--437, April
  2004.
\newblock ISSN 0003-4932.
\newblock \doi{10.1097/01.sla.0000120071.75691.1f}.

\bibitem[Barbaros et~al.(2007)Barbaros, Asoglu, Seven, Erbil, Dinccag, Deveci,
  Ozarmagan, and Mercan]{barbarosComparisonLaparoscopicOpen2007}
Umut Barbaros, O.~Asoglu, R.~Seven, Y.~Erbil, A.~Dinccag, U.~Deveci,
  S.~Ozarmagan, and Sel{\c c}uk Mercan.
\newblock The comparison of laparoscopic and open ventral hernia repairs: A
  prospective randomized study.
\newblock \emph{Hernia}, 11\penalty0 (1):\penalty0 51--56, February 2007.
\newblock ISSN 1248-9204.
\newblock \doi{10.1007/s10029-006-0160-9}.

\bibitem[Pereira et~al.(2022)Pereira, Lima, Friedmann, {Romero-Velez},
  Mandujano, {Ramos-Santillan}, {Garcia-Cabrera}, and
  Malcher]{pereiraRoboticAbdominalWall2022}
Xavier Pereira, Diego~L. Lima, Patricia Friedmann, Gustavo {Romero-Velez},
  Cosman~C. Mandujano, Vicente {Ramos-Santillan}, Ana {Garcia-Cabrera}, and
  Flavio Malcher.
\newblock Robotic abdominal wall repair: Adoption and early outcomes in a large
  academic medical center.
\newblock \emph{Journal of Robotic Surgery}, 16\penalty0 (2):\penalty0
  383--392, April 2022.
\newblock ISSN 1863-2491.
\newblock \doi{10.1007/s11701-021-01251-2}.

\bibitem[Schumpelick et~al.(1996)Schumpelick, Conze, and
  Klinge]{schumpelickPraeperitonealeNetzplastikReparation1996}
Volker Schumpelick, Joachim Conze, and Uwe Klinge.
\newblock {Die pr{\"a}peritoneale Netzplastik in der Reparation der
  NarbenhernieEine vergleichende retrospektive Studie an 272 operierten
  Narbenhernien}.
\newblock \emph{Der Chirurg}, 67\penalty0 (10):\penalty0 1028--1035, October
  1996.
\newblock ISSN 1433-0385.
\newblock \doi{10.1007/s001040050099}.

\bibitem[Anthony et~al.(2000)Anthony, Bergen, Kim, Henderson, Fahey, Rege, and
  Turnage]{anthonyFactorsAffectingRecurrence2000}
Thomas Anthony, Patricia~C. Bergen, L.~T. Kim, M.~Henderson, T.~Fahey,
  Robert~V. Rege, and Richard~H. Turnage.
\newblock Factors affecting recurrence following incisional herniorrhaphy.
\newblock \emph{World Journal of Surgery}, 24\penalty0 (1):\penalty0
  95--100;discussion 101, January 2000.
\newblock ISSN 0364-2313.
\newblock \doi{10.1007/s002689910018}.

\bibitem[Urschel et~al.(1988{\natexlab{b}})Urschel, Scott, and
  Williams]{urschelEffectMechanicalStress1988}
John~D. Urschel, Paul~G. Scott, and Henry Thomas~G. Williams.
\newblock The effect of mechanical stress on soft and hard tissue repair; a
  review.
\newblock \emph{British Journal of Plastic Surgery}, 41\penalty0 (2):\penalty0
  182--186, March 1988{\natexlab{b}}.
\newblock ISSN 0007-1226.
\newblock \doi{10.1016/0007-1226(88)90049-5}.

\bibitem[Ma et~al.(2023)Ma, Jing, Liu, Liu, Liu, and
  Chen]{maGlobalRegionalNational2023}
Qiuyue Ma, Wenzhan Jing, Xiaoli Liu, Jue Liu, Min Liu, and Jie Chen.
\newblock The global, regional, and national burden and its trends of inguinal,
  femoral, and abdominal hernia from 1990 to 2019: Findings from the 2019
  {{Global Burden}} of {{Disease Study}} -- a cross-sectional study.
\newblock \emph{International Journal of Surgery}, 109\penalty0 (3):\penalty0
  333, March 2023.
\newblock \doi{10.1097/JS9.0000000000000217}.

\bibitem[Rankin et~al.(2006)Rankin, Stokes, and
  Newham]{rankinAbdominalMuscleSize2006}
Gabrielle Rankin, Maria Stokes, and Dianne~J. Newham.
\newblock Abdominal muscle size and symmetry in normal subjects.
\newblock \emph{Muscle \& Nerve}, 34\penalty0 (3):\penalty0 320--326, 2006.
\newblock ISSN 1097-4598.
\newblock \doi{10.1002/mus.20589}.

\bibitem[Schlosser et~al.(2020)Schlosser, Maloney, Thielan, Prasad, Kercher,
  Colavita, Heniford, and Augenstein]{schlosserOutcomesSpecificPatient2020}
Kathryn~A. Schlosser, Sean~R. Maloney, Otto Thielan, Tanushree Prasad, Kent
  Kercher, Paul~D. Colavita, B.~Todd Heniford, and Vedra~A. Augenstein.
\newblock Outcomes specific to patient sex after open ventral hernia repair.
\newblock \emph{Surgery}, 167\penalty0 (3):\penalty0 614--619, March 2020.
\newblock ISSN 0039-6060, 1532-7361.
\newblock \doi{10.1016/j.surg.2019.11.016}.

\bibitem[Cox et~al.(2016)Cox, Huntington, Blair, Prasad, Lincourt, Heniford,
  and Augenstein]{coxPredictiveModelingChronic2016}
Tiffany~C. Cox, Ciara~R. Huntington, Laurel~J. Blair, Tanushree Prasad, Amy~E.
  Lincourt, Brant~T. Heniford, and Vedra~A. Augenstein.
\newblock Predictive modeling for chronic pain after ventral hernia repair.
\newblock \emph{The American Journal of Surgery}, 212\penalty0 (3):\penalty0
  501--510, September 2016.
\newblock ISSN 0002-9610, 1879-1883.
\newblock \doi{10.1016/j.amjsurg.2016.02.021}.

\bibitem[Craig et~al.(2016)Craig, Parikh, Markert, and
  Saxe]{craigPrevalencePredictorsHernia2016}
Paul Craig, Priti~P. Parikh, Ronald Markert, and Jonathan Saxe.
\newblock Prevalence and {{Predictors}} of {{Hernia Infection}}: {{Does Gender
  Matter}}?
\newblock \emph{The American Surgeon™}, 82\penalty0 (4):\penalty0 93--95,
  April 2016.
\newblock ISSN 0003-1348.
\newblock \doi{10.1177/000313481608200408}.

\bibitem[Holt et~al.(1992)Holt, Kirk, Regan, Hurson, Lindblad, and
  Barbul]{holtEffectAgeWound1992}
David~R. Holt, Stephen~J. Kirk, Mark~C. Regan, Moira Hurson, William~J.
  Lindblad, and Adrian Barbul.
\newblock Effect of age on wound healing in healthy human beings.
\newblock \emph{Surgery}, 112\penalty0 (2):\penalty0 293--298, August 1992.
\newblock ISSN 0039-6060, 1532-7361.
\newblock \doi{10.5555/uri:pii:003960609290223M}.

\bibitem[Goswami et~al.(2022)Goswami, Basu, and
  Shukla]{goswamiWoundHealingGolden2022}
Aakansha~Giri Goswami, Somprakas Basu, and Vijay~Kumar Shukla.
\newblock Wound {{Healing}} in the {{Golden Agers}}: {{What We Know}} and the
  {{Possible Way Ahead}}.
\newblock \emph{The International Journal of Lower Extremity Wounds},
  21\penalty0 (3):\penalty0 264--271, September 2022.
\newblock ISSN 1534-7346.
\newblock \doi{10.1177/15347346211037841}.

\bibitem[Antoniou et~al.(2011)Antoniou, Georgiadis, Antoniou, Granderath,
  Giannoukas, and Lazarides]{antoniouAbdominalAorticAneurysm2011}
George~A. Antoniou, George~S. Georgiadis, Stavros~A. Antoniou, Frank~A.
  Granderath, Athanasios~D. Giannoukas, and Miltos~K. Lazarides.
\newblock Abdominal aortic aneurysm and abdominal wall hernia as manifestations
  of a connective tissue disorder.
\newblock \emph{Journal of Vascular Surgery}, 54\penalty0 (4):\penalty0
  1175--1181, October 2011.
\newblock ISSN 0741-5214.
\newblock \doi{10.1016/j.jvs.2011.02.065}.

\bibitem[Rath et~al.(1996)Rath, Attali, Dumas, Goldlust, Zhang, and
  Chevrel]{rathAbdominalLineaAlba1996}
Ana Rath, P.~Attali, {\relax JL}.~Dumas, D.~Goldlust, J.~Zhang, and Jean~Paul
  Chevrel.
\newblock The abdominal linea alba: An anatomo-radiologic and biomechanical
  study.
\newblock \emph{Surgical and Radiologic Anatomy}, 18\penalty0 (4):\penalty0
  281--288, December 1996.
\newblock ISSN 1279-8517.
\newblock \doi{10.1007/BF01627606}.

\bibitem[Rath et~al.(1997)Rath, Zhang, and
  Chevrel]{rathSheathRectusAbdominis1997}
Ana Rath, J.~Zhang, and Jean~Paul Chevrel.
\newblock The sheath of the rectus abdominis muscle: An anatomical and
  biomechanical study.
\newblock \emph{Hernia}, 1997.
\newblock \doi{10.1007/BF02426420}.

\bibitem[Sanchez et~al.(2001)Sanchez, Tenofsky, Dort, Shen, Helmer, and
  Smith]{sanchezWhatNormalIntraabdominal2001}
Noel~C. Sanchez, P.~L. Tenofsky, J.~M. Dort, L.~Y. Shen, S.~D. Helmer, and
  R.~Stephen Smith.
\newblock What is normal intra-abdominal pressure?
\newblock \emph{The American Surgeon}, 67\penalty0 (3):\penalty0 243--248,
  March 2001.
\newblock ISSN 0003-1348.

\bibitem[Jansen et~al.(2004)Jansen, Mertens, Klinge, and
  Schumpelick]{jansenBiologyHerniaFormation2004}
Petra~Lynen Jansen, P.~R. Mertens, Uwe Klinge, and Volker Schumpelick.
\newblock The biology of hernia formation.
\newblock \emph{Surgery}, 136\penalty0 (1):\penalty0 1--4, July 2004.
\newblock ISSN 0039-6060.
\newblock \doi{10.1016/j.surg.2004.01.004}.

\bibitem[Henriksen et~al.(2015)Henriksen, Mortensen, Sorensen, {Bay-Jensen},
  {\AA}gren, Jorgensen, and Karsdal]{henriksenCollagenTurnoverProfile2015}
Nadia~A. Henriksen, Joachim~H. Mortensen, Lars~T. Sorensen, Anne~C.
  {Bay-Jensen}, Magnus~S. {\AA}gren, Lars~N. Jorgensen, and Morten~A. Karsdal.
\newblock The collagen turnover profile is altered in patients with inguinal
  and incisional hernia.
\newblock \emph{Surgery}, 157\penalty0 (2):\penalty0 312--321, February 2015.
\newblock ISSN 0039-6060.
\newblock \doi{10.1016/j.surg.2014.09.006}.

\bibitem[{Al-Khan} et~al.(2011){Al-Khan}, Shah, Altabban, Kaul, Dyer, Alvarez,
  and Saber]{al-khanMeasurementIntraabdominalPressure2011}
Abdulla {Al-Khan}, Manan Shah, Mohamed Altabban, Sanjeev Kaul, Keisha~Y. Dyer,
  Manuel Alvarez, and Shelley Saber.
\newblock Measurement of intraabdominal pressure in pregnant women at term.
\newblock \emph{The Journal of Reproductive Medicine}, 56\penalty0
  (1-2):\penalty0 53--57, 2011.
\newblock ISSN 0024-7758.

\bibitem[Owei et~al.(2017)Owei, Swendiman, Kelz, Dempsey, and
  Dumon]{oweiImpactBodyMass2017}
Lily Owei, Robert~A. Swendiman, Rachel~R. Kelz, Daniel~T. Dempsey, and
  Kristoffel~R. Dumon.
\newblock Impact of body mass index on open ventral hernia repair: {{A}}
  retrospective review.
\newblock \emph{Surgery}, 162\penalty0 (6):\penalty0 1320--1329, December 2017.
\newblock ISSN 0039-6060.
\newblock \doi{10.1016/j.surg.2017.07.025}.

\bibitem[Gignoux et~al.(2021)Gignoux, Bayon, Martin, Phan, Augusto, Darnis, and
  Sarazin]{gignouxIncidenceRiskFactors2021}
Benoit Gignoux, Yves Bayon, Damien Martin, Raksmey Phan, Vincent Augusto,
  Benjamin Darnis, and Marianne Sarazin.
\newblock Incidence and risk factors for incisional hernia and recurrence:
  {{Retrospective}} analysis of the {{French}} national database.
\newblock \emph{Colorectal Disease}, 23\penalty0 (6):\penalty0 1515--1523,
  2021.
\newblock ISSN 1463-1318.
\newblock \doi{10.1111/codi.15581}.

\bibitem[{van Silfhout} et~al.(2021){van Silfhout}, Leenders, Heisterkamp, and
  Ibelings]{vansilfhoutRecurrentIncisionalHernia2021}
Lysanne {van Silfhout}, L.~A.~M. Leenders, J.~Heisterkamp, and Maaike~S.
  Ibelings.
\newblock Recurrent incisional hernia repair: Surgical outcomes in correlation
  with body-mass index.
\newblock \emph{Hernia}, 25\penalty0 (1):\penalty0 77--83, February 2021.
\newblock ISSN 1248-9204.
\newblock \doi{10.1007/s10029-020-02320-5}.

\bibitem[Tastaldi et~al.(2019)Tastaldi, Krpata, Prabhu, Petro, Rosenblatt,
  Haskins, Olson, Stewart, Rosen, and
  Greenberg]{tastaldiEffectIncreasingBody2019}
Luciano Tastaldi, David~M. Krpata, Ajita~S. Prabhu, Clayton~C. Petro, Steven
  Rosenblatt, Ivy~N. Haskins, Molly~A. Olson, Thomas~G. Stewart, Michael~J.
  Rosen, and Jacob~A. Greenberg.
\newblock The effect of increasing body mass index on wound complications in
  open ventral hernia repair with mesh.
\newblock \emph{The American Journal of Surgery}, 218\penalty0 (3):\penalty0
  560--566, September 2019.
\newblock ISSN 0002-9610.
\newblock \doi{10.1016/j.amjsurg.2019.01.022}.

\bibitem[Frommer et~al.(2024)Frommer, Faderani, Kanapathy,
  {P{\'e}russeau-Lambert}, Shankar, Malhotra, Khosh~Zaban, Floyd, Butler, and
  Ghali]{frommerPreoperativeCTImaging2024}
Marvin~L. Frommer, R.~Faderani, M.~Kanapathy, A.~{P{\'e}russeau-Lambert},
  A.~Shankar, A.~Malhotra, M.~Khosh~Zaban, D.~Floyd, P.~E.~M. Butler, and Shadi
  Ghali.
\newblock Preoperative {{CT}} imaging as a tool to predict incisional hernia
  outcomes following abdominal wall reconstruction: {{A}} retrospective cohort
  analysis.
\newblock \emph{Journal of Plastic, Reconstructive \& Aesthetic Surgery},
  88:\penalty0 369--377, January 2024.
\newblock ISSN 1748-6815.
\newblock \doi{10.1016/j.bjps.2023.11.007}.

\bibitem[Yamamoto et~al.(2018)Yamamoto, Takakura, Ikeda, Itamoto, Urushihara,
  and Egi]{yamamotoVisceralObesitySignificant2018}
Masateru Yamamoto, Yuji Takakura, Satoshi Ikeda, Toshiyuki Itamoto, Takashi
  Urushihara, and Hiroyuki Egi.
\newblock Visceral obesity is a significant risk factor for incisional hernia
  after laparoscopic colorectal surgery: {{A}} single-center review.
\newblock \emph{Asian Journal of Endoscopic Surgery}, 11\penalty0 (4):\penalty0
  373--377, 2018.
\newblock ISSN 1758-5910.
\newblock \doi{10.1111/ases.12466}.

\bibitem[Kotidis et~al.(2011)Kotidis, Papavramidis, Ioannidis, Cheva, Lazou,
  Michalopoulos, Karkavelas, and
  Papavramidis]{kotidisEffectChronicallyIncreased2011}
Efstathios~V. Kotidis, Theodosis~S. Papavramidis, Kostas Ioannidis, Angeliki
  Cheva, Thomai Lazou, Nikolaos Michalopoulos, George Karkavelas, and Spiros~T.
  Papavramidis.
\newblock The {{Effect}} of {{Chronically Increased Intra-Abdominal Pressure}}
  on {{Rectus Abdominis Muscle Histology}} an {{Experimental Study}} on
  {{Rabbits}}.
\newblock \emph{Journal of Surgical Research}, 171\penalty0 (2):\penalty0
  609--614, December 2011.
\newblock ISSN 0022-4804, 1095-8673.
\newblock \doi{10.1016/j.jss.2010.06.034}.

\bibitem[Schachtrupp et~al.(2019)Schachtrupp, Wetter, and
  H{\"o}er]{schachtruppInfluenceElevatedIntraabdominal2019}
Alexander Schachtrupp, Oliver Wetter, and J{\"o}rg H{\"o}er.
\newblock Influence of {{Elevated Intra-abdominal Pressure}} on {{Suture
  Tension Dynamics}} in a {{Porcine Model}}.
\newblock \emph{Journal of Surgical Research}, 233:\penalty0 207--212, January
  2019.
\newblock ISSN 0022-4804.
\newblock \doi{10.1016/j.jss.2018.07.043}.

\bibitem[Mure{\c s}an et~al.(2016)Mure{\c s}an, Mure{\c s}an, Bara, Neagoe,
  Sala, and Suciu]{muresanHerniaRecurrenceLong2016}
Mircea Mure{\c s}an, Simona Mure{\c s}an, Tivadar Bara, Radu Neagoe, Daniela
  Sala, and Bogdan Suciu.
\newblock Hernia recurrence long term follow-up after open procedures of
  abdominal wall plasty -- {{Prospective}} study including 142 patients.
\newblock \emph{Cirug{\'i}a y Cirujanos (English Edition)}, 84\penalty0
  (5):\penalty0 376--383, September 2016.
\newblock ISSN 2444-0507.
\newblock \doi{10.1016/j.circen.2016.08.007}.

\bibitem[{Ibarra-Hurtado} et~al.(2009){Ibarra-Hurtado}, {Nu{\~n}o-Guzm{\'a}n},
  {Echeagaray-Herrera}, {Robles-V{\'e}lez}, and {de Jes{\'u}s
  Gonz{\'a}lez-Jaime}]{ibarra-hurtadoUseBotulinumToxin2009}
Tomas~R. {Ibarra-Hurtado}, Carlos~M. {Nu{\~n}o-Guzm{\'a}n}, Jorge~E.
  {Echeagaray-Herrera}, Everardo {Robles-V{\'e}lez}, and Jos{\'e} {de Jes{\'u}s
  Gonz{\'a}lez-Jaime}.
\newblock Use of {{Botulinum Toxin Type A Before Abdominal Wall Hernia
  Reconstruction}}.
\newblock \emph{World Journal of Surgery}, 33\penalty0 (12):\penalty0 2014,
  2009.
\newblock ISSN 1432-2323.
\newblock \doi{10.1007/s00268-009-0203-3}.

\bibitem[Motz et~al.(2018)Motz, Schlosser, and
  Heniford]{motzChemicalComponentsSeparation2018}
Benjamin~M. Motz, Kathryn~A. Schlosser, and B.~Todd Heniford.
\newblock Chemical {{Components Separation}}: {{Concepts}}, {{Evidence}}, and
  {{Outcomes}}.
\newblock \emph{Plastic and Reconstructive Surgery}, 142\penalty0
  (3S):\penalty0 58S, September 2018.
\newblock ISSN 0032-1052.
\newblock \doi{10.1097/PRS.0000000000004856}.

\bibitem[{Whitehead-Clarke} and
  Windsor(2021)]{whitehead-clarkeUseBotulinumToxin2021}
Thomas {Whitehead-Clarke} and Alastair Windsor.
\newblock The {{Use}} of {{Botulinum Toxin}} in {{Complex Hernia Surgery}}:
  {{Achieving}} a {{Sense}} of {{Closure}}.
\newblock \emph{Frontiers in Surgery}, 8, October 2021.
\newblock ISSN 2296-875X.
\newblock \doi{10.3389/fsurg.2021.753889}.

\bibitem[Zhang et~al.(2016)Zhang, Liu, Tang, Sun, Ai, Yang, Jiang, and
  Zhang]{zhangEffectDifferentTypes2016}
Hua-Yu Zhang, Dong Liu, Hao Tang, Shi-Jin Sun, Shan-Mu Ai, Wen-Qun Yang,
  Dong-Po Jiang, and Lian-Yang Zhang.
\newblock The effect of different types of abdominal binders on intra-abdominal
  pressure.
\newblock \emph{Saudi Medical Journal}, 37\penalty0 (1):\penalty0 66--72,
  January 2016.
\newblock ISSN 0379-5284, 1658-3175.
\newblock \doi{10.15537/smj.2016.1.12865}.

\bibitem[S{\o}rensen et~al.(2005)S{\o}rensen, Hemmingsen, Kirkeby, Kallehave,
  and J{\o}rgensen]{sorensenSmokingRiskFactor2005}
Lars~Tue S{\o}rensen, Ulla~B. Hemmingsen, Lene~T. Kirkeby, Finn Kallehave, and
  Lars~Nannestad J{\o}rgensen.
\newblock Smoking {{Is}} a {{Risk Factor}} for {{Incisional Hernia}}.
\newblock \emph{Archives of Surgery}, 140\penalty0 (2):\penalty0 119--123,
  February 2005.
\newblock ISSN 0004-0010.
\newblock \doi{10.1001/archsurg.140.2.119}.

\bibitem[Loftus et~al.(2017)Loftus, Go, Jordan, Croft, Smith, Moore, Efron,
  Mohr, and Brakenridge]{loftusCTEvidenceFluid2017}
Tyler~J. Loftus, Kristina~L. Go, Janeen~R. Jordan, Chasen~A. Croft, R.~Stephen
  Smith, Frederick~A. Moore, Philip~A. Efron, Alicia~M. Mohr, and Scott~C.
  Brakenridge.
\newblock {{CT}} evidence of fluid in the hernia sac predicts surgical site
  infection following mesh repair of acutely incarcerated ventral and groin
  hernias.
\newblock \emph{The journal of trauma and acute care surgery}, 83\penalty0
  (1):\penalty0 170--174, July 2017.
\newblock ISSN 2163-0755.
\newblock \doi{10.1097/TA.0000000000001503}.

\bibitem[Pollock and Evans(1989)]{pollockEarlyPredictionLate1989}
Alan~V. Pollock and Mary Evans.
\newblock Early prediction of late incisional hernias.
\newblock \emph{The British Journal of Surgery}, 76\penalty0 (9):\penalty0
  953--954, September 1989.
\newblock ISSN 0007-1323.
\newblock \doi{10.1002/bjs.1800760926}.

\bibitem[K{\"o}hler et~al.(2015)K{\"o}hler, Luketina, and
  Emmanuel]{kohlerSuturedRepairPrimary2015}
Gernot K{\"o}hler, Ruzica-Rosalia Luketina, and Klaus Emmanuel.
\newblock Sutured {{Repair}} of {{Primary Small Umbilical}} and {{Epigastric
  Hernias}}: {{Concomitant Rectus Diastasis Is}} a {{Significant Risk Factor}}
  for {{Recurrence}}.
\newblock \emph{World Journal of Surgery}, 39\penalty0 (1):\penalty0 121--126,
  January 2015.
\newblock ISSN 1432-2323.
\newblock \doi{10.1007/s00268-014-2765-y}.

\bibitem[Elfanagely et~al.(2020)Elfanagely, Mellia, Othman, Basta, Mauch, and
  Fischer]{elfanagelyComputedTomographyImage2020}
Omar Elfanagely, Joseph~A. Mellia, Sammy Othman, Marten~N. Basta, Jaclyn~T.
  Mauch, and John~P. Fischer.
\newblock Computed {{Tomography Image Analysis}} in {{Abdominal Wall
  Reconstruction}}: {{A Systematic Review}}.
\newblock \emph{Plastic and Reconstructive Surgery Global Open}, 8\penalty0
  (12):\penalty0 e3307, December 2020.
\newblock ISSN 2169-7574.
\newblock \doi{10.1097/GOX.0000000000003307}.

\bibitem[Mitchell et~al.(2013)Mitchell, Lange, and Brus]{mitchell2013gendered}
Sara~McLaughlin Mitchell, Samantha Lange, and Holly Brus.
\newblock Gendered citation patterns in international relations journals.
\newblock \emph{International Studies Perspectives}, 14\penalty0 (4):\penalty0
  485--492, 2013.

\bibitem[Dion et~al.(2018)Dion, Sumner, and Mitchell]{dion2018gendered}
Michelle~L Dion, Jane~Lawrence Sumner, and Sara~McLaughlin Mitchell.
\newblock Gendered citation patterns across political science and social
  science methodology fields.
\newblock \emph{Political Analysis}, 26\penalty0 (3):\penalty0 312--327, 2018.

\bibitem[Caplar et~al.(2017)Caplar, Tacchella, and
  Birrer]{caplar2017quantitative}
Neven Caplar, Sandro Tacchella, and Simon Birrer.
\newblock Quantitative evaluation of gender bias in astronomical publications
  from citation counts.
\newblock \emph{Nature Astronomy}, 1\penalty0 (6):\penalty0 0141, 2017.

\bibitem[Maliniak et~al.(2013)Maliniak, Powers, and Walter]{maliniak2013gender}
Daniel Maliniak, Ryan Powers, and Barbara~F Walter.
\newblock The gender citation gap in international relations.
\newblock \emph{International Organization}, 67\penalty0 (4):\penalty0
  889--922, 2013.

\bibitem[Dworkin et~al.(2020)Dworkin, Linn, Teich, Zurn, Shinohara, and
  Bassett]{Dworkin2020.01.03.894378}
Jordan~D. Dworkin, Kristin~A. Linn, Erin~G. Teich, Perry Zurn, Russell~T.
  Shinohara, and Danielle~S. Bassett.
\newblock The extent and drivers of gender imbalance in neuroscience reference
  lists.
\newblock \emph{bioRxiv}, 2020.
\newblock \doi{10.1101/2020.01.03.894378}.
\newblock URL
  \url{https://www.biorxiv.org/content/early/2020/01/11/2020.01.03.894378}.

\bibitem[Bertolero et~al.(2020)Bertolero, Dworkin, David, Lloreda, Srivastava,
  Stiso, Zhou, Dzirasa, Fair, Kaczkurkin, Marlin, Shohamy, Uddin, Zurn, and
  Bassett]{bertolero2021racial}
Maxwell~A. Bertolero, Jordan~D. Dworkin, Sophia~U. David, Claudia~López
  Lloreda, Pragya Srivastava, Jennifer Stiso, Dale Zhou, Kafui Dzirasa,
  Damien~A. Fair, Antonia~N. Kaczkurkin, Bianca~Jones Marlin, Daphna Shohamy,
  Lucina~Q. Uddin, Perry Zurn, and Danielle~S. Bassett.
\newblock Racial and ethnic imbalance in neuroscience reference lists and
  intersections with gender.
\newblock \emph{bioRxiv}, 2020.

\bibitem[Wang et~al.(2021)Wang, Dworkin, Zhou, Stiso, Falk, Bassett, Zurn, and
  Lydon-Staley]{wang2021gendered}
Xinyi Wang, Jordan~D. Dworkin, Dale Zhou, Jennifer Stiso, Emily~B Falk,
  Danielle~S. Bassett, Perry Zurn, and David~M. Lydon-Staley.
\newblock Gendered citation practices in the field of communication.
\newblock \emph{Annals of the International Communication Association}, 2021.
\newblock \doi{10.1080/23808985.2021.1960180}.

\bibitem[Chatterjee and Werner(2021)]{chatterjee2021gender}
Paula Chatterjee and Rachel~M Werner.
\newblock Gender disparity in citations in high-impact journal articles.
\newblock \emph{JAMA Netw Open}, 4\penalty0 (7):\penalty0 e2114509, 2021.

\bibitem[Fulvio et~al.(2021)Fulvio, Akinnola, and Postle]{fulvio2021imbalance}
Jacqueline~M Fulvio, Ileri Akinnola, and Bradley~R Postle.
\newblock Gender (im)balance in citation practices in cognitive neuroscience.
\newblock \emph{J Cogn Neurosci}, 33\penalty0 (1):\penalty0 3--7, 2021.

\bibitem[Zhou et~al.(2020)Zhou, Cornblath, Stiso, Teich, Dworkin, Blevins, and
  Bassett]{zhou_dale_2020_3672110}
Dale Zhou, Eli~J. Cornblath, Jennifer Stiso, Erin~G. Teich, Jordan~D. Dworkin,
  Ann~S. Blevins, and Danielle~S. Bassett.
\newblock Gender diversity statement and code notebook v1.0, February 2020.
\newblock URL \url{https://doi.org/10.5281/zenodo.3672110}.

\bibitem[Ambekar et~al.(2009)Ambekar, Ward, Mohammed, Male, and
  Skiena]{ambekar2009name}
Anurag Ambekar, Charles Ward, Jahangir Mohammed, Swapna Male, and Steven
  Skiena.
\newblock Name-ethnicity classification from open sources.
\newblock In \emph{Proceedings of the 15th ACM SIGKDD international conference
  on Knowledge Discovery and Data Mining}, pages 49--58, 2009.

\bibitem[Sood and Laohaprapanon(2018)]{sood2018predicting}
Gaurav Sood and Suriyan Laohaprapanon.
\newblock Predicting race and ethnicity from the sequence of characters in a
  name.
\newblock \emph{arXiv preprint arXiv:1805.02109}, 2018.

\end{thebibliography}

\end{document}